\documentclass[useAMS,usenatbib]{mn2e}

\usepackage{graphicx,color}
\usepackage{amssymb,amsmath}
\usepackage{amsmath} %Alinhamento das equações
\usepackage{multicol}
\usepackage{latexsym}
\usepackage{amsfonts}
\usepackage{amssymb}
\usepackage{wasysym}
\usepackage{textcomp}
\usepackage{longtable,lscape}
\usepackage{rotating}
\usepackage{mathtools}
\usepackage{aas_macros}
\usepackage[dvipsnames]{xcolor}

\usepackage{natbib}
\usepackage{amssymb,amsmath}
\title[Dynamical states of Galactic OCs with \textit{Gaia} DR2]{Characterizing dynamical states of Galactic open clusters with \textit{Gaia} DR2}
\author[M. S. Angelo et al.]{M. S. Angelo$^{1}$\thanks{E-mail:
mateusangelo@cefetmg.br}, W. J. B. Corradi$^{2,3}$, J. F. C. Santos Jr.$^{2,4}$, F. F. S. Maia$^{5}$ \newauthor and F. A. Ferreira$^{2}$ \\ %\newauthor and  \\ \\ 
\noindent
$^1$Centro Federal de Educa\c{c}\~ao Tecnol\'ogica de Minas Gerais, Av. Monsenhor Luiz de Gonzaga, 103, 37250-000 Nepomuceno, MG, Brazil\\
$^2$Departamento de F\'isica, ICEx, Universidade Federal de Minas Gerais, Av. Ant\^onio Carlos 6627, 31270-901 Belo Horizonte, MG, Brazil\\
$^3$Laborat\'orio Nacional de Astrof\'isica, R. Estados Unidos 154, 37530-000 Itajub\'a, MG, Brazil\\
$^4$Departamento de Astronom\'ia, Universidad de La Serena, Av. Juan Cisternas 1200, La Serena, Chile\\
$^5$Universidade Federal do Rio de Janeiro, Instituto de F\'isica, 21941-972, Brazil}

\begin{document}

\date{Accepted XXX. Received XXX; in original form XXX}

\pagerange{\pageref{firstpage}--\pageref{lastpage}} \pubyear{XXXX}

\maketitle

\label{firstpage}

\begin{abstract}

In this work, we investigate the dynamical properties of 38 Galactic open clusters: 34
of them are located at low Galactic latitudes ($\vert b\vert < 10^{\circ}$) and are projected against dense stellar fields; the other 4 comparison objects present clearer contrasts with the field population. We determine structural and time-related parameters that are associated with the clusters' dynamical evolution: core ($r_c$), tidal ($r_t$) and half-mass ($r_{hm}$) radii, ages ($t$) and crossing times ($t_{cr}$). We have also incorporated results for 27 previously studied clusters, creating a sample of 65, spanning the age and Galactocentric distance ($R_G$) ranges: $7.0 < \textrm{log}\,t < 9.7$ and $6 < R_G (\textrm{kpc}) < 13$. We employ a uniform analysis method which incorporates photometric and astrometric data from the Gaia DR2 catalogue. Member stars are identified by employing a decontamination algorithm which operates on the 3D astrometric space of parallax and proper motion and attributes membership likelihoods for stars in the cluster region. Our results show that the internal relaxation causes $r_c$ to correlate negatively with the dynamical ratio $\tau_{\textrm{dyn}} = t/t_{cr}$. This implies that dynamically older systems tend to be more centrally concentrated. The more concentrated ones tend to present smaller $r_{hm}/r_t$ ratios, which means that they are less subject to tidal disruption. The analysis of coeval groups at compatible $R_G$ suggests that the inner structure of clusters is reasonably insensitive to variations in the external tidal field. Additionally, our results confirm, on average, an increase in $r_t$ for regions with less intense Galactic gravitational field.

\end{abstract}


\section{Introduction}

Investigating Galactic open clusters (OCs) is a fundamental task for a proper comprehension of the Milky Way structure and its evolution. Young OCs are important to investigate the intricate process of stellar formation (e.g., \citeauthor{Krumholz:2019}\,\,\citeyear{Krumholz:2019}) and also the recent disc history, while older ones allow to draw statements regarding the chemical, structural and kinematic evolution of the Milky Way (\citeauthor{Carraro:1994}\,\,\citeyear{Carraro:1994}; \citeauthor{Dias:2019}\,\,\citeyear{Dias:2019}). 

The OCs that survive the initial gas expulsion phase end up suffering mass loss due to: (a) stellar evolution (e.g., \citeauthor{Vink:2001}\,\,\citeyear{Vink:2001}; \citeauthor{Smith:2014}\,\,\citeyear{Smith:2014}), (b) internal interactions, which lead the system to energy equipartition and cause preferential evaporation of low-mass stars, in a process that is regulated by the external tidal field (\citeauthor{de-La-Fuente-Marcos:1997}\,\,\citeyear{de-La-Fuente-Marcos:1997}; \citeauthor{Portegies-Zwart:2010}\,\,\citeyear{Portegies-Zwart:2010}), (c) external interactions, such as disc shocking \citep{Ostriker:1972} and collisions with giant molecular clouds (\citeauthor{Spitzer:1958}\,\,\citeyear{Spitzer:1958}; \citeauthor{Theuns:1991}\,\,\citeyear{Theuns:1991}).

%It is widely accepted that most stars form in stellar aggregates embedded in molecular clouds and then gradually dissolve, with member stars sucessively becoming part of the general Galactic field population (e.g., \citeauthor{Lada:2003}\,\,\citeyear{Lada:2003}). 

The interplay among the above mentioned disruption processes lead to variations in the OCs structural  parameters, which can be employed as indicators of the evolutionary/dynamical states (e.g., \citeauthor{Piatti:2017a}\,\,\citeyear{Piatti:2017a} and references therein). Relations among parameters associated with the OCs evolution serve as observational constraints for theoretical studies aimed at detailing the physical processes that lead clusters to dissolution \citep{Bonatto:2004a}. 

%In this context, it is desirable to characterize large samples of OCs in different evolutionary stages, spanning wide age ranges and located at different Galactocentric distances ($R_G$). Ideally, such characterizations should be performed by employing uniform databases and analysis methods, in order to avoid possible biases among the studied objects. The availability of \textit{Gaia} DR2 catalogue \citep{Gaia-Collaboration:2018} allows this kind of investigation. The unprecedent high precision reached in astrometric and photometric data has allowed not only the discovery of new OCs (\citeauthor{Ryu:2018}\,\,\citeyear{Ryu:2018}; \citeauthor{Castro-Ginard:2018}\,\,\citeyear{Castro-Ginard:2018}; \citeauthor{Cantat-Gaudin:2018b}\,\,2018b, hereafter CJV2018; \citeauthor{Torrealba:2019}\,\,\citeyear{Torrealba:2019}; \citeauthor{Sim:2019}\,\,\citeyear{Sim:2019}; \citeauthor{Ferreira:2019}\,\,\citeyear{Ferreira:2019}), but also improvements in the lists of member stars of already known OCs (\citeauthor{Cantat-Gaudin:2018a}\,\,2018a), besides significant refinements on their fundamental astrophysical parameters (e.g., \citeauthor{Monteiro:2019}\,\,\citeyear{Monteiro:2019}; \citeauthor{Piatti:2019}\,\,\citeyear{Piatti:2019}).

In this context, it is desirable the characterization of large samples of OCs in different evolutionary stages, spanning wide age ranges and located at different Galactocentric distances ($R_G$). This is a growing need, given the increasing number of recently discovered OCs (\citeauthor{Ryu:2018}\,\,\citeyear{Ryu:2018}; \citeauthor{Castro-Ginard:2018}\,\,\citeyear{Castro-Ginard:2018}; \citeauthor{Cantat-Gaudin:2018b}\,\,2018b, hereafter CJV2018; \citeauthor{Torrealba:2019}\,\,\citeyear{Torrealba:2019}; \citeauthor{Sim:2019}\,\,\citeyear{Sim:2019}; \citeauthor{Liu:2019}\,\,\citeyear{Liu:2019}; \citeauthor{Ferreira:2019}\,\,\citeyear{Ferreira:2019}). Ideally, the characterization of OCs should be performed by employing uniform databases and analysis methods, in order to avoid possible biases among the studied objects. 

In \citeauthor{Angelo:2020}\,\,(\citeyear{Angelo:2020}, hereafter Paper I), we employed data from the \textit{Gaia} DR2 catalogue \citep{Gaia-Collaboration:2018} to investigate dynamical properties of a sample of 16 low-contrast OCs, complemented with other 11 comparison ones (see references therein). The observed trends among the derived parameters indicated a general disruption scenario in which OCs tend to be more centrally concentrated as they evolve dynamically, therefore being successively less subject to mass loss due to tidal effects. We also observed that the OCs' external structure is, in fact, influenced by the Galactic gravitational field since, on average, a positive correlation was identified between tidal radius and $R_G$.    

%The present work is a contribution towards enlarging the number of dynamically investigated systems. Our procedures are analogous to those employed in Paper I. In this paper, 38 Galactic OCs have been incorporated into our database, which now totalizes 65 objects. Among this set of 38 OCs explored here, 34 were previously characterized in the following series of papers: \cite{Bica:2004}, Bonatto \& Bica\,(\citeyear{Bonatto:2007}, \citeyear{Bonatto:2008}), Camargo et al. (\citeyear{Camargo:2009}, \citeyear{Camargo:2010}), \cite{Bonatto:2010}. For comparison purposes, other 4 OCs were included in the complementary sample, composed by well characterized objects presenting clearer contrast with the field population. This complementary sample contributes to enlarge the parameters space coverage comprised by our cluster sample and also to confirm the efficacy of our methods in dealing with low-contrast OCs. 

The present work is a contribution towards enlarging the number of dynamically investigated systems. Our procedures are analogous to those employed in Paper I. In the present paper, 38 Galactic OCs have been analysed and included in our database. Among this set of 38 OCs explored here, 34 were previously characterized in the following series of papers (hereafter BBC): \cite{Bica:2004}, Bonatto \& Bica\,(\citeyear{Bonatto:2007}, \citeyear{Bonatto:2008}), Camargo et al. (\citeyear{Camargo:2009}, \citeyear{Camargo:2010}), \cite{Bonatto:2010}. For comparison purposes, other 4 OCs presenting clearer contrast with the field population were included in the complementary sample. This complementary sample contributes to enlarge the parameters space coverage comprised by our cluster sample and also to confirm the efficacy of our methods in dealing with low-contrast OCs.

Our main goal is to explore relations among parameters associated to the clusters dynamical evolution, such as core ($r_c$), tidal ($r_t$) and half-mass radii ($r_{hm}$), age and crossing times ($t_\textrm{cr}$), and to draw some evolutionary connections. Here we incorporate the results obtained previously for 27 OCs and revisit the discussions presented in Paper I, but now with a more significant sample. Increasing the number of objects allowed us to establish better constrained relations and also a closer investigation of coeval OCs located at compatible $R_G$. The analysis of our complete database represents an intermediate stage in a long-term objective which is to shed light on the debated topic of clusters dissolution, the role of internal interactions and the influence of the Galactic tidal field on this process. 

In BBC, from which most of our sample is taken, the OCs are typically located at low Galactic latitudes ($\vert b\vert\lesssim10{\degr}$) and were uniformly analysed using 2MASS \citep{Skrutskie:2006} photometry after application of a decontamination algorithm to each object colour-magnitude diagram (CMD). However, since the clusters are projected against dense stellar fields and considerably affected by interstellar absorption, often the decontaminated sequences in their CMDs become somewhat dubious. As in Paper I, with the use of \textit{Gaia} DR2 photometry and astrometry, here we have been able to establish unambigously the physical nature of the investigated OCs, improving significantly the lists of member stars and thus providing a critical review of their fundamental parameters.

%The recent compilation by \cite{Bica:2019} lists star clusters, associations and candidates with 10978 entries. To date, about 3200 objects have been characterized.

%\citeauthor{Cantat-Gaudin:2018b}\,\,2018b, hereafter CJV2018

This paper is organized as follows: in Section~\ref{sec:data_description}, we describe our sample and the collected data. In Section~\ref{sec:method} we briefly describe our analysis method. Results are presented in Section~\ref{sec:results} and discussed in Section~\ref{sec:discussion}. In Section~\ref{sec:conclusions} we summarize the main conclusions.

%%%%%%%%%%%%%%%%%%MEUS INPUTS%%%%%%%%%%%%%%%%%%%
\section{Data and sample description}
\label{sec:data_description}

Table \ref{tab:investig_sample} lists the sample of 38 objects investigated in the present paper, organized in ascending order of right ascension. Thirty-four of them are part of our main sample, which includes OCs typically located at low Galactic latitudes ($\vert b\vert\lesssim10^{\circ}$) and presenting low-contrast with the general Galactic disc population. For reasons of clarity, in Table\,\ref{tab:previous_lit_information} we can find additional information regarding previous data taken from the literature. Four comparison clusters (namely, NGC\,5617, Pismis\,19, Trumpler\,22 and Dias\,6) were included in the complementary sample.

%For completeness, the 27 OCs investigated in Paper I (16 in the main sample and 11 in the complementary one) were compiled in the same Table.
%For reasons of clarity, in Table\,\ref{tab:previous_lit_information} we can find additional information regarding previous data taken from the literature and details of the samples in both Paper I and the present one.} 

\begin{table*}
\centering
\tiny
\rotcaption{ Coordinates, Galactocentric distances, structural and fundamental parameters, mean proper motion components and crossing times ($t_{\textrm{cr}}$; Section~\ref{sec:investig_struct_params}) for the studied sample. }
\begin{sideways}
\begin{minipage}{240mm}

\begin{tabular}{lllccccccccccccc}

 Cluster   &$\rmn{RA}$     &$\rmn{DEC}$    &$\ell$    &$b$    & R$_G^{(*)}$        & $r_c$    & $r_{hm}$    & $r_t$   & $(m-M)_0$  & $E(B-V)$  & log\,$t$  & $[Fe/H]$  & $\langle\mu_{\alpha}\,\textrm{cos}\,\delta\rangle$   & $\langle\mu_{\delta}\rangle$  & $t_{\textrm{cr}}$  \\
               &($\rmn{h}$:$\rmn{m}$:$\rmn{s}$) & ($\degr$:$\arcmin$:$\arcsec$)   &$(\rmn{^\circ})$  &$(\rmn{^\circ})$   &  (kpc)     &  (pc)     &     (pc)     &  (pc)  & (mag)   & (mag)  & (dex)   & (dex)  & (mas yr$^{-1}$) & (mas yr$^{-1}$)  & (Myr)   \\

\hline
\multicolumn{16}{c}{Main sample}  \\
\hline

Czernik\,7                     & 02:02:58   & +62:15:12   & 131.1   &  0.5   & 10.2\,$\pm$\,0.6   & 0.53\,$\pm$\,0.22   & 1.02\,$\pm$\,0.11   &  3.15\,$\pm$\,0.87   & 12.39\,$\pm$\,0.30   & 0.68\,$\pm$\,0.10   & 8.60\,$\pm$\,0.15   &  0.22\,$\pm$\,0.21   & -0.76\,$\pm$\,0.39   & -0.14\,$\pm$\,0.12   & 0.20\,$\pm$\,0.04  \\
Berkeley\,63                   & 02:19:25   & +63:42:42   & 132.5   &  2.5   & 11.2\,$\pm$\,0.6   & 1.46\,$\pm$\,0.36   & 3.00\,$\pm$\,0.39   & 10.43\,$\pm$\,3.03   & 13.10\,$\pm$\,0.30   & 0.97\,$\pm$\,0.10   & 7.90\,$\pm$\,0.20   &  0.05\,$\pm$\,0.30   & -0.98\,$\pm$\,0.13   &  0.20\,$\pm$\,0.14   & 0.89\,$\pm$\,0.13  \\
Czernik\,12                    & 02:39:20   & +54:54:51   & 138.1   & -4.7   &  9.4\,$\pm$\,0.5   & 0.76\,$\pm$\,0.10   & 1.31\,$\pm$\,0.10   &  3.54\,$\pm$\,0.76   & 11.20\,$\pm$\,0.30   & 0.41\,$\pm$\,0.05   & 9.05\,$\pm$\,0.10   & -0.13\,$\pm$\,0.16   & -0.22\,$\pm$\,0.07   &  0.91\,$\pm$\,0.04   & 2.76\,$\pm$\,0.53  \\
Czernik\,22                    & 05:49:02   & +30:12:14   & 179.3   &  1.3   & 11.0\,$\pm$\,0.6   & 3.18\,$\pm$\,0.52   & 3.60\,$\pm$\,0.39   &  5.24\,$\pm$\,0.52   & 12.35\,$\pm$\,0.20   & 0.70\,$\pm$\,0.07   & 8.00\,$\pm$\,0.10   & -0.06\,$\pm$\,0.26   &  0.82\,$\pm$\,0.12   & -1.81\,$\pm$\,0.11   & 1.75\,$\pm$\,0.24  \\
Czernik\,23                    & 05:50:06   & +28:53:15   & 180.5   &  0.8   & 10.6\,$\pm$\,0.6   & 1.53\,$\pm$\,0.27   & 2.14\,$\pm$\,0.25   &  4.44\,$\pm$\,0.99   & 12.10\,$\pm$\,0.30   & 0.43\,$\pm$\,0.10   & 8.95\,$\pm$\,0.10   & -0.13\,$\pm$\,0.23   &  0.08\,$\pm$\,0.18   & -1.76\,$\pm$\,0.17   & 0.80\,$\pm$\,0.11  \\
Czernik\,24                    & 05:55:25   & +20:53:11   & 188.0   & -2.3   & 11.4\,$\pm$\,0.7   & 1.61\,$\pm$\,0.25   & 2.62\,$\pm$\,0.26   &  6.46\,$\pm$\,1.01   & 12.70\,$\pm$\,0.30   & 0.69\,$\pm$\,0.07   & 9.25\,$\pm$\,0.15   & -0.22\,$\pm$\,0.28   &  0.28\,$\pm$\,0.20   & -2.67\,$\pm$\,0.24   & 0.50\,$\pm$\,0.06  \\
Ruprecht\,1                    & 06:36:20   & -14:09:04   & 224.0   & -9.7   &  9.0\,$\pm$\,0.5   & 0.42\,$\pm$\,0.10   & 0.82\,$\pm$\,0.12   &  3.07\,$\pm$\,1.15   & 10.60\,$\pm$\,0.30   & 0.12\,$\pm$\,0.05   & 8.85\,$\pm$\,0.10   &  0.00\,$\pm$\,0.17   & -0.31\,$\pm$\,0.12   & -0.88\,$\pm$\,0.15   & 0.82\,$\pm$\,0.15  \\
Ruprecht\,10                   & 07:06:25   & -20:07:08   & 232.5   & -5.8   &  9.5\,$\pm$\,0.5   & 1.50\,$\pm$\,0.33   & 1.93\,$\pm$\,0.21   &  3.32\,$\pm$\,0.65   & 11.75\,$\pm$\,0.35   & 0.35\,$\pm$\,0.10   & 8.30\,$\pm$\,0.25   &  0.10\,$\pm$\,0.23   & -0.85\,$\pm$\,0.10   &  0.59\,$\pm$\,0.11   & 1.47\,$\pm$\,0.26  \\
%Haffner\,9                     & 07:24:43   & -17:00:00   & 231.8   & -0.6   &  9.9\,$\pm$\,0.6   & 2.19\,$\pm$\,0.31   & 3.60\,$\pm$\,0.25   &  9.07\,$\pm$\,1.54   & 12.11\,$\pm$\,0.40   & 0.80\,$\pm$\,0.10   & 8.55\,$\pm$\,0.15   & -0.13\,$\pm$\,0.23   & -0.94\,$\pm$\,0.23   &  0.66\,$\pm$\,0.44   & 0.87\,$\pm$\,0.09  \\
Ruprecht\,23                   & 07:30:37   & -23:22:41   & 238.1   & -2.4   &  9.3\,$\pm$\,0.5   & 1.34\,$\pm$\,0.24   & 2.02\,$\pm$\,0.24   &  4.44\,$\pm$\,0.91   & 11.60\,$\pm$\,0.30   & 0.63\,$\pm$\,0.05   & 8.80\,$\pm$\,0.10   & -0.06\,$\pm$\,0.20   & -2.07\,$\pm$\,0.08   &  1.59\,$\pm$\,0.09   & 1.69\,$\pm$\,0.24  \\
%%%CEFET-MG1$^{a,\dag\dag}$      & 07:30:42   & -23:08:36   & 237.9   & -2.3   &  9.6\,$\pm$\,0.6   & 1.43\,$\pm$\,0.36   & 1.85\,$\pm$\,0.28   &  3.21\,$\pm$\,0.44   & 12.00\,$\pm$\,0.50   & 0.65\,$\pm$\,0.15   & 8.80\,$\pm$\,0.20   &  0.00\,$\pm$\,0.29   & -1.68\,$\pm$\,0.12   &  1.73\,$\pm$\,0.09   & 1.44\,$\pm$\,0.34  \\
%Haffner\,11                    & 07:35:22   & -27:42:03   & 242.4   & -3.5   & 11.1\,$\pm$\,0.7   & 3.85\,$\pm$\,0.57   & 5.46\,$\pm$\,0.56   & 11.54\,$\pm$\,2.14   & 13.45\,$\pm$\,0.40   & 0.59\,$\pm$\,0.10   & 8.90\,$\pm$\,0.10   &  0.00\,$\pm$\,0.17   & -1.49\,$\pm$\,0.18   &  3.17\,$\pm$\,0.18   & 1.07\,$\pm$\,0.12  \\
%NGC\,2421$^a$                  & 07:36:12   & -20:37:48   & 236.3   &  0.1   &  9.6\,$\pm$\,0.5   & 3.43\,$\pm$\,0.50   & 4.64\,$\pm$\,0.46   &  8.28\,$\pm$\,1.43   & 11.95\,$\pm$\,0.30   & 0.55\,$\pm$\,0.10   & 7.75\,$\pm$\,0.30   & -0.13\,$\pm$\,0.23   & -3.15\,$\pm$\,0.11   &  3.09\,$\pm$\,0.11   & 2.75\,$\pm$\,0.29  \\   
%Czernik\,31                    & 07:37:00   & -20:31:05   & 236.2   &  0.2   &  9.9\,$\pm$\,0.5   & 1.76\,$\pm$\,0.29   & 2.35\,$\pm$\,0.27   &  4.20\,$\pm$\,0.67   & 12.30\,$\pm$\,0.30   & 0.60\,$\pm$\,0.10   & 8.00\,$\pm$\,0.40   &  0.05\,$\pm$\,0.20   & -1.95\,$\pm$\,0.14   &  3.03\,$\pm$\,0.13   & 0.98\,$\pm$\,0.14  \\
L19\,2326$^a$                  & 07:37:01   & -15:35:24   & 232.0   &  2.7   &  8.6\,$\pm$\,0.5   & 0.83\,$\pm$\,0.10   & 1.30\,$\pm$\,0.11   &  3.06\,$\pm$\,0.56   &  9.90\,$\pm$\,0.30   & 0.10\,$\pm$\,0.10   & 7.55\,$\pm$\,0.30   &  0.00\,$\pm$\,0.29   & -3.19\,$\pm$\,0.10   &  0.07\,$\pm$\,0.10   & 2.36\,$\pm$\,0.32  \\
Ruprecht\,26                   & 07:37:09   & -15:41:40   & 232.1   &  2.7   &  9.7\,$\pm$\,0.5   & 2.63\,$\pm$\,0.44   & 3.33\,$\pm$\,0.28   &  5.41\,$\pm$\,0.58   & 12.00\,$\pm$\,0.30   & 0.46\,$\pm$\,0.05   & 8.90\,$\pm$\,0.10   & -0.06\,$\pm$\,0.20   & -1.24\,$\pm$\,0.14   &  2.79\,$\pm$\,0.13   & 1.53\,$\pm$\,0.17  \\
Ruprecht\,27                   & 07:37:39   & -26:30:43   & 241.6   & -2.5   &  8.9\,$\pm$\,0.5   & 1.06\,$\pm$\,0.14   & 1.76\,$\pm$\,0.16   &  4.44\,$\pm$\,0.82   & 11.10\,$\pm$\,0.30   & 0.35\,$\pm$\,0.05   & 8.85\,$\pm$\,0.10   & -0.13\,$\pm$\,0.23   & -1.24\,$\pm$\,0.09   &  0.45\,$\pm$\,0.14   & 1.74\,$\pm$\,0.19  \\
%Ruprecht\,30                   & 07:42:32   & -31:27:20   & 246.4   & -4.0   & 10.1\,$\pm$\,0.6   & 1.66\,$\pm$\,0.33   & 2.44\,$\pm$\,0.29   &  5.53\,$\pm$\,1.11   & 12.90\,$\pm$\,0.40   & 0.63\,$\pm$\,0.10   & 7.50\,$\pm$\,0.20   &  0.00\,$\pm$\,0.29   & -2.03\,$\pm$\,0.19   &  3.09\,$\pm$\,0.26   & 0.62\,$\pm$\,0.18  \\
Ruprecht\,34                   & 07:45:57   & -20:22:42   & 237.2   &  2.2   &  9.7\,$\pm$\,0.5   & 0.86\,$\pm$\,0.16   & 1.52\,$\pm$\,0.20   &  3.98\,$\pm$\,1.17   & 12.14\,$\pm$\,0.30   & 0.42\,$\pm$\,0.10   & 8.25\,$\pm$\,0.15   & -0.13\,$\pm$\,0.16   & -2.16\,$\pm$\,0.15   &  1.03\,$\pm$\,0.12   & 0.75\,$\pm$\,0.12  \\
Ruprecht\,35                   & 07:46:16   & -31:16:49   & 246.7   & -3.3   & 10.0\,$\pm$\,0.5   & 1.07\,$\pm$\,0.32   & 1.95\,$\pm$\,0.35   &  5.67\,$\pm$\,1.61   & 12.83\,$\pm$\,0.30   & 0.83\,$\pm$\,0.10   & 7.40\,$\pm$\,0.25   &  0.00\,$\pm$\,0.29   & -2.26\,$\pm$\,0.12   &  3.13\,$\pm$\,0.14   & 0.62\,$\pm$\,0.12  \\
%Herschel\,1$^\dag$             & 07:47:04   & +00:01:16   & 219.4   & 12.4   &  8.2\,$\pm$\,0.5   &       $-$           &       $-$           &  0.86\,$\pm$\,0.09   &  7.35\,$\pm$\,0.40   & 0.06\,$\pm$\,0.15   & 8.65\,$\pm$\,0.65   &  0.00\,$\pm$\,0.23   &  0.57\,$\pm$\,0.27   & -4.15\,$\pm$\,0.19   &       $-$          \\
Ruprecht\,37                   & 07:49:49   & -17:15:02   & 234.9   &  4.5   & 11.2\,$\pm$\,0.5   & 1.43\,$\pm$\,0.39   & 2.47\,$\pm$\,0.34   &  6.89\,$\pm$\,1.95   & 13.25\,$\pm$\,0.20   & 0.25\,$\pm$\,0.05   & 9.40\,$\pm$\,0.05   & -0.47\,$\pm$\,0.17   & -1.72\,$\pm$\,0.11   &  2.40\,$\pm$\,0.11   & 0.79\,$\pm$\,0.12  \\
%Czernik\,32                    & 07:50:31   & -29:51:12   & 245.9   & -1.7   &  9.7\,$\pm$\,0.5   & 2.02\,$\pm$\,0.28   & 3.47\,$\pm$\,0.36   & 10.39\,$\pm$\,1.10   & 12.50\,$\pm$\,0.15   & 0.85\,$\pm$\,0.05   & 9.10\,$\pm$\,0.10   & -0.13\,$\pm$\,0.23   & -2.95\,$\pm$\,0.14   &  2.49\,$\pm$\,0.12   & 1.35\,$\pm$\,0.15  \\
%$[\textrm{FSR}2007]$\,1325$^a$ & 07:50:33   & -29:57:15   & 245.9   & -1.7   & 10.7\,$\pm$\,0.6   & 2.66\,$\pm$\,0.40   & 4.32\,$\pm$\,0.43   & 11.44\,$\pm$\,1.99   & 13.30\,$\pm$\,0.40   & 1.14\,$\pm$\,0.10   & 7.00\,$\pm$\,0.20   &  0.22\,$\pm$\,0.17   & -2.72\,$\pm$\,0.07   &  3.44\,$\pm$\,0.08   & 1.97\,$\pm$\,0.41  \\
NGC\,2477$^a$                  & 07:52:18   & -38:31:48   & 253.6   & -5.8   &  8.5\,$\pm$\,0.5   & 2.29\,$\pm$\,0.42   & 4.42\,$\pm$\,0.40   & 17.60\,$\pm$\,2.29   & 10.59\,$\pm$\,0.40   & 0.40\,$\pm$\,0.10   & 9.05\,$\pm$\,0.15   & -0.13\,$\pm$\,0.23   & -2.46\,$\pm$\,0.09   &  0.87\,$\pm$\,0.10   & 6.45\,$\pm$\,0.64  \\
Ruprecht\,41                   & 07:53:46   & -26:58:09   & 243.8   &  0.4   &  9.6\,$\pm$\,0.5   & 2.54\,$\pm$\,0.41   & 3.25\,$\pm$\,0.37   &  5.08\,$\pm$\,0.82   & 12.25\,$\pm$\,0.30   & 0.18\,$\pm$\,0.05   & 9.00\,$\pm$\,0.10   &  0.05\,$\pm$\,0.20   & -2.52\,$\pm$\,0.09   &  3.55\,$\pm$\,0.11   & 1.46\,$\pm$\,0.20  \\
Ruprecht\,152                  & 07:54:28   & -38:14:14   & 253.5   & -5.3   & 13.0\,$\pm$\,0.8   & 2.15\,$\pm$\,0.48   & 3.57\,$\pm$\,0.37   &  9.78\,$\pm$\,2.15   & 14.57\,$\pm$\,0.30   & 0.67\,$\pm$\,0.10   & 8.75\,$\pm$\,0.10   & -0.22\,$\pm$\,0.28   & -1.31\,$\pm$\,0.11   &  2.22\,$\pm$\,0.13   & 0.60\,$\pm$\,0.10  \\
Ruprecht\,54                   & 08:11:23   & -31:56:15   & 250.0   &  1.0   & 10.1\,$\pm$\,0.5   & 1.54\,$\pm$\,0.30   & 2.77\,$\pm$\,0.31   &  8.77\,$\pm$\,1.78   & 13.05\,$\pm$\,0.30   & 0.45\,$\pm$\,0.10   & 8.75\,$\pm$\,0.10   & -0.06\,$\pm$\,0.20   & -2.62\,$\pm$\,0.16   &  3.11\,$\pm$\,0.12   & 0.75\,$\pm$\,0.09  \\
Ruprecht\,60                   & 08:24:27   & -47:12:51   & 264.1   & -5.5   &  9.9\,$\pm$\,0.6   & 2.33\,$\pm$\,0.36   & 3.98\,$\pm$\,0.47   &  9.91\,$\pm$\,1.46   & 13.50\,$\pm$\,0.30   & 0.65\,$\pm$\,0.05   & 8.65\,$\pm$\,0.10   &  0.14\,$\pm$\,0.16   & -3.77\,$\pm$\,0.11   &  5.38\,$\pm$\,0.15   & 0.94\,$\pm$\,0.12  \\
Ruprecht\,63                   & 08:32:40   & -48:18:05   & 265.8   & -5.0   &  9.1\,$\pm$\,0.5   & 1.71\,$\pm$\,0.22   & 3.31\,$\pm$\,0.29   & 13.49\,$\pm$\,3.87   & 12.90\,$\pm$\,0.30   & 0.61\,$\pm$\,0.01   & 8.60\,$\pm$\,0.10   & -0.06\,$\pm$\,0.20   & -2.51\,$\pm$\,0.11   &  3.28\,$\pm$\,0.13   & 1.15\,$\pm$\,0.12  \\
Ruprecht\,66                   & 08:40:34   & -38:05:03   & 258.5   &  2.3   &  9.3\,$\pm$\,0.5   & 1.51\,$\pm$\,0.25   & 2.56\,$\pm$\,0.26   &  7.16\,$\pm$\,1.51   & 12.70\,$\pm$\,0.20   & 0.60\,$\pm$\,0.05   & 9.15\,$\pm$\,0.05   & -0.22\,$\pm$\,0.19   & -3.09\,$\pm$\,0.16   &  3.07\,$\pm$\,0.16   & 0.70\,$\pm$\,0.08  \\
%Pismis\,12                     & 09:20:02   & -45:06:54   & 268.6   &  3.2   &  8.3\,$\pm$\,0.5   & 1.15\,$\pm$\,0.14   & 2.02\,$\pm$\,0.15   &  5.92\,$\pm$\,1.44   & 11.48\,$\pm$\,0.40   & 0.55\,$\pm$\,0.05   & 9.20\,$\pm$\,0.10   &  0.00\,$\pm$\,0.17   & -6.68\,$\pm$\,0.21   &  4.80\,$\pm$\,0.17   & 0.93\,$\pm$\,0.11  \\
%Trumpler\,13                   & 10:23:50   & -60:07:41   & 285.5   & -2.4   &  8.0\,$\pm$\,0.6   & 2.30\,$\pm$\,0.36   & 3.78\,$\pm$\,0.47   &  9.70\,$\pm$\,1.82   & 13.10\,$\pm$\,0.40   & 0.63\,$\pm$\,0.10   & 8.10\,$\pm$\,0.10   & -0.06\,$\pm$\,0.13   & -6.74\,$\pm$\,0.03   &  4.10\,$\pm$\,0.09   & 3.32\,$\pm$\,1.26  \\
%NGC\,3519                      & 11:04:15   & -61:24:01   & 290.4   & -1.1   &  7.6\,$\pm$\,0.5   & 0.59\,$\pm$\,0.15   & 1.03\,$\pm$\,0.16   &  2.96\,$\pm$\,0.99   & 11.15\,$\pm$\,0.40   & 0.17\,$\pm$\,0.10   & 8.70\,$\pm$\,0.10   &  0.18\,$\pm$\,0.15   & -6.44\,$\pm$\,0.08   &  3.12\,$\pm$\,0.07   & 1.24\,$\pm$\,0.27  \\
%Lynga\,2$^\dag$                & 14:24:14   & -61:21:30   & 313.8   & -0.4   &  7.4\,$\pm$\,0.5   &       $-$           &       $-$           &  2.92\,$\pm$\,0.42   &  9.90\,$\pm$\,0.30   & 0.40\,$\pm$\,0.10   & 7.90\,$\pm$\,0.30   & -0.13\,$\pm$\,0.23   & -6.68\,$\pm$\,0.09   & -4.68\,$\pm$\,0.10   &       $-$          \\
UBC\,296$^a$                   & 14:24:36   & -61:03:51   & 313.8   & -0.4   &  6.8\,$\pm$\,0.6   & 1.84\,$\pm$\,0.46   & 2.69\,$\pm$\,0.37   &  6.04\,$\pm$\,1.73   & 11.48\,$\pm$\,0.40   & 0.63\,$\pm$\,0.10   & 8.00\,$\pm$\,0.30   &  0.00\,$\pm$\,0.23   & -3.72\,$\pm$\,0.14   & -2.24\,$\pm$\,0.15   & 1.77\,$\pm$\,0.27  \\                      
NGC\,5715                      & 14:43:27   & -57:33:49   & 317.5   &  2.1   &  7.0\,$\pm$\,0.5   & 0.83\,$\pm$\,0.10   & 1.35\,$\pm$\,0.14   &  3.40\,$\pm$\,0.62   & 10.77\,$\pm$\,0.30   & 0.60\,$\pm$\,0.05   & 8.90\,$\pm$\,0.10   &  0.05\,$\pm$\,0.15   & -3.49\,$\pm$\,0.13   & -2.29\,$\pm$\,0.07   & 1.16\,$\pm$\,0.13  \\
Lynga\,4                       & 15:33:19   & -55:14:11   & 324.7   &  0.7   &  6.5\,$\pm$\,0.5   & 0.74\,$\pm$\,0.11   & 1.33\,$\pm$\,0.15   &  4.20\,$\pm$\,1.02   & 11.45\,$\pm$\,0.30   & 1.78\,$\pm$\,0.10   & 8.05\,$\pm$\,0.10   &  0.14\,$\pm$\,0.20   & -4.02\,$\pm$\,0.26   & -2.99\,$\pm$\,0.23   & 0.46\,$\pm$\,0.06  \\
Trumpler\,23                   & 16:00:54   & -53:32:31   & 328.9   & -0.5   &  6.6\,$\pm$\,0.5   & 1.26\,$\pm$\,0.13   & 1.94\,$\pm$\,0.16   &  4.75\,$\pm$\,0.76   & 11.20\,$\pm$\,0.30   & 0.75\,$\pm$\,0.05   & 9.00\,$\pm$\,0.10   &  0.05\,$\pm$\,0.15   & -4.20\,$\pm$\,0.17   & -4.72\,$\pm$\,0.08   & 1.19\,$\pm$\,0.11  \\
Lynga\,9                       & 16:20:43   & -48:32:14   & 334.6   &  1.1   &  6.5\,$\pm$\,0.5   & 1.06\,$\pm$\,0.13   & 1.84\,$\pm$\,0.13   &  5.26\,$\pm$\,0.76   & 11.20\,$\pm$\,0.30   & 0.99\,$\pm$\,0.10   & 9.00\,$\pm$\,0.10   & -0.13\,$\pm$\,0.23   & -2.54\,$\pm$\,0.18   & -2.44\,$\pm$\,0.11   & 1.22\,$\pm$\,0.10  \\
%Lynga\,12                      & 16:46:08   & -50:45:04   & 335.7   & -3.5   &  6.2\,$\pm$\,0.6   & 1.03\,$\pm$\,0.24   & 1.66\,$\pm$\,0.24   &  4.13\,$\pm$\,1.22   & 11.60\,$\pm$\,0.40   & 0.88\,$\pm$\,0.15   & 8.70\,$\pm$\,0.15   &  0.00\,$\pm$\,0.29   & -2.32\,$\pm$\,1.00   & -3.75\,$\pm$\,0.85   & 0.12\,$\pm$\,0.02  \\
Trumpler\,26                   & 17:28:36   & -29:29:33   & 357.5   &  2.8   &  6.8\,$\pm$\,0.5   & 1.05\,$\pm$\,0.22   & 1.55\,$\pm$\,0.19   &  3.25\,$\pm$\,0.72   & 10.47\,$\pm$\,0.20   & 0.73\,$\pm$\,0.10   & 8.30\,$\pm$\,0.20   &  0.00\,$\pm$\,0.17   & -0.88\,$\pm$\,0.09   & -3.10\,$\pm$\,0.09   & 2.16\,$\pm$\,0.37  \\
%Ruprecht\,130                  & 17:47:35   & -30:05:12   & 359.2   & -1.0   &  6.0\,$\pm$\,0.6   & 1.04\,$\pm$\,0.23   & 1.43\,$\pm$\,0.23   &  3.08\,$\pm$\,0.41   & 11.50\,$\pm$\,0.40   & 1.22\,$\pm$\,0.10   & 8.40\,$\pm$\,0.10   &  0.14\,$\pm$\,0.20   &  0.41\,$\pm$\,0.22   & -1.84\,$\pm$\,0.06   & 0.57\,$\pm$\,0.10  \\
Bica\,3                        & 18:26:05   & -13:02:54   &  18.4   & -0.4   &  6.1\,$\pm$\,0.7   & 0.64\,$\pm$\,0.14   & 1.19\,$\pm$\,0.11   &  3.60\,$\pm$\,0.76   & 11.50\,$\pm$\,0.50   & 2.33\,$\pm$\,0.15   & 7.45\,$\pm$\,0.20   &  0.28\,$\pm$\,0.18   & -0.50\,$\pm$\,0.17   & -1.55\,$\pm$\,0.18   & 0.55\,$\pm$\,0.07  \\
Ruprecht\,144                  & 18:33:34   & -11:25:22   &  20.7   & -1.3   &  6.8\,$\pm$\,0.5   & 0.59\,$\pm$\,0.11   & 0.93\,$\pm$\,0.10   &  2.09\,$\pm$\,0.55   & 10.50\,$\pm$\,0.30   & 0.77\,$\pm$\,0.15   & 8.10\,$\pm$\,0.40   &  0.00\,$\pm$\,0.23   &  0.19\,$\pm$\,0.14   & -0.91\,$\pm$\,0.13   & 0.94\,$\pm$\,0.14  \\
$[\textrm{FSR}2007]$\,0101     & 18:49:19   & +02:46:06   &  35.1   &  1.8   &  6.5\,$\pm$\,0.6   & 1.11\,$\pm$\,0.19   & 1.64\,$\pm$\,0.18   &  3.66\,$\pm$\,0.55   & 11.40\,$\pm$\,0.40   & 2.40\,$\pm$\,0.15   & 8.70\,$\pm$\,0.15   &  0.05\,$\pm$\,0.30   & -1.88\,$\pm$\,0.28   & -2.49\,$\pm$\,0.22   & 0.53\,$\pm$\,0.10  \\
Berkeley\,84                   & 20:04:39   & +33:54:26   &  70.9   &  1.3   &  7.6\,$\pm$\,0.5   & 0.83\,$\pm$\,0.18   & 1.47\,$\pm$\,0.19   &  4.12\,$\pm$\,1.19   & 11.56\,$\pm$\,0.30   & 0.76\,$\pm$\,0.10   & 8.55\,$\pm$\,0.15   &  0.00\,$\pm$\,0.29   & -1.98\,$\pm$\,0.11   & -5.53\,$\pm$\,0.12   & 1.06\,$\pm$\,0.27  \\
Ruprecht\,172                  & 20:11:36   & +35:36:18   &  73.1   &  1.0   &  7.7\,$\pm$\,0.5   & 1.05\,$\pm$\,0.21   & 2.00\,$\pm$\,0.27   &  6.70\,$\pm$\,1.40   & 11.90\,$\pm$\,0.30   & 0.63\,$\pm$\,0.05   & 9.15\,$\pm$\,0.10   & -0.32\,$\pm$\,0.24   & -2.05\,$\pm$\,0.04   & -3.65\,$\pm$\,0.09   & 1.56\,$\pm$\,0.25  \\
Ruprecht\,174                  & 20:43:29   & +37:01:00   &  78.0   & -3.4   &  7.8\,$\pm$\,0.5   & 0.97\,$\pm$\,0.23   & 1.48\,$\pm$\,0.18   &  3.41\,$\pm$\,0.69   & 11.00\,$\pm$\,0.30   & 0.63\,$\pm$\,0.10   & 8.80\,$\pm$\,0.15   & -0.32\,$\pm$\,0.24   & -3.17\,$\pm$\,0.11   & -4.69\,$\pm$\,0.13   & 1.17\,$\pm$\,0.16  \\
%IC\,1434                       & 22:10:29   & +52:51:09   &  99.9   & -2.7   &  9.0\,$\pm$\,0.5   & 2.05\,$\pm$\,0.45   & 3.02\,$\pm$\,0.29   &  6.51\,$\pm$\,1.34   & 12.43\,$\pm$\,0.30   & 0.49\,$\pm$\,0.10   & 8.50\,$\pm$\,0.10   & -0.06\,$\pm$\,0.20   & -3.90\,$\pm$\,0.04   & -3.35\,$\pm$\,0.16   & 1.07\,$\pm$\,0.11  \\
%King\,20                       & 23:33:17   & +58:28:33   & 112.8   & -2.9   &  8.8\,$\pm$\,0.5   & 2.11\,$\pm$\,0.35   & 3.21\,$\pm$\,0.33   &  7.30\,$\pm$\,1.26   & 11.19\,$\pm$\,0.40   & 0.89\,$\pm$\,0.10   & 8.20\,$\pm$\,0.20   &  0.28\,$\pm$\,0.18   & -2.67\,$\pm$\,0.13   & -2.63\,$\pm$\,0.16   & 2.06\,$\pm$\,0.24  \\

\hline
%\multicolumn{16}{c}{Complementary sample$^{(**)}$}  \\
   \multicolumn{16}{c}{Complementary sample}  \\
\hline

NGC\,5617          & 14:29:44   & -60:42:42   & 314.7   & -0.1   &  6.9\,$\pm$\,0.5   & 3.50\,$\pm$\,0.36   & 3.92\,$\pm$\,0.34   &  5.46\,$\pm$\,0.52   & 11.24\,$\pm$\,0.30   & 0.62\,$\pm$\,0.10   & 8.30\,$\pm$\,0.15   & -0.13\,$\pm$\,0.31   & -5.64\,$\pm$\,0.10   & -3.18\,$\pm$\,0.16   & 4.65\,$\pm$\,0.58  \\
Pismis\,19         & 14:30:40   & -60:53:30   & 314.7   & -0.3   &  6.8\,$\pm$\,0.5   & 0.57\,$\pm$\,0.12   & 1.05\,$\pm$\,0.15   &  3.56\,$\pm$\,0.57   & 11.48\,$\pm$\,0.30   & 1.40\,$\pm$\,0.10   & 9.00\,$\pm$\,0.10   & -0.06\,$\pm$\,0.26   & -5.46\,$\pm$\,0.05   & -3.25\,$\pm$\,0.20   & 0.57\,$\pm$\,0.08  \\
Trumpler\,22       & 14:31:13   & -61:10:02   & 314.6   & -0.6   &  6.9\,$\pm$\,0.5   & 3.00\,$\pm$\,0.49   & 3.64\,$\pm$\,0.38   &  5.31\,$\pm$\,0.49   & 11.14\,$\pm$\,0.30   & 0.67\,$\pm$\,0.10   & 8.20\,$\pm$\,0.10   & -0.22\,$\pm$\,0.28   & -5.09\,$\pm$\,0.09   & -2.68\,$\pm$\,0.15   & 5.38\,$\pm$\,1.07  \\
%NGC\,6216          & 16:49:24   & -44:43:17   & 340.7   &  0.0   &  6.1\,$\pm$\,0.7   & 1.82\,$\pm$\,0.27   & 2.73\,$\pm$\,0.24   &  6.08\,$\pm$\,0.73   & 11.60\,$\pm$\,0.50   & 1.04\,$\pm$\,0.10   & 7.85\,$\pm$\,0.30   &  0.00\,$\pm$\,0.23   & -1.26\,$\pm$\,0.21   & -2.53\,$\pm$\,0.14   & 1.48\,$\pm$\,0.14  \\
%BH\,200            & 16:49:55   & -44:11:15   & 341.1   &  0.2   &  6.2\,$\pm$\,0.5   & 1.08\,$\pm$\,0.17   & 1.79\,$\pm$\,0.22   &  4.82\,$\pm$\,0.85   & 11.45\,$\pm$\,0.20   & 1.00\,$\pm$\,0.10   & 7.90\,$\pm$\,0.30   &  0.00\,$\pm$\,0.23   & -0.07\,$\pm$\,0.18   & -1.18\,$\pm$\,0.13   & 1.19\,$\pm$\,0.17  \\
%Collinder\,351     & 17:49:02   & -28:44:27   &   0.6   & -0.5   &  6.3\,$\pm$\,0.6   & 0.59\,$\pm$\,0.15   & 0.96\,$\pm$\,0.13   &  2.62\,$\pm$\,0.74   & 11.15\,$\pm$\,0.40   & 0.87\,$\pm$\,0.15   & 8.85\,$\pm$\,0.10   & -0.06\,$\pm$\,0.33   & -0.24\,$\pm$\,0.16   & -1.70\,$\pm$\,0.56   & 0.18\,$\pm$\,0.03  \\
%Czernik\,37        & 17:53:17   & -27:22:10   &   2.2   & -0.6   &  6.6\,$\pm$\,0.6   & 0.57\,$\pm$\,0.08   & 1.20\,$\pm$\,0.11   &  5.34\,$\pm$\,1.64   & 10.75\,$\pm$\,0.50   & 1.50\,$\pm$\,0.10   & 8.70\,$\pm$\,0.15   &  0.00\,$\pm$\,0.29   &  0.43\,$\pm$\,0.20   & -0.45\,$\pm$\,0.14   & 0.81\,$\pm$\,0.09  \\
Dias\,6            & 18:30:27   & -12:20:01   &  19.6   & -1.0   &  6.1\,$\pm$\,0.6   & 0.65\,$\pm$\,0.12   & 1.31\,$\pm$\,0.15   &  5.05\,$\pm$\,1.49   & 11.55\,$\pm$\,0.30   & 0.99\,$\pm$\,0.05   & 8.85\,$\pm$\,0.10   &  0.00\,$\pm$\,0.17   &  0.47\,$\pm$\,0.14   & -0.57\,$\pm$\,0.06   & 0.79\,$\pm$\,0.10  \\

\hline
\end{tabular}

%$^a$ Not originally in Bica, Bonatto \& Camargo's series of papers. They have been included in the main sample since they are projected in the same direction of other OCs studied in these papers. See also Table\,\ref{tab:previous_lit_information} for more details. The OC UBC\,296 has been recently catalogued by \cite{Castro-Ginard:2020}. L19\,2326 refers to the cluster 2326 catalogued by \cite{Liu:2019}. \\

$^a$ Not originally in BBC. They have been included in the main sample since they are projected in the same direction of other OCs studied in these papers. See also Table\,\ref{tab:previous_lit_information} for more details. The OC UBC\,296 has been recently catalogued by \cite{Castro-Ginard:2020}. L19\,2326 refers to the cluster 2326 catalogued by \cite{Liu:2019}. \\

%$^\dag$ No profile fits could be performed for Herschel\,1 and Lynga\,2. The $r_t$ for both OCs was assumed as being equal to these clusters' limiting radius (Section~\ref{sec:method}).  \\
%%%$^{\dag\dag}$ New OC candidate: see Section~\ref{sec:CEFET1}. \\
$^{(*)}$ The $R_G$ values were obtained assuming that the Sun is located at 8.0\,$\pm$\,0.5\,kpc from the Galactic centre \citep{Reid:1993a}. \\
%$^{(**)}$ Pismis\,19, NGC\,5617, Trumpler\,22 and Dias\,6 were analysed in the present paper. Parameters for the other 11 OCs in the complementary sample were obtained in Paper I. \\

\end{minipage}
\end{sideways}
\label{tab:investig_sample}
\end{table*}

\begin{table*}
%\centering
\scriptsize
\caption{Cluster, fundamental parameters from the literature and references for OCs in the main sample.}

\begin{tabular}{lcccl}

\hline

 Cluster          & $(m-M)_0$             & $E(B-V)$            & log $t$            & Reference    \\
                      &   (mag)                   & (mag)                  & (dex)               &                     \\
                                                                                                               
\hline

Czernik\,7                         & 12.59\,$\pm$\,0.07    & 0.70\,$\pm$\,0.03   & 8.34\,$\pm$\,0.10  & \cite{Camargo:2009}   \\
Berkeley\,63                       & 13.78\,$\pm$\,0.04    & 0.96\,$\pm$\,0.03   & 7.48\,$\pm$\,0.14  & \cite{Camargo:2009}   \\
Czernik\,12                        & 11.51\,$\pm$\,0.11    & 0.26\,$\pm$\,0.03   & 9.10\,$\pm$\,0.14  & \cite{Camargo:2009}   \\
Czernik\,22                        & 12.07\,$\pm$\,0.08    & 0.64\,$\pm$\,0.03   & 8.30\,$\pm$\,0.11  & \cite{Camargo:2010}   \\
Czernik\,23                        & 11.99\,$\pm$\,0.09    & 0.00\,$\pm$\,0.03   & 9.70\,$\pm$\,0.09  & \cite{Bonatto:2008}   \\
Czernik\,24$^{\dag}$               & 13.31                 & 0.26                & 9.40               & \cite{Camargo:2010}   \\
Ruprecht\,1                        & 11.18\,$\pm$\,0.20    & 0.26\,$\pm$\,0.06   & 8.70\,$\pm$\,0.09  & \cite{Bonatto:2010}   \\
Ruprecht\,10                       & 11.85\,$\pm$\,0.20    & 0.64\,$\pm$\,0.06   & 8.70\,$\pm$\,0.09  & \cite{Bonatto:2010}   \\
Ruprecht\,23                       & 12.43\,$\pm$\,0.21    & 0.54\,$\pm$\,0.06   & 8.78\,$\pm$\,0.07  & \cite{Bonatto:2010}   \\
%%%CEFET-MG1$^a$                   &        $-$            &       $-$           &       $-$          &                       \\
L19\,2326$^a$                      &        $-$            &       $-$           &       $-$          &                       \\
Ruprecht\,26                       & 11.30\,$\pm$\,0.20    & 0.35\,$\pm$\,0.06   & 8.60\,$\pm$\,0.05  & \cite{Bonatto:2010}   \\
Ruprecht\,27                       & 10.87\,$\pm$\,0.20    & 0.03\,$\pm$\,0.06   & 8.95\,$\pm$\,0.05  & \cite{Bonatto:2010}   \\
Ruprecht\,34                       & 12.10\,$\pm$\,0.21    & 0.00\,$\pm$\,0.06   & 9.00\,$\pm$\,0.04  & \cite{Bonatto:2010}   \\
Ruprecht\,35                       & 12.96\,$\pm$\,0.31    & 0.45\,$\pm$\,0.10   & 8.60\,$\pm$\,0.11  & \cite{Bonatto:2010}   \\
Ruprecht\,37                       & 13.60\,$\pm$\,0.31    & 0.00\,$\pm$\,0.06   & 9.48\,$\pm$\,0.14  & \cite{Bonatto:2010}   \\
NGC\,2477$^a$                      &        $-$            &       $-$           &       $-$          &                       \\
Ruprecht\,41                       & 12.49\,$\pm$\,0.31    & 0.13\,$\pm$\,0.10   & 8.85\,$\pm$\,0.06  & \cite{Bonatto:2010}   \\
Ruprecht\,152                      & 14.52\,$\pm$\,0.31    & 0.67\,$\pm$\,0.10   & 8.78\,$\pm$\,0.07  & \cite{Bonatto:2010}   \\
Ruprecht\,54                       & 13.69\,$\pm$\,0.31    & 0.13\,$\pm$\,0.10   & 8.90\,$\pm$\,0.05  & \cite{Bonatto:2010}   \\
Ruprecht\,60                       & 13.95\,$\pm$\,0.31    & 0.64\,$\pm$\,0.10   & 8.60\,$\pm$\,0.11  & \cite{Bonatto:2010}   \\
Ruprecht\,63                       & 12.88\,$\pm$\,0.21    & 0.61\,$\pm$\,0.10   & 8.70\,$\pm$\,0.09  & \cite{Bonatto:2010}   \\
Ruprecht\,66                       & 12.88\,$\pm$\,0.31    & 0.90\,$\pm$\,0.10   & 8.78\,$\pm$\,0.07  & \cite{Bonatto:2010}   \\
UBC\,296$^a$                       &        $-$            &       $-$           &       $-$          &                       \\                      
NGC\,5715                          & 10.88\,$\pm$\,0.15    & 0.42\,$\pm$\,0.03   & 8.90\,$\pm$\,0.05  & \cite{Bonatto:2007}   \\
Lynga\,4                           & 10.21\,$\pm$\,0.20    & 0.70\,$\pm$\,0.07   & 9.11\,$\pm$\,0.07  & \cite{Bonatto:2007}   \\
Trumpler\,23                       & 11.39\,$\pm$\,0.11    & 0.58\,$\pm$\,0.03   & 8.95\,$\pm$\,0.05  & \cite{Bonatto:2007}   \\
Lynga\,9                           & 11.15\,$\pm$\,0.26    & 1.18\,$\pm$\,0.11   & 8.85\,$\pm$\,0.06  & \cite{Bonatto:2007}   \\
Trumpler\,26                       & 10.00\,$\pm$\,0.22    & 0.35\,$\pm$\,0.03   & 8.85\,$\pm$\,0.06  & \cite{Bonatto:2007}   \\
Bica\,3                            & 11.07\,$\pm$\,0.25    & 2.18\,$\pm$\,0.03   & 7.40\,$\pm$\,0.09  & \cite{Bica:2004}      \\
Ruprecht\,144                      & 11.02\,$\pm$\,0.14    & 0.77\,$\pm$\,0.10   & 8.65\,$\pm$\,0.10  & \cite{Camargo:2009}   \\
$[\textrm{FSR}2007]$\,0101         & 11.39\,$\pm$\,0.11    & 2.37\,$\pm$\,0.03   & 8.95\,$\pm$\,0.10  & \cite{Camargo:2009}   \\
Berkeley\,84                       & 11.15\,$\pm$\,0.13    & 0.58\,$\pm$\,0.06   & 8.56\,$\pm$\,0.06  & \cite{Camargo:2009}   \\
Ruprecht\,172                      & 12.46\,$\pm$\,0.07    & 0.64\,$\pm$\,0.06   & 8.95\,$\pm$\,0.10  & \cite{Camargo:2009}   \\
Ruprecht\,174                      & 11.62\,$\pm$\,0.21    & 0.32\,$\pm$\,0.06   & 8.90\,$\pm$\,0.05  & \cite{Bonatto:2010}   \\

\hline

%\multicolumn{5}{l}{\textit{Note:} Paper I is \cite{Angelo:2020}.} \\

\multicolumn{5}{l}{$^\dag$ This cluster is present in the sample analysed by \cite{Camargo:2010}, but its} \\
\multicolumn{5}{l}{parameters (no uncertainties informed and no CMD available) were determined} \\
 
%\multicolumn{5}{l}{$^a$ NGC\,2421, [FSR2007]\,1325, CEFET-MG1, L19\,2326, NGC\,2477 and UBC\,296 are not part of the  }    \\ 

\multicolumn{5}{l}{by \cite{Koposov:2008}.}    \\ 

%\multicolumn{5}{l}{$^a$ NGC\,2421, L19\,2326, [FSR2007]\,1325, NGC\,2477 and UBC\,296 are not part of the}  \\

\multicolumn{5}{l}{$^a$ L19\,2326, NGC\,2477 and UBC\,296 are not part of the original samples in BBC. }  \\

\multicolumn{5}{l}{They have been included in the main sample since they are projected in the same} \\

%\multicolumn{5}{l}{included in the main sample since they are projected in the same regions of, respectively: Czernik\,31, Ruprecht\,26, Czernik\,32, Ruprecht\,152 and Lynga\,2.} 

\multicolumn{5}{l}{regions of, respectively: Ruprecht\,26, Ruprecht\,152 and Lynga\,2 (this third OC has}   \\

\multicolumn{5}{l}{been investigated in Paper I.}

\label{tab:previous_lit_information}
\end{tabular}

\end{table*}

The separation of our complete sample in two groups (main and complementary ones) was based on previous information from the literature. OCs in the main sample were uniformly investigated in BBC (see Table~\ref{tab:previous_lit_information}) by means of 2MASS photometry and a decontamination algorithm applied to $J\times(J-H)$ and $J\times(J-K_S)$ CMDs. This allows more objective comparisons between our results and the literature ones (Section~\ref{compara_previous_studies}), thus avoiding biases due to heterogeneous analysis methods.

%\textbf{Part of our complementary sample is composed by well-known objects, promptly distinguishable from the Galactic field and extensively employed in previous studies, as is the case of NGC\,188 (e.g., \citeauthor{Bonatto:2005}\,\,\citeyear{Bonatto:2005}), Collinder\,110 (e.g., \citeauthor{Pancino:2010}\,\,\citeyear{Pancino:2010}; \citeauthor{Netopil:2016}\,\,\citeyear{Netopil:2016}), NGC\,2439 (e.g., \citeauthor{Dias:2019}\,\,\citeyear{Dias:2019}; \citeauthor{Conrad:2017}\,\,\citeyear{Conrad:2017}), M\,67 (e.g., \citeauthor{Bonatto:2003}\,\,\citeyear{Bonatto:2003}), NGC\,4337 (e.g., \citeauthor{Carraro:2014}\,\,\citeyear{Carraro:2014}) and Collinder\,261 (e.g., \citeauthor{Rain:2020}\,\,\citeyear{Rain:2020}; \citeauthor{Drazdauskas:2016}\,\,\citeyear{Drazdauskas:2016}). NGC\,5617, Pismis\,19 and Trumpler\,22 constitute a mutiple system candidate according to \cite{de-La-Fuente-Marcos:2009}. NGC\,3680 is a prototype of an OC in advanced stage of dissolution and resembles in several aspects what is expected for open cluster remnants (\citeauthor{Pavani:2007}\,\,\citeyear{Pavani:2007}; \citeauthor{Pavani:2011}\,\,\citeyear{Pavani:2011}; \citeauthor{Angelo:2019a}\,\,2019a).}

Our complementary sample is composed by OCs presenting clearer contrasts against the field population in comparison to those in the main sample. NGC\,5617, Pismis\,19 and Trumpler\,22 are projected in the same area and constitute a multiple system candidate according to \cite{de-La-Fuente-Marcos:2009}. The study of these objects is useful to investigate the impact of close encounters on the clusters' structural parameters (Section~\ref{sec:investig_struct_params}). In turn, Dias\,6 is a moderately rich OC whose astrophysical parameters were determined by \cite{Dias:2018} using unprecedented deep CCD $UBVRI$ photometry combined with \textit{Gaia} DR2, as part of the \textit{OPD photometric survey of open clusters} \citep{Caetano:2015}. This OC is among the most dynamically evolved ones in our sample (Section~\ref{sec:discussion}).

%\textbf{The other part of our complementary sample includes less populated OCs, presenting significantly smaller contrasts with the field. The investigation of these objects is useful for determining their physical nature and also contributes to enlarge the parameters space coverage of our study, since some of them present signals of being in advanced stages of dissolution. NGC\,6216 is a relatively young OC ($t\sim70\,$Myr, Section~\ref{sec:fundamental_params_determ} and Table~\ref{tab:investig_sample}) projected very close to the Galactic plane. It had its physical nature previously confirmed by \cite{Piatti:2000} through CCD $BVI$ photometry and integrated spectra. It is projected $\sim30\arcmin$ southwards of BH\,200, whose physical nature was determined from the analysis of deep CCD $UBVRI$ photometry in \cite{Monteiro:2017} as part of the \textit{OPD photometric survey of open clusters} \citep{Caetano:2015}. This survey also includes Dias\,6, whose parameters were determined by means of deep CCD $UBVRI$ photometry combined with \textit{GAIA} DR2 in \cite{Dias:2018}.} 

%\textbf{Czernik\,37 was previously characterized by means of 2MASS photometry \citep{Tadross:2008}, CCD Washington $CT_1$ data (\citeauthor{Marcionni:2014}\,\,\citeyear{Marcionni:2014}; \citeauthor{Angelo:2019b}\,\,2019b) and proper motions \citep{Sampedro:2017}. Collinder\,351, in turn, is a sparse and poorly-studied object, barely distinguishable from the field. Considering our complete sample (Table~\ref{tab:investig_sample}), it is among the most evolved OCs (Section~\ref{sec:discussion}).}

For each OC, we extracted astrometric and photometric data from the \textit{Gaia} DR2 catalogue in circular regions with radius $r$\,=\,1$^{\circ}$ centered on the equatorial coordinates as informed in the catalogue of \citeauthor{Dias:2002}\,\,(\citeyear{Dias:2002}, hereafter DAML02) or in the SIMBAD database. The Vizier tool\footnote[1]{http://vizier.u-strasbg.fr/viz-bin/VizieR} was used to accomplish this task. The extraction radius is typically grater than $\sim10$ times the apparent  radius informed in DAML02, thus encompassing the whole clusters' region and part of the adjacent field population. The original data were then filtered according to equations 1 and 2 of \cite{Arenou:2018}, in order to ensure the best quality of the photometric and astrometric information employed throughout our analysis.

The central coordinates obtained from the literature were refined based on the procedure described in Section~\ref{sec:method}. Table \ref{tab:investig_sample} contains the redetermined $\rm{RA}$ and $\rmn{DEC}$ values. The determination of other parameters is also described in Section~\ref{sec:method}.

\section{Method}
\label{sec:method}

\begin{figure*}
\begin{center}

\parbox[c]{1.0\textwidth}
  {
   
   \begin{center}
    \includegraphics[width=0.40\textwidth]{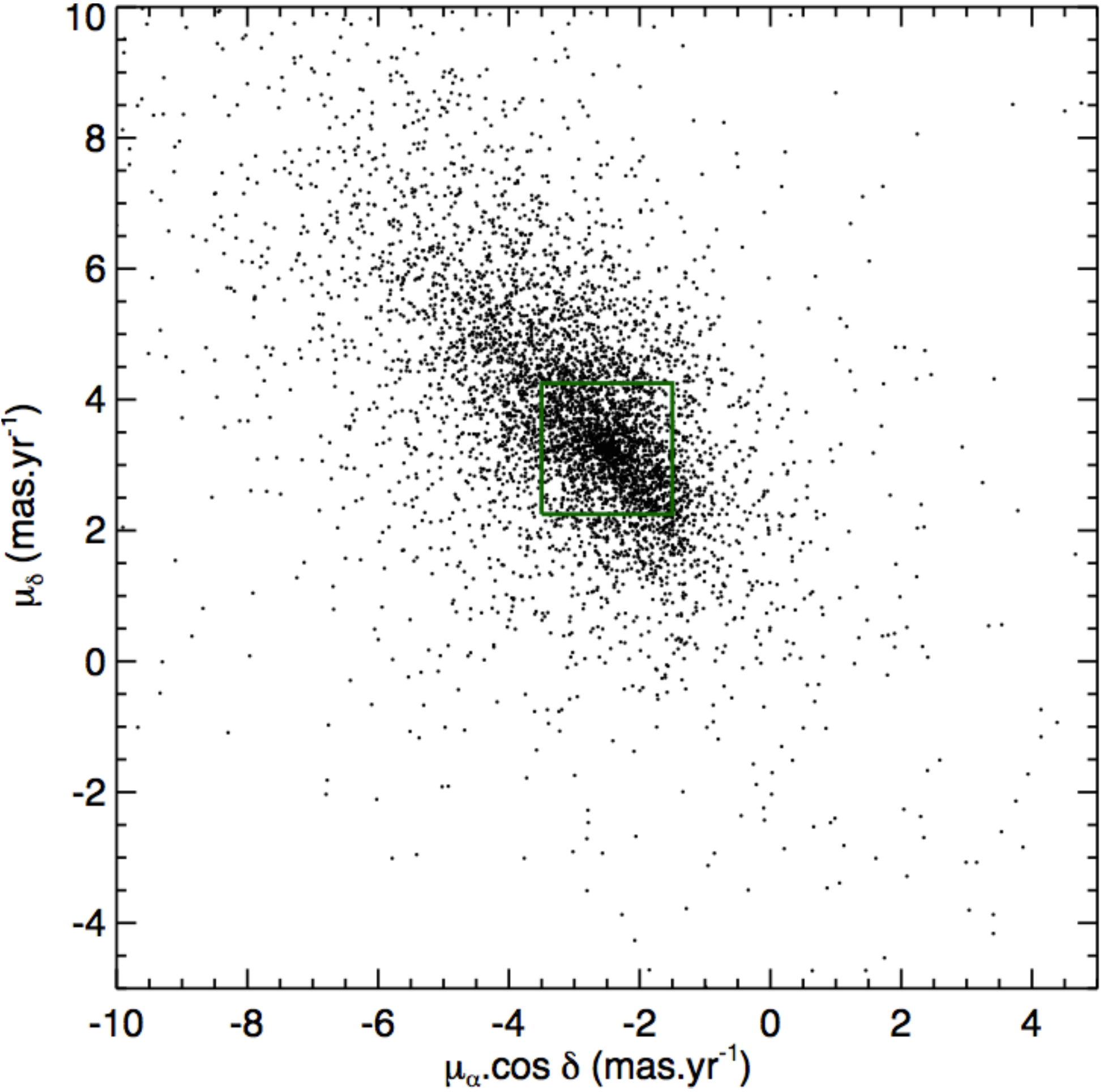}    
    \includegraphics[width=0.40\textwidth]{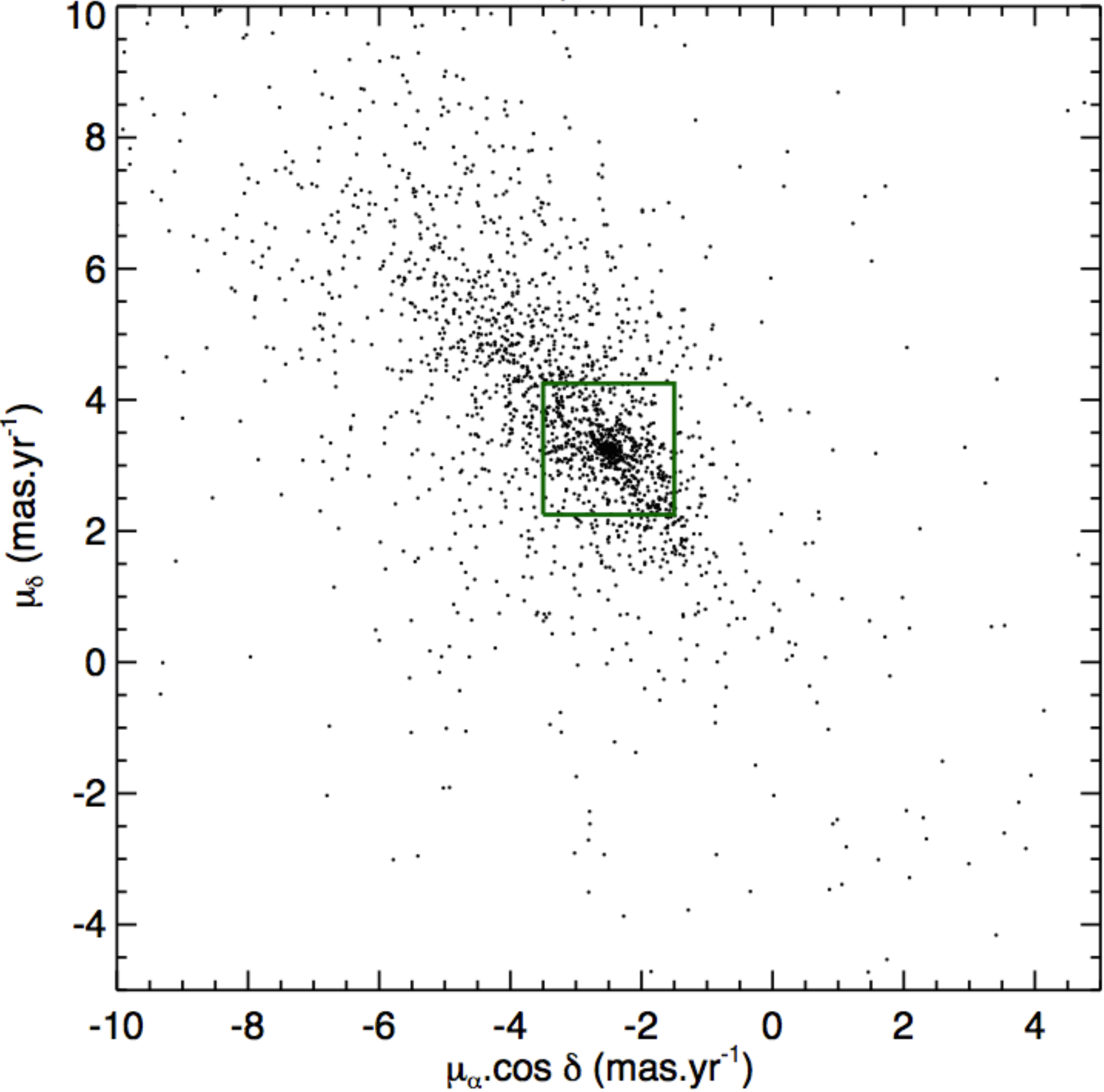}   
    \end{center}
  }
\caption{VPD for stars in an area of $18\arcmin\times18\arcmin$ centered on Ruprecht\,63 before (left panel) and after (right panel) restricting stars to the interval $G\,\leq18\,$mag. The green box marks a detached overdensity, which corresponds mostly to Ruprecht\,63's member stars.}

\label{fig:VPD_Ruprecht63_original_and_filtered}
\end{center}
\end{figure*}

The methodology employed in the present work is the same as that of Paper I, where the procedures are described in detail. In what follows, we give a summarized description of the analysis steps.

%%%%%%%%%%%%%%%%%%%%%%%%%%%%
\subsection{Preliminary analysis}
\label{sec:preanalysis}
%%%%%%%%%%%%%%%%%%%%%%%%%%%%

In this preliminary stage, we successively apply filters (besides Arenou et al.'s ones) on our data in order to limit the field contamination and look for detached concentration of stars in each cluster vector-point diagram (VPD), which is an indicative of a possible physical system. To accomplish this, firstly a spatial filter is applied: we restrict our analysis to stars inside a projected square area whose size corresponds to $\sim4\times$ the objects' radius as informed in DAML02. Then we apply a magnitude filter: stars are restricted to those with $G\leq18\,$mag. This magnitude limit ensures completeness levels greater than 90\% with respect to HST data even in crowded regions with $\sim10^5\,$stars/deg$^2$, representative of globular clusters (see figure 7 of \citeauthor{Arenou:2018}\,\,\citeyear{Arenou:2018}). 

Fig.~\ref{fig:VPD_Ruprecht63_original_and_filtered} illustrates the relevance of the magnitude filter. Both panels exhibit stars in an area of $18\arcmin\times18\arcmin$ centered on Ruprecht\,63's coordinates as informed in DAML02. Left panel is the VPD without magnitude restriction. The right panel corresponds to stars with $G\leq18\,$mag. The green square box (side equal to 2\,mas\,yr$^{-1}$) delimits the conspicuous concentration of Ruprecht\,63's stars, which is only noticeable after applying the magnitude filter. This VPD box is large enough to encompass the OC's member stars, but small enough to limit considerably the contamination by field stars.   %Se o referee criticar o tamanho dos VPD box, podemos justificar usando a Figura 10 do paper do Filipe (arquivo original: "New_Galactic_star_clusters_discovered_with_Gaia_DR2_GRANDE.pdf").

%%%%%%%%%%%%%%%%%%%%%%%%%%%%
\subsection{Structural analysis}
%%%%%%%%%%%%%%%%%%%%%%%%%%%%

In this step, we firstly redetermined the cluster's central coordinates. To accomplish this, we restricted our analysis to those stars inside the VPD box as shown in the right panel of Fig.~\ref{fig:VPD_Ruprecht63_original_and_filtered}. Then a uniform grid of tentative central coordinates was constructed surrounding the literature coordinates. Typically, $\sim200-300$ pairs of ($\alpha,\delta$), evenly spaced ($\Delta\sim0.25\arcmin-0.50\arcmin$), were employed.    

%For each ($\alpha,\delta$) pair in the grid, a radial density profile (RDP) was built by counting the number of stars in concentric rings and dividing this number by the ring's area to obtain $\sigma(r)$. Different bin widths were employed and overplotted on the same RDP. The mean background density ($\sigma_{\textrm{bg}}$) was determined by averaging the $\sigma(r)$ values beyond the limiting radius ($R_{\textrm{lim}}$), which is defined as the $r$ value from which the density values are reasonably constant. Core and tidal radii ($r_c$ and $r_t$, respectively) are determined from the best fitted \cite{King:1962} model to the cluster's background subtracted RDP via $\chi^2$ minimization. The cluster centre is defined as the ($\alpha$, $\delta$) pair which resulted in the best model fit and, at the same time, the highest central density. 

For each ($\alpha,\delta$) pair in the grid, a radial density profile (RDP) was built by counting the number of stars in concentric rings and dividing this number by the ring's area to obtain $\sigma(r)$. Different bin widths were employed and overplotted on the same RDP. The mean background density ($\sigma_{\textrm{bg}}$) was determined by averaging the $\sigma(r)$ values beyond the limiting radius ($R_{\textrm{lim}}$), which is defined as the $r$ value from which the density values are reasonably constant. The cluster's background subtracted RDP was fitted via $\chi^2$ minimization using \cite{King:1962}'s model through the expression:

\begin{equation}
   \sigma(r)\propto\left(\frac{1}{\sqrt{1+(r/r_c)^2}} - \frac{1}{\sqrt{1+(r_t/r_c)^2}} \right)^2
,\end{equation}

\noindent
where $r_c$ and $r_t$ are the core and tidal radii, respetively. The $r_c$ radius provides a length scale of the cluster's central structure and $r_t$ is defined as the truncation radius parameter of the King model. Observationally, $r_t$ provides information about the overall cluster size. The cluster centre (see Table~\ref{tab:investig_sample}) is defined as the ($\alpha$, $\delta$) pair which resulted in the best model fit and, at the same time, the highest central density. In order to check the robustness of our central coordinates determination procedure, we took only member stars with high membership likelihoods (above 70\%; Sections \ref{sec:membership_assignment} and \ref{sec:fundamental_params_determ}) and averaged their coordinates. The resulting centre is tipically within 1\,arcmin from those given in Table~\ref{tab:investig_sample}.

\begin{figure*}
\begin{center}

\parbox[c]{1.0\textwidth}
  {
   \begin{center}
      \includegraphics[width=0.75\textwidth]{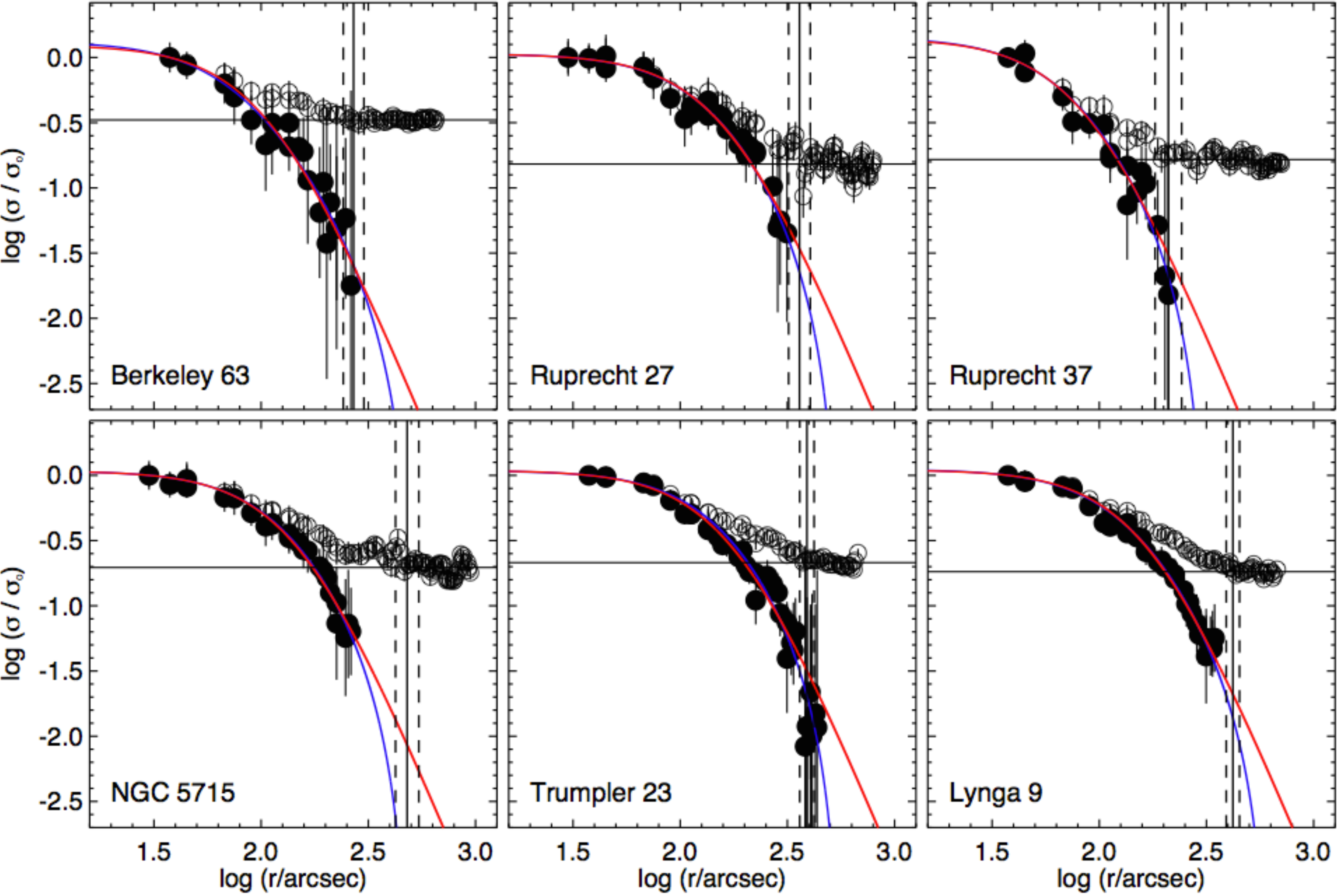} 
   \end{center}
    
  }
\caption{ RDPs for 6 of our studied OCs. The background subtracted and normalized profiles are represented by filled circles, to which we fitted King (blue line) and Plummer profiles (red lines). The profiles are normalized to unity at the innermost radial bin. Non-background subtracted RDPs are represented by open circles and the mean backround density ($\sigma_{\textrm{bg}}$) is indicated by the continuous horizontal line. The continuous and dashed vertical lines show, respectively, the cluster limiting radius ($R_{\textrm{lim}}$) and its uncertainty. For better convergence of the solution during parameters determination, in some cases (e.g., NGC\,5715) the fit domain in log\,($r$) was truncated to values smaller than $R_{\textrm{lim}}$, due to counts fluctuations in the cluster's external regions. Error bars correspond to Poisson statistics. }
%In both cases, rt ~ R_{lim}

\label{fig:RDPs_part1}
\end{center}
\end{figure*}

Fig.~\ref{fig:RDPs_part1} shows the result of this procedure for 6 of the investigated OCs. Figures for the rest of our sample are shown in Appendix\,B of the supplementary material, which is available online. From now on, the same 6 clusters are represented for other Figures. Additionally, we perfomed indepent fits of \cite{Plummer:1911} profile (red lines in Fig.~\ref{fig:RDPs_part1})

\begin{equation}
   \sigma(r)\propto\frac{1}{[1+(r/a)]^2}.
\end{equation}

%and derived half-mass radius from the Plummer's \textit{a} parameter through the relation $r_{hm}\sim1.3\,a$.  Table~\ref{tab:investig_sample} contains the derived structural parameters, which were converted to pc by employing the distance moduli determined in Section~\ref{sec:fundamental_params_determ}. 
\noindent
The Plummer's \textit{a} parameter is proportional to the half-mass radius through the relation $r_{hm}\sim1.3\,a$. Table~\ref{tab:investig_sample} contains the derived structural parameters, which were converted to pc by employing the distance moduli determined in Section~\ref{sec:fundamental_params_determ}.

%%%%%%%%%%%%%%%%%%%%%%%%%%%%
\subsection{Membership assignment}
\label{sec:membership_assignment}
%%%%%%%%%%%%%%%%%%%%%%%%%%%%

In this step, we employed a decontamination algorithm that performs statistical comparisons between cluster and field stars in different parts of the 3D astrometric space ($\varpi$, $\mu_{\alpha}\,\textrm{cos}\,\delta$, $\mu_{\delta}$) and assigns membership likelihoods. The method is fully described and tested in \citeauthor{Angelo:2019a}\,\,(2019a). Here we describe its main procedures. 

Firstly, the parameters space is defined from the astrometric information for stars within the cluster $r_t$ and in an annular control field (inner radius equal to 3\,$r_t$), centered on the cluster coordinates informed in Table~\ref{tab:investig_sample}, and with area equal to 3 times the cluster area. This provides statistical representativity of the field population in both proper motions and parallax domains. For a better perfomance of the method, cluster and control field stars are restricted to those inside the cluster VPD box as illustrated in Fig.~\ref{fig:VPD_Ruprecht63_original_and_filtered} (right panel). At this stage, the magnitude filter employed in the preliminary analysis (Section~\ref{sec:preanalysis}) was dismissed.    

Then the parameters space is divided in a uniform grid of cells with varying sizes. Inside each one, a membership likelihood is determined for each star in the cluster area using multivariate gaussians, which properly incorporate measurement uncertainties and correlations among the astrometric parameters (equations 1 and 2 of \citeauthor{Angelo:2019a}\,\,2019a). Analogous calculations are performed for stars in the control field within the same 3D cell. Both sets of likelihoods values are objectively compared by employing an \textit{entropy-like} function $S$ (equation 3 of \citeauthor{Angelo:2019a}\,\,2019a). Stars within cells for which $S_{\textrm{cluster}}<S_{\textrm{field}}$ are considered possible members.

\begin{figure*}
\begin{center}

\parbox[c]{1.0\textwidth}
  {
   \begin{center}
      \includegraphics[width=0.497\textwidth]{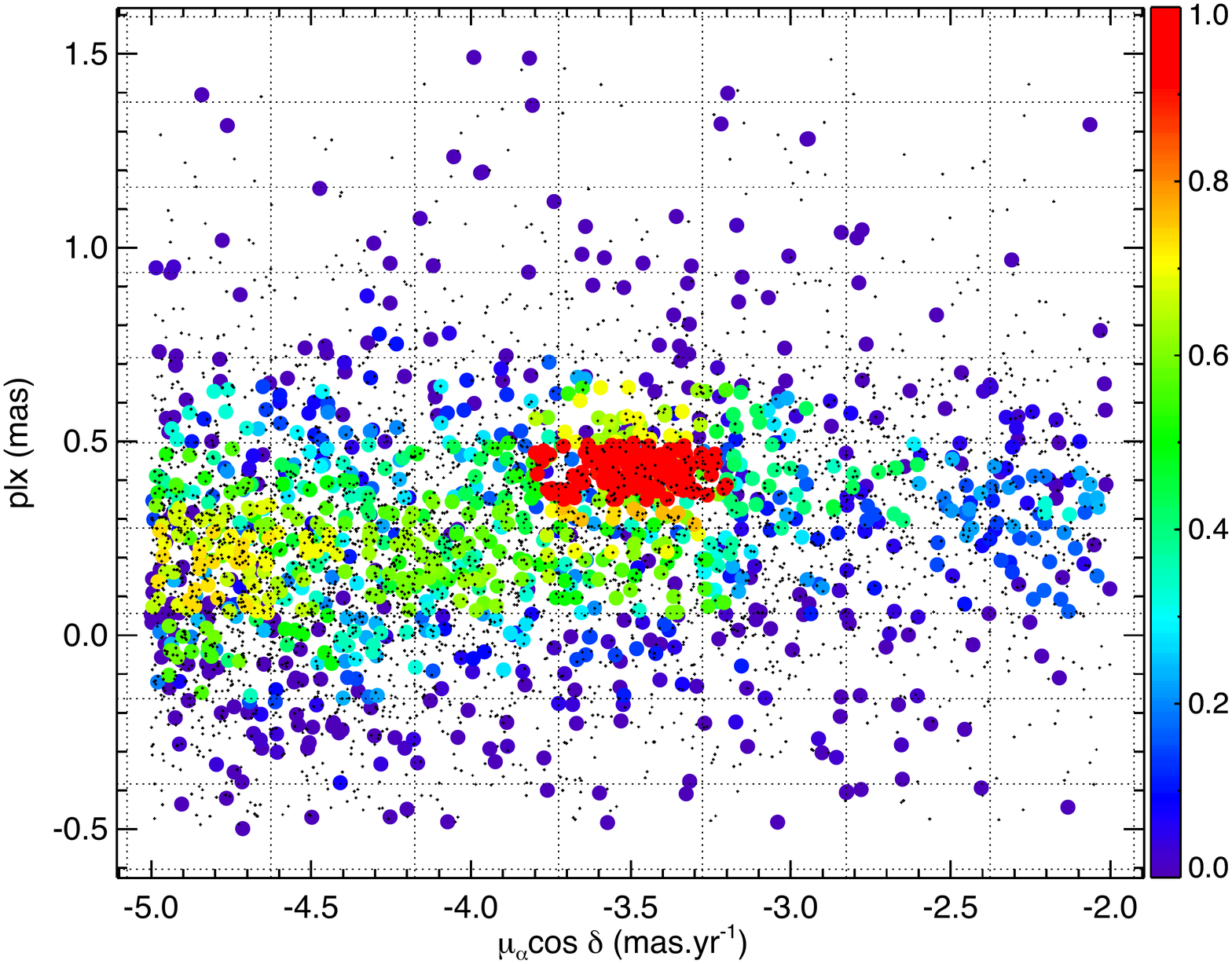}
      \includegraphics[width=0.497\textwidth]{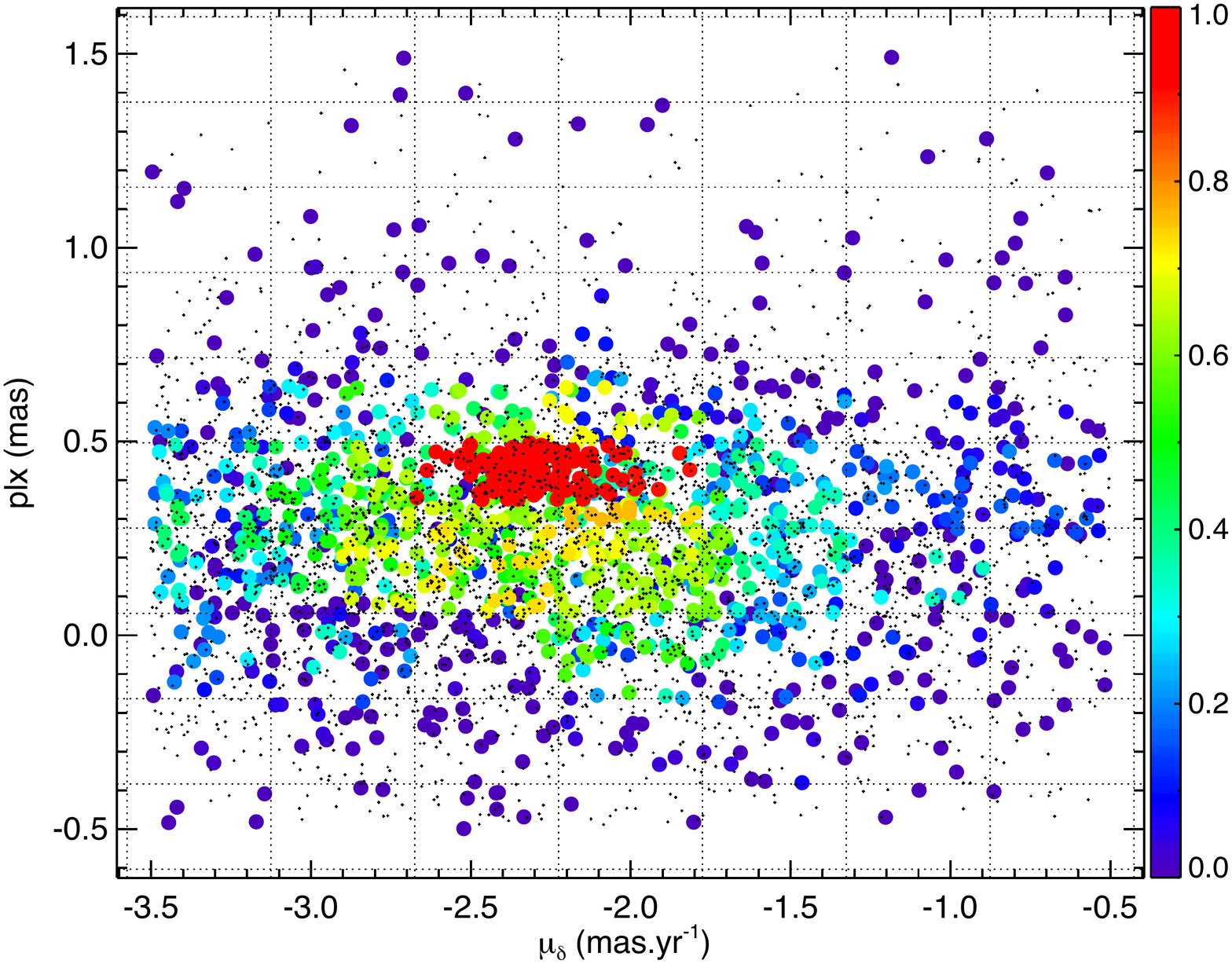}       
   \end{center}
   
   \begin{center}
      \includegraphics[width=0.35\textwidth]{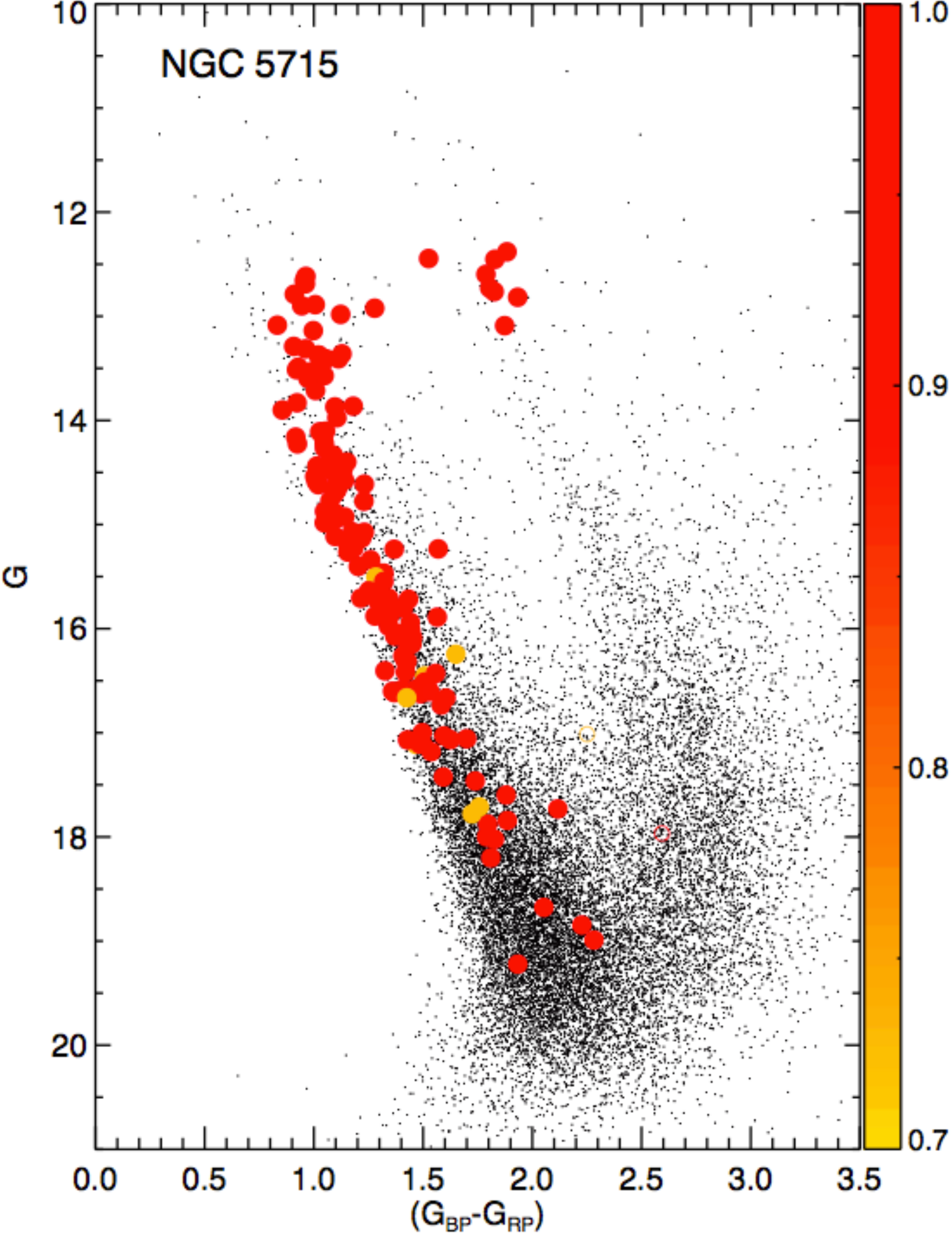}   
   \end{center}
    
  }
\caption{ Top panels: astrometric space $\varpi$\,versus\,$\mu_{\alpha}\textrm{cos}\,\delta$ (left) and $\varpi$\,versus\,$\mu_{\delta}$ (right). Symbol colours represent membership likelihoods, as indicated by the colourbars. The grid of cells is also represented (dotted lines). Bottom plot: decontaminated CMD. In all plots, the small black dots represent control field stars.}

\label{fig:ilustra_decontam_method}
\end{center}
\end{figure*}

For those cells whose stars were flagged as possible members, an additional factor was determined, which evaluates the overdensity of cluster stars relatively to the complete grid of cells (equation 4 of \citeauthor{Angelo:2019a}\,\,2019a). With this procedure, we ensure that appreciable likelihoods will be assigned only to those significant  overdensities that are statistically more concentrated than the local distribution of control field stars. The dependence of the results on the initial choice of cells sizes is alleviated by varying their widths in each dimension. The final likelihood of each star corresponds to the median of the set of values obtained with the whole grid configurations.  

Fig.~\ref{fig:ilustra_decontam_method} illustrates the outcomes of the decontamination method applied on NGC\,5715 data. The top panels show the astrometric space ($\varpi$\,versus\,$\mu_{\alpha}\textrm{cos}\,\delta$ and $\varpi$\,versus\,$\mu_{\delta}$ plots) defined by stars in the cluster area ($r\leq r_t$; coloured symbols) and in the control field (small black dots). Symbol colours indicate membership likelihoods, as shown by the colourbars. The decontaminated CMD in the bottom was restricted to stars with likelihoods greater than $\sim70\%$, which define recognizable evolutionary sequences: an extended main sequence, the main sequence turnoff (around $(G_{\textrm{BP}}-G_{\textrm{RP}})\sim1.0\,$mag, $G\sim13\,$mag) and the red clump (($G_{\textrm{BP}}-G_{\textrm{RP}})\sim1.8\,$mag, $G\sim12.5\,$mag). It is noticeable that NGG\,5715 is projected against a very dense stellar field, which would make challenging the task of disentangling cluster and field populations through purely photometric methods. This reinforces the importance of combining the high precise astrometric and photometric information provided by the \textit{Gaia} DR2 catalogue.

%%%%%%%%%%%%%%%%%%%%%%%%%%%%%%%%
\subsection{Fundamental parameters determination}
\label{sec:fundamental_params_determ}
%%%%%%%%%%%%%%%%%%%%%%%%%%%%%%%%

In order to construct the decontaminated CMDs, we restricted each cluster sample to stars with membership likelihoods greater than $\sim$70\%. This threshold allowed to identify clear evolutionary sequences, to which we fitted theoretical isochrones computed from the PARSEC models \citep{Bressan:2012} and convoluted with \textit{Gaia}'s filters bandpasses \citep{Evans:2018}. The procedure of isochrone fitting was performed in two parts: firstly, we obtained initial guesses for the fundamental parameters $(m-M)_0$, $E(B-V)$ and log\,$t$ by means of a visual fit. Solar metallicity isochrones and $E(B-V)$ values informed in DAML02 were considered at this stage. Then we applied vertical shifts on the isochrone until matching the clusters' main sequence. When necessary, the literature value for $E(B-V)$ was gradually modified in order to improve the match. The log\,$t$ was then estimated from the disposal of high membership stars along the more evolved sequences: turnoff, subgiant and red giant branches, besides the red clump (if present). Further refinements were possible by changing the isochrone overall metallicity $Z$. In this preliminary fit, the relative distance between the red clump and the main sequence turnoff was a very useful observational constraint.     

These initial guesses for the fundamental parameters ($(m-M)_{0,\textrm{ini}}$, $E(B-V)_{\textrm{ini}}$, log\,$t_{\textrm{ini}}$ and $Z_{\textrm{ini}}$) were then refined by means of an automatic isochrone fitting as performed by the ASteCA code \citep{Perren:2015}. In few words, it employs a genetic algorithm to look for the best possible match between the observed CMD and a set of synthetic ones, based on PARSEC isochrones, generated for clusters with different masses and metallicities. To improve the ASteCA performance, we filtered the parameters space according to the initial guesses determined previously: the models were restricted to 0.5\,mag above and below $(m-M)_{0,\textrm{ini}}$, in steps of 0.1\,mag; for log\,$t$, we considered models with 0.3\,dex above and below log\,$t_{\textrm{ini}}$, with steps of 0.05\,dex; for $E(B-V)$, a range of 1.0\,mag centered on $E(B-V)_{\textrm{ini}}$ was employed, in steps of 0.02\,mag; for the metallicity, the allowed interval for the models corresponds to $Z_{\textrm{ini}}-0.01\leq Z\leq Z_{\textrm{ini}}+0.02$, in steps of 0.002. The cluster metallicity $[Fe/H]$ (Table~\ref{tab:investig_sample}) was determined from $Z$ following the approximate relation $[Fe/H]\sim\textrm{log}\,(Z/Z_{\odot})$ \citep{Bonfanti:2016}, where $Z_{\odot}=0.0152$ \citep{Bressan:2012}.

%%%%%%%%%%%%%%%%%%%%%%%%%%%%%%%%%%%
\section{Results}
\label{sec:results}
%%%%%%%%%%%%%%%%%%%%%%%%%%%%%%%%%%%

\begin{figure*}
\begin{center}

\parbox[c]{0.75\textwidth}
  {
   \begin{center}
     \includegraphics[width=0.75\textwidth]{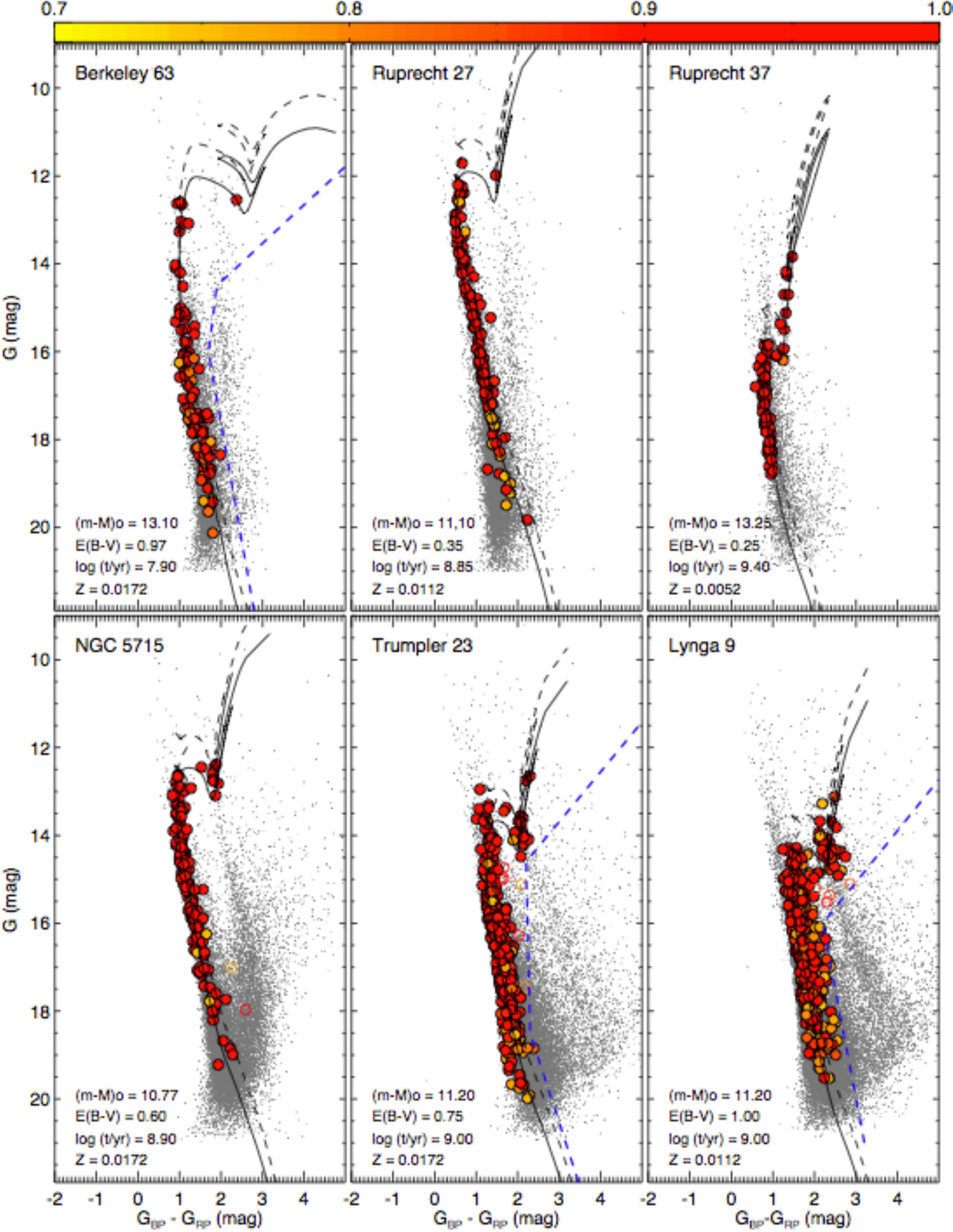}    
    \end{center}
    
  }
\caption{ Decontaminated $G\times(G_{\textrm{BP}}-G_{\textrm{RP}})$ CMDs for 6 investigated OCs. Symbol colours represent membership likelihoods and filled circles are member stars. Small grey dots are stars in a control field. The continuous black lines are PARSEC isochrones fitted to the data (fundamental parameters are indicated; see also Table~\ref{tab:investig_sample}), while the dashed ones (same isochrone, but vertically shifted by -0.75\,mag) represent the loci of unresolved binaries with equal mass components. The colour filters (Berkeley\,63, Trumpler\,23 and Lynga\,9's CMDs) are shown as blue dashed lines. }

\label{fig:CMDs_part1}
\end{center}
\end{figure*}

\begin{figure*}
\begin{center}

\parbox[c]{0.75\textwidth}
  {
   \begin{center}
    \includegraphics[width=0.75\textwidth]{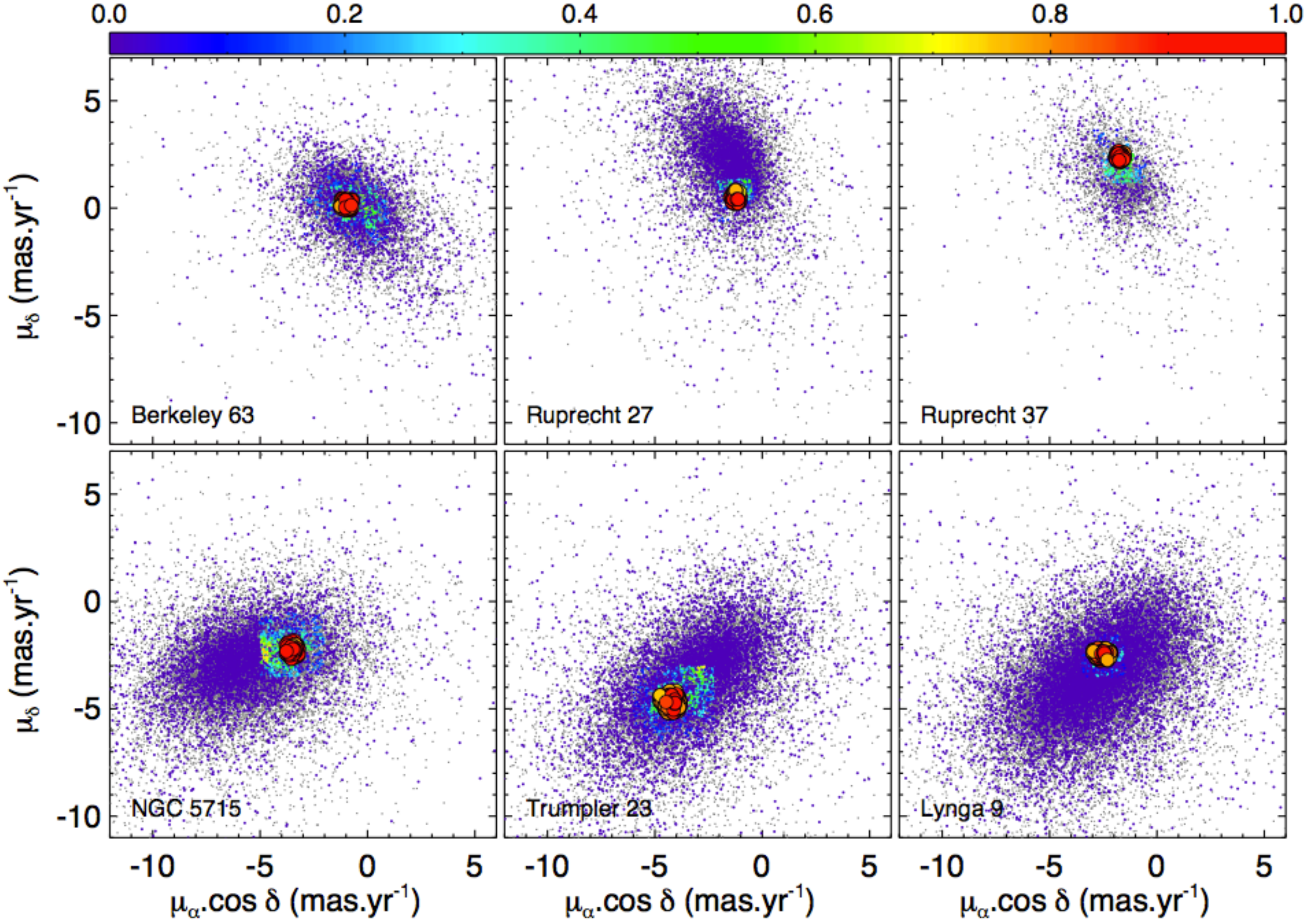}    
    \end{center}
    
  }
\caption{ VPDs for stars in the clusters' areas ($r\,\leq\,r_t$; coloured symbols) and in the control field (grey dots) for 6 of our investigated OCs. Symbols convention is the same as that of Fig.~\ref{fig:CMDs_part1}. }

\label{fig:VPDs_part1}
\end{center}
\end{figure*}

\begin{figure*}
\begin{center}

\parbox[c]{0.75\textwidth}
  {
   \begin{center}
    \includegraphics[width=0.75\textwidth]{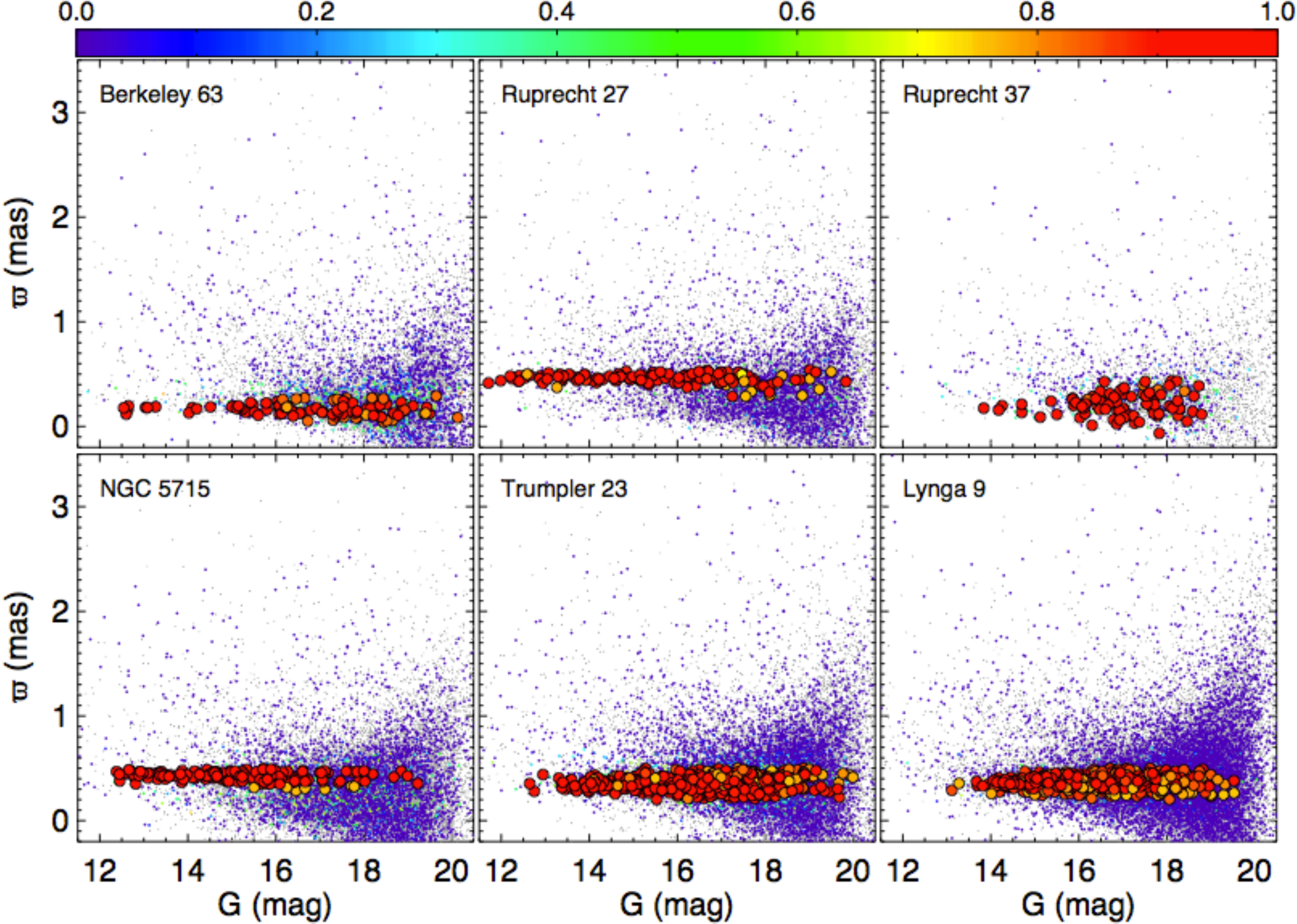}    
    \end{center}
    
  }
\caption{ Parallax versus $G$ magnitude for 6 investigated OCs. Symbols convention is the same of Fig.~\ref{fig:CMDs_part1}. }

\label{fig:plx_vs_G_part1}
\end{center}
\end{figure*}

The four analysis steps described above were applied to our complete sample and the results are illustrated in Figs.~\ref{fig:CMDs_part1} to \ref{fig:plx_vs_G_part1} for six investigated OCs. In some cases (e.g., Berkeley\,63, Trumpler\,23 and Lynga\,9), the centroid location of the member stars in the clusters' VPD is compatible with the bulk motion of the field (see Fig.~\ref{fig:VPDs_part1}), which makes the disentanglement between cluster and field populations more difficult due to lower contrasts. This results in a higher number of outliers, that is, stars with appreciable membership likelihoods but with $G$ and $(G_{\textrm{BP}}-G_{\textrm{RP}})$ values incompatible with the evolutionary sequences defined in the decontaminated CMDs. To alleviate this behavior, in these cases we employed colour filters (blue dashed lines in Fig.~\ref{fig:CMDs_part1}) to both cluster and control field samples (in addition to the VPD box; see Section~\ref{sec:preanalysis} and Fig.~\ref{fig:VPD_Ruprecht63_original_and_filtered}) previously to the run of the decontamination method. In the cases of Berkeley\,63, Trumpler\,23 and Lynga\,9, the colour filters are useful to remove very reddened stars, which results in clearer evolutionary sequences.

Figs.~\ref{fig:VPDs_part1} and \ref{fig:plx_vs_G_part1} exhibit separately the astrometric space of proper motions and parallax (the latter is plotted as function of $G$ magnitude) corresponding to member and non-member stars inside cluster's $r_t$. Stars in the control field are also shown. The same symbol convention of Figs.~\ref{fig:ilustra_decontam_method} and \ref{fig:CMDs_part1} was employed. We can see that, as expected, member stars  form conspicuous clumps in the VPDs, present compatible parallaxes and define recognizable sequences in the  CMDs.

%%%%%%%%%%%%%%%%%%%%%%%%%%%%%%%%%%%
\subsection{Comparison with previous studies}
\label{compara_previous_studies}
%%%%%%%%%%%%%%%%%%%%%%%%%%%%%%%%%%%

%%%%%%%%%%%%%%%%%%%%%%%
\subsubsection{Mean astrometric values}
%%%%%%%%%%%%%%%%%%%%%%%

Using \textit{Gaia} DR2, CJV2018 obtained a list of members and mean astromeric parameters for 1212 OCs. Their method (detailed in \citeauthor{Cantat-Gaudin:2018a}\,\,2018a) is based on applying an unsupervised membership assignment code, UPMASK \citep{Krone-Martins:2014}, to the astrometric data contained within the fields of those clusters. The main assumption is that member stars of a physical system must be more tightly distributed in the astrometric space than a random distribution. In this way, two main steps are executed: (1) the $k-$means clustering algorithm (e.g., \citeauthor{Lloyd:1982}\,\,\citeyear{Lloyd:1982}) is used to identify groups of stars with similar parallaxes and proper motion components; (2) it is then verified whether the distribution of stars in each of these groups is more concentrated than a random distribution. After that, random offsets are applied to each datapoint in both parallax and proper motions components and the overall procedure is repeated. After 10 iterations, a membership probability is assigned to each star in the cluster region.
%No paper do Cantat fala que são 100 iterações, mas o referee acha que são 10.....

\begin{figure*}
\begin{center}

\parbox[c]{0.59\textwidth}
  {
   \begin{center}
    \includegraphics[width=0.59\textwidth]{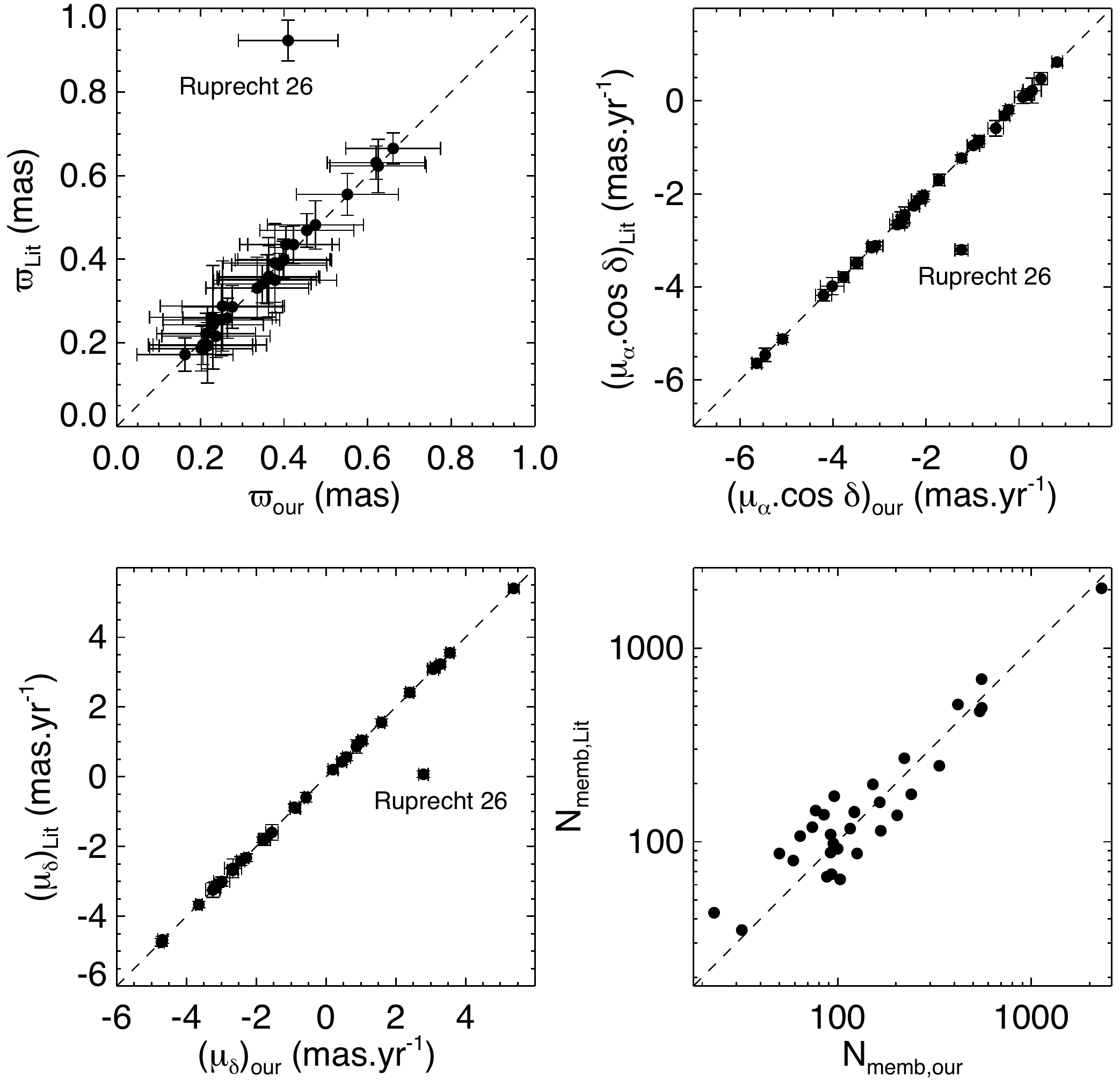}    
    \end{center}
    
  }
\caption{ Comparisons between the mean astrometric parameters $\langle\varpi\rangle, \langle\mu_{\alpha}\,\textrm{cos}\,\delta\rangle$, $\langle\mu_{\delta}\rangle$ obtained in this work ($X$-axis) and in CJV2018 ($Y$-axis). The outlier is the OC Ruprecht\,26 (see text for details). The rightmost plot in the second line compares the number of member stars determined for each cluster. Thirty-two OCs are common to both studies. In all panels, the dashed line is the identity locus. }

\label{fig:compara_com_CGaudin}
\end{center}
\end{figure*}

%In step (2), the comparisons between cluster and random distributions are performed using the total branch length of the minimum spanning tree connecting all stars in each group.

Compared to our decontamination method (Section~\ref{sec:membership_assignment}), the main differences in relation to Cantat-Gaudin et al.'s procedure are the sampling of the astrometric space (in our case, we employ uniform grids with cells of varying sizes) and the selection of control field stars to be statistically compared with stars in the OC region; in our case, we take into account the real movement and parallax distribution of the field instead of using random samples. Despite these differences, both methods return very similar results. Thirty-two OCs in our complete sample (Table~\ref{tab:investig_sample}) were also investigated by CJV2018 (6 of our OCs are absent in their catalogue, namely: Czernik\,7, L19\,2326, Ruprecht\,152, UBC\,296, $[\textrm{FSR}2007]\,0101$ and Berkeley\,84).

Fig.~\ref{fig:compara_com_CGaudin} exhibits a comparison between the mean astrometric parameters derived in both studies for coincidental OCs. We also compare the number of member stars in our study ($N_{\textrm{memb,our}}$) and in CJV2018 ($N_{\textrm{memb,Lit}}$). In order to determine $N_{\textrm{memb,Lit}}$, we considered stars with membership probabilities $P\ge50\%$ in CJV2018's tables. We can see that almost the same mean values of $\langle\varpi\rangle, \langle\mu_{\alpha}\,\textrm{cos}\,\delta\rangle$ and $\langle\mu_{\delta}\rangle$ were obtained, except for Ruprecht\,26. This OC is projected in the same region of the OC L19\,2326, recently catalogued by \cite{Liu:2019}. As discussed in more detail in Appendix A of the online  supplementary material, we believe that CJV2018 has mistakenly identified Ruprecht\,26 as L19\,2326, leading to the discrepancy found in Fig.~\ref{fig:compara_com_CGaudin}. Their reported central coordinates for Ruprecht\,26 are displaced 5$\arcmin$ north in relation to the literature values, putting it much closer to L19\,2326 than to the original cluster. Furthermore, their reported astrometric parameters for the Ruprecht\,26 actually matches the values found by both our analysis and by \cite{Liu:2019} for L19\,2326 (not Ruprecht\,26). The rightmost panel in the second line of Fig.~\ref{fig:compara_com_CGaudin} exhibits a general agreement regarding the number of member stars for each OC as determined in the present study and in CJV2018.

%As discussed in detail in Appendix\,A of the online Supplementary material, the object named as Ruprecht\,26 in CJV2018 catalogue is, in reality, the OC L19\,2326. This is the reason for the discrepancy found in Fig.~\ref{fig:compara_com_CGaudin}. Also in Appendix\,A, comments on some individual clusters are presented. The rightmost panel in the second line of Fig.~\ref{fig:compara_com_CGaudin} exhibits a general agreement regarding the number of member stars for each OC as determined in the present study and in CJV2018.    }

%For our data ($X$-axis), the error bars correspond to the intrinsic dispersions (i.e., individual measurement errors have been considered) of the astrometric data for all member stars. For parallaxes, we summed in quadrature an uncertainty of 0.1\,mas systematically affecting the astrometric solution in \textit{Gaia} DR2 \citep{Luri:2018}. The uncertainties informed in CJV2018 ($Y-$axis) were determined from the standard deviation of the mean values after applying a thousand random redrawings where stars were picked according to their probability of being cluster members , after applying a 2\,$\sigma$ clipping algorithm to reject outliers among member stars (see section 3.1 of \citeauthor{Cantat-Gaudin:2018a}\,\,2018a).   

For our data ($X$-axis), the error bars correspond to the intrinsic dispersions (i.e., individual measurement errors have been considered) of the astrometric data for all member stars. For parallaxes, we summed in quadrature an uncertainty of 0.1\,mas systematically affecting the astrometric solution in \textit{Gaia} DR2 \citep{Luri:2018}. The uncertainties informed in CJV2018 ($Y-$axis) were determined from the standard deviation of the mean values after applying a thousand random redrawings where stars were picked according to their probability of being cluster members, after applying a 2\,$\sigma$ clipping algorithm to reject outliers among member stars (see section 3.1 of \citeauthor{Cantat-Gaudin:2018a}\,\,2018a).

%%%%%%%%%%%%%%%%%%%%%%%
\subsubsection{Fundamental parameters}
%%%%%%%%%%%%%%%%%%%%%%%

\begin{figure*}
\begin{center}

\parbox[c]{1.00\textwidth}
  {
   \begin{center}
    \includegraphics[width=1.00\textwidth]{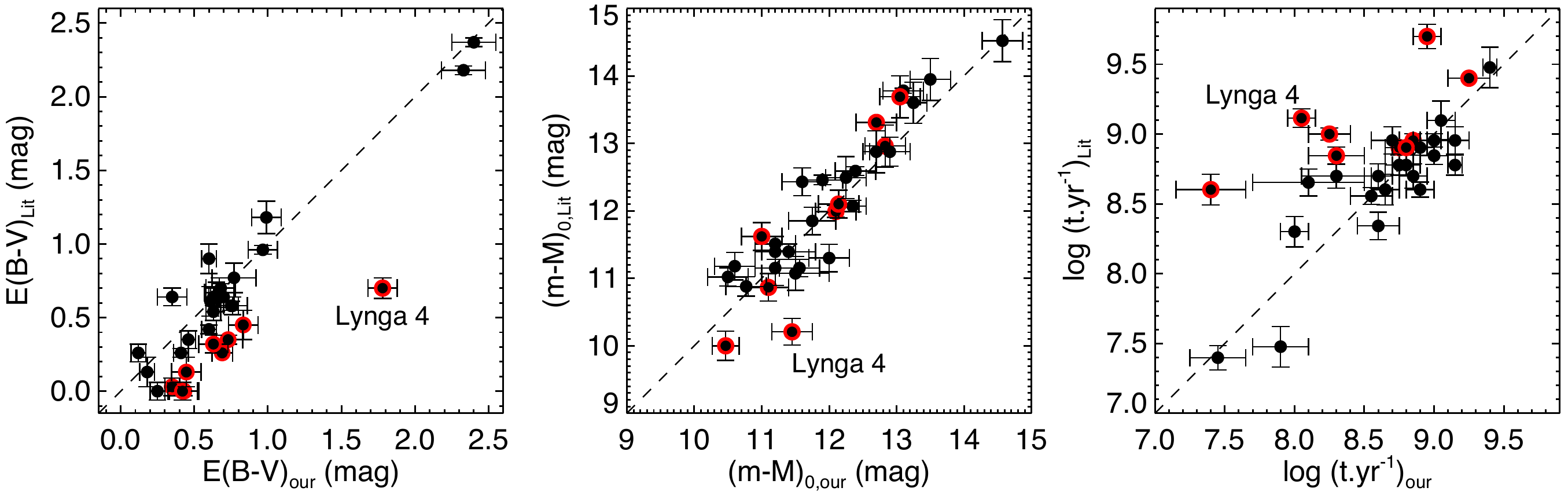}    
    \end{center}
    
  }
\caption{ Same of Fig.~\ref{fig:compara_com_CGaudin}, but comparing the fundamental parameters ($E(B-V)$, $(m-M)_0$ and log\,$t$) derived in the present paper and in the literature (see Table\,\ref{tab:previous_lit_information} for references). The 9 most discrepant clusters in the low-$E(B-V)$ domain are marked with red circles in the left panel. In this same panel, the most discrepant one, Lynga\,4, is identified (see text for details). Czernik\,23 and Ruprecht\,34 present almost the same reddening determined in the present study ($E(B-V)_{\textrm{our}}\sim0.4\,$mag) and $E(B-V)_{\textrm{lit}}=0.0$, therefore their symbols are overplotted in the left panel. The same 9 clusters are identified in the middle and right panels. }

\label{fig:compara_com_BicaBonatto}
\end{center}
\end{figure*}

Thirty-one clusters in our main sample (Table~\ref{tab:investig_sample}) had their fundamental parameters determined from near infrared $J\times(J-H)$ and $J\times(J-K_s)$ CMDs in BBC. These previous values and some additional observations are given in Table\,\ref{tab:previous_lit_information}. Three OCs are absent, namely: L19\,2326, NGC\,2477 and UBC\,296. Fig.~\ref{fig:compara_com_BicaBonatto} compares the results obtained in the present paper with the literature ones. We can see that, although there is a general rough agreement between both datasets, some severe discrepancies (greater than 1.0\,mag in $E(B-V)$ and $(m-M)_0$ and greater than 1.0\,dex in log\,$t$) are notable. 

In order to explore the differences between our parameters and the literature ones, we have highlighted in Fig.~\ref{fig:compara_com_BicaBonatto} nine OCs (namely, Czernik\,23, Czernik\,24, Ruprecht\,27, Ruprecht\,34, Ruprecht\,35, Ruprecht\,54, Lynga\,4, Trumpler\,26 and Ruprecht\,174) for which discrepancies larger than 0.3\,mag in $E(B-V)$ have been obtained. Lynga\,4 is discussed in more detail later in this section.

Particularly, in the low-$E(B-V)$ domain, the values derived in the present paper are systematically larger than those listed in the literature. A possible reason for this is that the near infrared colours are considerably less affected by variations in $E(B-V)$ compared to the optical ones. Considering the $(J-H)$ and $(J-K_s)$ colour indexes and the extinction relations of \cite{Rieke:1985}, even $E(B-V)$ values as large as $\sim$0.4\,mag results in $E(J-H)$ and $E(J-K_S)$ of about $\sim$0.15\,mag. Consequently, these variations in interstellar reddening do not severely impact the isochrone fitting procedure, which was implemented by means of visual inspection of photometrically decontaminated CMDs in BBC.

We can note that our derived $(m-M)_0$ are reasonably compatible with previous studies. With respect to log\,$t$ and considering the whole sample, there are not noticeable systematic differences between the values  derived here and those in the literature. Taking only the 9 highlighted OCs, the systematic differences in $E(B-V)$ resulted in our log\,$t$ values being smaller than those derived by BBC. This may be consequence of degeneracy among the fundamental parameters derived via isochrone fitting.

Beyond that, these discrepancies may also be attributed to differences in the process of member stars identification and the data used. BBC's decontamination method employs 2MASS photometry and consists on statistical comparisons between CMDs built for stars in the inner area of the investigated OCs and the corresponding ones built for stars in equal area offset fields. Outliers in the resulting decontaminated CMDs are then removed with the use of colour filters and the OC's fundamental parameters are derived with visual isochrone fitting of solar metallicity Padova \citep{Girardi:2002} isochrones.

\subsubsection*{Lynga\,4}
%%%%%%%%%%%%%%%%%%%%%%%%%%%%%%%%%%%%%%%%%%%%%%

In the left panel of Fig.~\ref{fig:compara_com_BicaBonatto}, we can see that Lynga\,4 is the most discrepant OC in comparison to the literature results. Figure 7 of \cite{Bonatto:2007} shows the $J\times(J-H)$ CMD of this OC, which is severely contaminated by red stars from the Galactic disc. Its cleaned CMD, after applying \citeauthor{Bonatto:2007}'s\,\,(\citeyear{Bonatto:2007}) decontamination method, presents considerable residual contamination, which are removed with the use of colour filters. In their CMD, we can note that 7 giant stars ($J\lesssim11.2$\,mag; 0.6  $\lesssim$ $(J-H)$ $\lesssim$ 1.1\,mag) provide critical constraints for isochrone fitting. Among them, the 4 bluest ones ($(J-H)<0.8\,$mag) are not part of our list of members. The astrometric data for these 4 stars are: ($\varpi$,\,$\mu_{\alpha}\,\textrm{cos}\,\delta$,\,$\mu_{\delta}$) = $(0.4181\,\pm\,0.0319,\,-2.544\,\pm\,0.062,\,-5.431\,\pm\,0.054)_{\#1}$, $(0.6791\,\pm\,0.0406,\,-4.934\,\pm\,0.079,\,-9.065\,\pm\,0.073)_{\#2}$, 
$(0.7702\,\pm\,0.0409,\,-5.337\,\pm\,0.078,\,-6.321\,\pm\,0.072)_{\#3}$, $(0.5171\,\pm\,0.0262,\,-9.293\,\pm\,0.050,\,-7.461\,\pm\,0.046)_{\#4}$. As usual, these above values for parallax and proper motion components are given in mas and mas\,yr$^{-1}$, respectively. These 4 stars have received null membership likelihoods after applying our decontamination method (Section~\ref{sec:membership_assignment}), since their proper motions are incompatible with the bulk movement of Lynga\,4: ($\langle\mu_{\alpha}\,\textrm{cos}\,\delta\rangle,\,\langle\mu_{\delta}\rangle$) $\sim$ (-4.0,\,-3.0; see Appendix B).

%($\langle\mu_{\alpha}\,\textrm{cos}\,\delta\rangle,\,\langle\mu_{\delta}\rangle$) $\sim$ ($
%-4.017\,\pm\,0.255,\,-2.990\,\pm\,0.228$; see Appendix B).}

The 3 giants located in the interval $(J-H)>0.8\,$mag in \citeauthor{Bonatto:2007}'s\,\,(\citeyear{Bonatto:2007}) CMD are coincident with our list of members (three member stars in the range $G<13.7$ shown in the CMD of Fig.~B7). This suggests that a larger reddening value would be necessary in the literature $J\times(J-H)$ CMD in order to properly fit this group of 3 stars. For Lynga\,4, in fact the severe difference in the derived $E(B-V)$ also leads to considerable discrepancies in other parameters, as shown in the middle and right panels of Fig.~\ref{fig:compara_com_BicaBonatto}. For comparison, we have overplotted in our decontaminated CMD (Fig.~B7) of Lynga\,4 a solar metallicity PARSEC isochrone, which have been shifted according to the fundamental parameters informed in \cite{Bonatto:2007}. It is noticeable a huge discrepancy between them, caused mainly by differences in the lists of member stars identified in both studies. 

%\textbf{A more detailed scrutiny about the origin of the discrepancies showed in Fig.~\ref{fig:compara_com_BicaBonatto} would require the lists of member stars identified in Bica, Bonatto \& Camargo's papers, which are not available.} 

The present paper has the advantage of relying on more recent high-precision data and combining astrometric and photometric information. The astrometric information provides stronger observational constraints for identification of member stars of an OC, since they share common distances and proper motions independently of their spectral types and reddening. This is particularly useful for OCs projected against crowded fields; in these cases, purely photometric decontamination methods are significantly affected by fluctuations in the luminosity function of the field population \citep{Maia:2010}, which may result in a considerable number of outliers. In general, our CMDs present clearer evolutionary sequences and $\sim2-4\,$mag deeper main sequences, which allows better constraints for isochrone fitting. Besides, BBC employ solar metallicity isochrones, for simplicity, while our method allows different metallicities. Discrepancies in relation to our results may also be attributed to different sets of isochrones and CMD fitting techniques.

%%%%%%%%%%%%%%%%%
\subsubsection*{Trumpler\,23}
%%%%%%%%%%%%%%%%%

The OC Trumpler\,23 is worth mentioning since, after \cite{Bonatto:2007}, it was investigated by \cite{Overbeek:2017}, who employed spectroscopic data obtained with the VLT FLAMES spectrograph, as part of the public \textit{Gaia}-ESO survey (GES; \citeauthor{Gilmore:2012}\,\,\citeyear{Gilmore:2012}; \citeauthor{Randich:2006}\,\,\citeyear{Randich:2006}). CMDs were built with the use of $VI$ photometric data from \cite{Carraro:2006}. Potential member stars were identified from their radial velocities ($V_{\textrm{rad}}$). After determining the systemic velocity of the cluster and membership of individual stars, 10 of them were identified as fiducial members, as inferred from the coherence between $V_{\textrm{rad}}$ and metallicity ($[Fe/H]$; see their figure 3). 

Nine of these stars are also present in our sample of member stars of Trumpler\,23. The only discrepant star receives the GES identification 16003885-5334507 (\textit{Gaia} DR2 designation 5980824255575234816), which received null astrometric likelihood in our method since it presents $\mu_{\alpha}\,\textrm{cos}\,\delta\,=\,-6.697\,\pm\,0.075\,$mas\,yr$^{-1}$. This value is discrepant with the bulk movement defined by Trumpler\,23 member stars (Fig.~\ref{fig:VPDs_part1}).

The fundamental parameters found by \cite{Overbeek:2017}, using PARSEC isochrones (their table 1), resulted: $(m-M)_{0}\,=\,11.61\,\pm\,0.21\,$mag, log\,$t$\,=\,$8.90\,\pm\,0.13$ and $E(B-V)=0.82\,\pm\,0.09\,$mag. Considering the fiducial members, they found an average cluster metallicity of $[Fe/H]=0.14\,\pm0.03\,$dex, with a typical stellar error of 0.10\,dex in $[Fe/H]$ (as measured from high-resolution spectra obtained with the UVES spectrograph). We can see that our results are consistent with the literature ones, considering uncertainties.

%%%%%%%%%%%%%%%%%
\subsubsection*{Lynga\,9}
%%%%%%%%%%%%%%%%%

Before \cite{Bonatto:2007}, Lynga\,9 was identified as an asterism by \cite{Carraro:2005a}, based on photometric $BVI$ data. They applied a decontamination method which consists on subtracting CMDs constructed for cluster and control field (same area) stars. For each star in the field region, their method identifies the closest cluster star in terms of magnitude and colour index and removes it from the cluster CMD.

Since Lynga\,9 is projected against a highly populated background, their method resulted in almost all stars being removed from their sample. Only a small group, which resembles a red clump, is present in their decontaminated CMDs (figures 11 and 12 of \citeauthor{Carraro:2005a}\,\,\citeyear{Carraro:2005a}). No proeminent main sequence is present, as would be expected in the case of a real stellar aggregate, considering the completeness level of their photometry. Therefore, \citeauthor{Carraro:2005a}'s\,\,(\citeyear{Carraro:2005a}) results point out the presence of a possible asterism.

Contrarily to their conclusions, the decontaminated CMD obtained in the present paper for Lynga\,9 (Fig.~\ref{fig:CMDs_part1}) exhibit clear sequences of a $\sim1\,$Gyr cluster with subsolar metallicity ($[Fe/H]\sim-0.1$). The concentration of its member stars on the VPD (Fig.~\ref{fig:VPDs_part1}) and $\varpi\,$versus $G\,$magnitude plot (Fig.~\ref{fig:plx_vs_G_part1}) is also characteristic of a genuine OC. Our main conclusion regarding the physical nature of Lynga\,9 is in agreement with \cite{Bonatto:2007}.

\begin{figure}
\begin{center}
 \includegraphics[width=8.5cm]{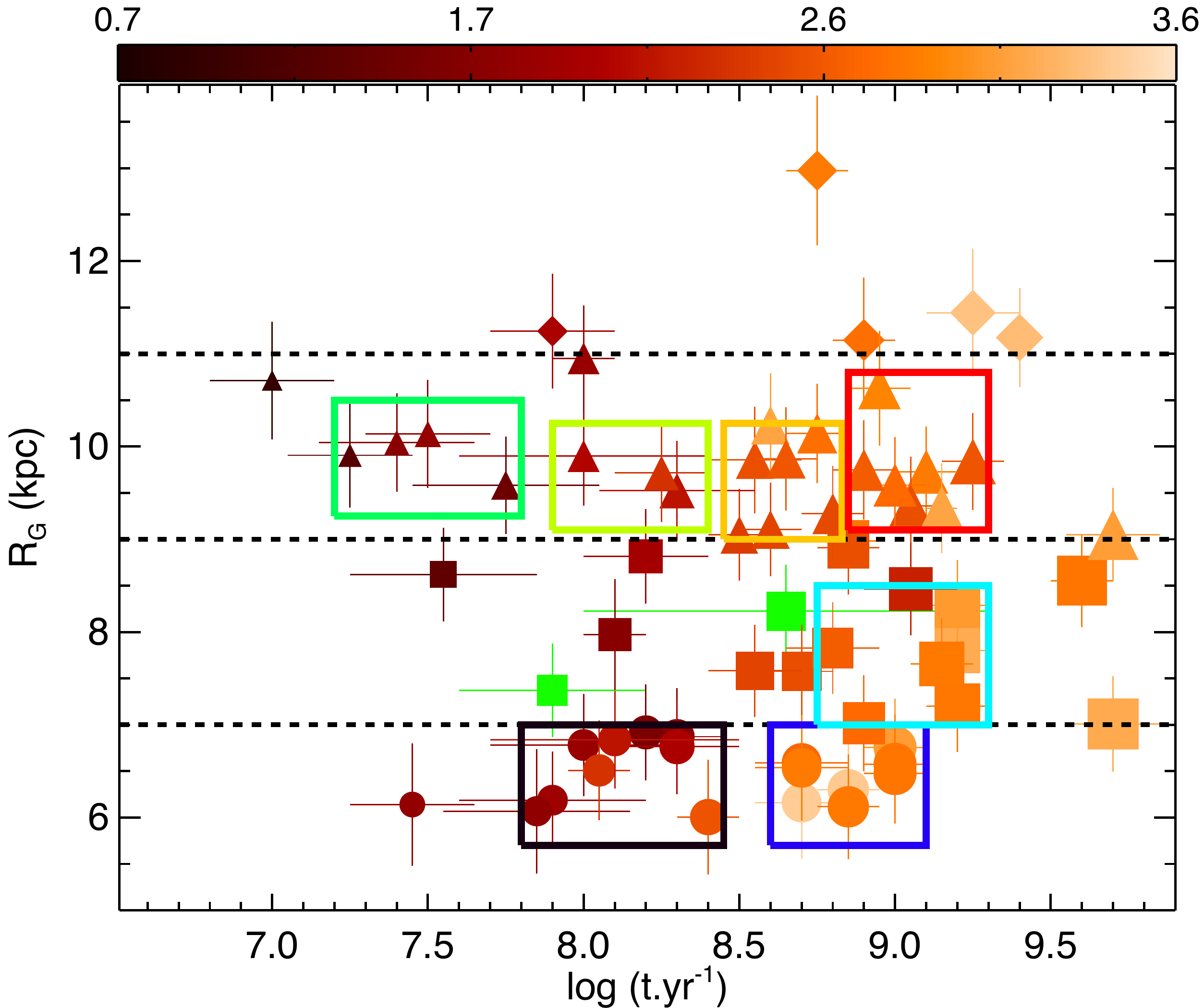}
 \caption{ Galactocentric distance versus age for the our complete sample of 65 OCs. Symbol colours were given according to the clusters evolutionary stage (see text for details), as indicated by the colourbar. Symbol sizes are proportional to log\,$t$. The green symbols represent the OCs Herschel\,1 and Lynga\,2, for which no $r_c$ and $r_{hm}$ could be derived in Paper I. The horizontal dashed lines are plotted for reference. The coloured boxes identify 7 groups (black, red, yellow, dark green, light green, dark blue and light blue rectangles) of coeval OCs with compatible $R_G$. }
   \label{Fig:R_gal_versus_age_groups}
\end{center}
\end{figure}

%%%%%%%%%%%%%%%%%%%%%%%%%%%%%%%%%%%%
\section{Discussion}
\label{sec:discussion}
%%%%%%%%%%%%%%%%%%%%%%%%%%%%%%%%%%%%

%%%%%%%%%%%%%%%%%%%%%%%%%%%%%%%%%%%%
\subsection{The investigated sample in the Milky Way context}
%%%%%%%%%%%%%%%%%%%%%%%%%%%%%%%%%%%%

Fig.~\ref{Fig:R_gal_versus_age_groups} shows the $R_G$ and ages for our complete investigated sample of 65 objects (27 analysed in Paper I and 38 in the present work). Our $R_G$ values vary from $\sim$6 to 13\,kpc and clusters ages vary from 10\,Myr to $\sim5\,$Gyr. Different symbols represent different $R_G$ bins, separated by the horizontal dashed lines in this Figure. Symbol sizes are proportional to log\,$t$. Colours were attributed according to the clusters evolutionary stage, as determined by the dynamical ratio $\tau_{\textrm{dyn}}$=age/t$_{\textrm{cr}}$, following the scheme indicated by the colourbar in the top of Fig.~\ref{Fig:R_gal_versus_age_groups}: darker colours for dynamically younger objects and lighter ones for those dynamically older OCs. 
%Values on the colourbar (see also Fig.~\ref{Fig:discussions_bloco2}) indicate log\,($\tau_{\textrm{dyn}}$)}. 

The crossing time $t_{\textrm{cr}}=r_{hm}/\sigma_{\textrm{v}}$ is the dynamical timescale for cluster stars to perform an orbit across the system. $\sigma_{\textrm{v}}$ is the 3D velocity dispersion of member stars, which was determined from the dispersion in proper motions (Table \ref{tab:investig_sample}) assuming that velocity components relative to each cluster centre are isotropically distributed. Based on this assumption, $\sigma_{\textrm{v}}=\sqrt{3/2}\,\sigma_{\mu}$, where $\sigma_{\mu}$ is the dispersion of the projected angular velocities $\mu=\sqrt{\mu_{\alpha}^2\textrm{cos}^2\,\delta+\mu_{\delta}^2}$. We have employed the procedure  described in \cite{Sagar:1989} and in section 4 of \cite{van-Altena:2013} to properly take into account the uncertainties in the proper motion components when deriving $\sigma_{\mu}$. The coloured rectangles in Fig.~\ref{Fig:R_gal_versus_age_groups} delimit groups of coeval OCs with compatible $R_G$ values, which will be employed in the following discussions. %(Figs.~\ref{Fig:discussions_bloco2} and \ref{Fig:discussions_bloco3}).      

\begin{figure*}
\begin{center}

\parbox[c]{1.0\textwidth}
  {
   \begin{center}
    \includegraphics[width=1.00\textwidth]{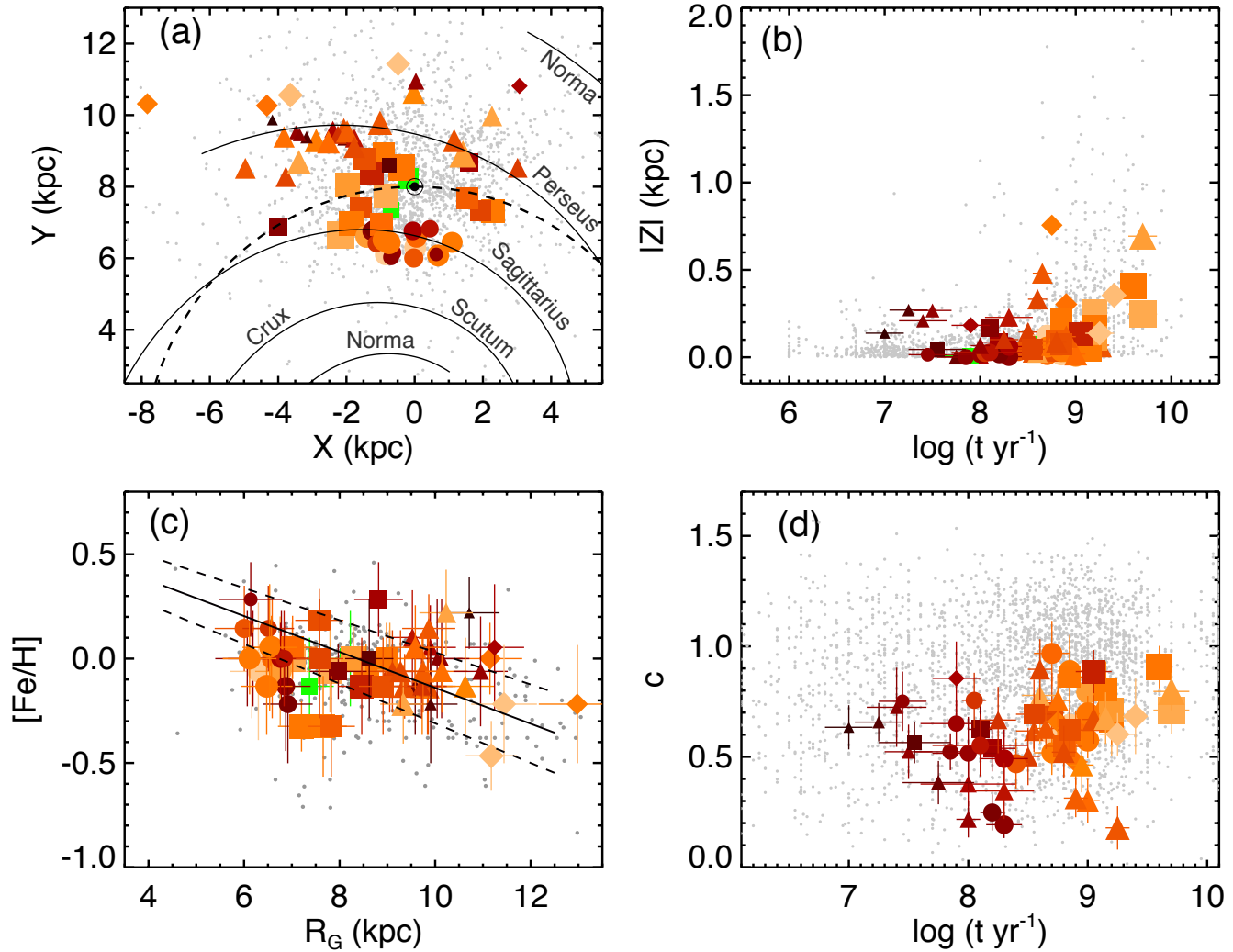}    
    \end{center}    
  }
\caption{ Panel (a): Position of the investigated OCs along the Galactic plane. The schematic position of the spiral arms, together with their identifiers, were obtained from \citeauthor{Vallee:2008}\,\,(\citeyear{Vallee:2008}). Colours are given according to the OCs dynamical stages (see text for details): darker colours for dynamically younger objects and lighter for more evolved ones. The location of the Sun ($R_G$$\,\sim\,$8\,kpc) and the solar circle are identified. Small grey symbols are OCs taken from the DAML02 catalogue. Panel (b): Vertical distance to the Galactic disc versus age plot. Panel (c): Metallicity versus $R_G$ plot. The continuous and dashed lines represent, respectively, the radial metallicty gradient and its uncertainty as derived by \citeauthor{Netopil:2016}\,\,(\citeyear{Netopil:2016}). Panel (d): Concentration parameter ($c=\textrm{log}\,(r_t/r_c)$) versus age plot. The grey symbols represent OCs taken from \citeauthor{Kharchenko:2013}\,\,(\citeyear{Kharchenko:2013}). }

\label{Fig:discussions_bloco1}
\end{center}
\end{figure*}

Fig.~\ref{Fig:discussions_bloco1}, panel (a), exhibits the location of our investigated OCs along the Galactic plane. The position of the Sun is indicated and the solar circle is represented by the dashed line, together with the schematic representation of the spiral arms \citep{Vallee:2008}. As in DAML02 catalogue (whose OCs are plotted as small grey circles), most of our sample is located close to the Sagittarius and Perseus arms or in the interarm region. Panel (b) exhibits their distance perpendicularly to the Galactic plane ($\vert Z\vert$) as function of age. Twenty-eight investigated OCs ($\sim43$\% of our sample) are located within the vertical scale-height of 60\,pc \citep{Bonatto:2006}; 10 OCs ($\sim15$\%) are located at vertical distances between 60$-$100\,pc and 27 ($\sim42$\%) of them are more than 100\,pc (and less than $\sim800\,$pc) distant from the Galactic disc. As expected, there is a general trend in which the older clusters tend to be located farther away from the disc.

Panel (c) of Fig.~\ref{Fig:discussions_bloco1} exhibits the derived $[Fe/H]$ (Section \ref{sec:fundamental_params_determ}) plotted as function of $R_G$. Considering uncertainties, most of our $[Fe/H]$ values located within the limits of the radial gradient derived by \citeauthor{Netopil:2016}\,\,(\citeyear{Netopil:2016}; the continuous and dashed lines represent, respectively, the fit and its uncertainties) and the dispersion of our $[Fe/H]$ is compatible with that exhibited by OCs taken from the literature (DAML02; grey symbols). The concentration parameters ($c=$log$(r_t/r_c)$) are shown in panel (d) of Fig.~\ref{Fig:discussions_bloco1}. Compared to  OCs taken from the literature (in this panel, the $c$ values were determined from \citeauthor{Kharchenko:2013}\,\,\citeyear{Kharchenko:2013}, since DAML02 do not provide $r_t$ or $r_c$ values), our clusters present low ($c\sim0.2$) to moderately high concentration parameters ($c\sim1.0$).

%%%%%%%%%%%%%%%%%%%%%%%%%%%%%%%%%%%%%%
\subsection{Investigating structural and time-related parameters}
\label{sec:investig_struct_params}
%%%%%%%%%%%%%%%%%%%%%%%%%%%%%%%%%%%%%%

\begin{figure*}
\begin{center}

\parbox[c]{1.0\textwidth}
  {
   \begin{center}
    \includegraphics[width=1.00\textwidth]{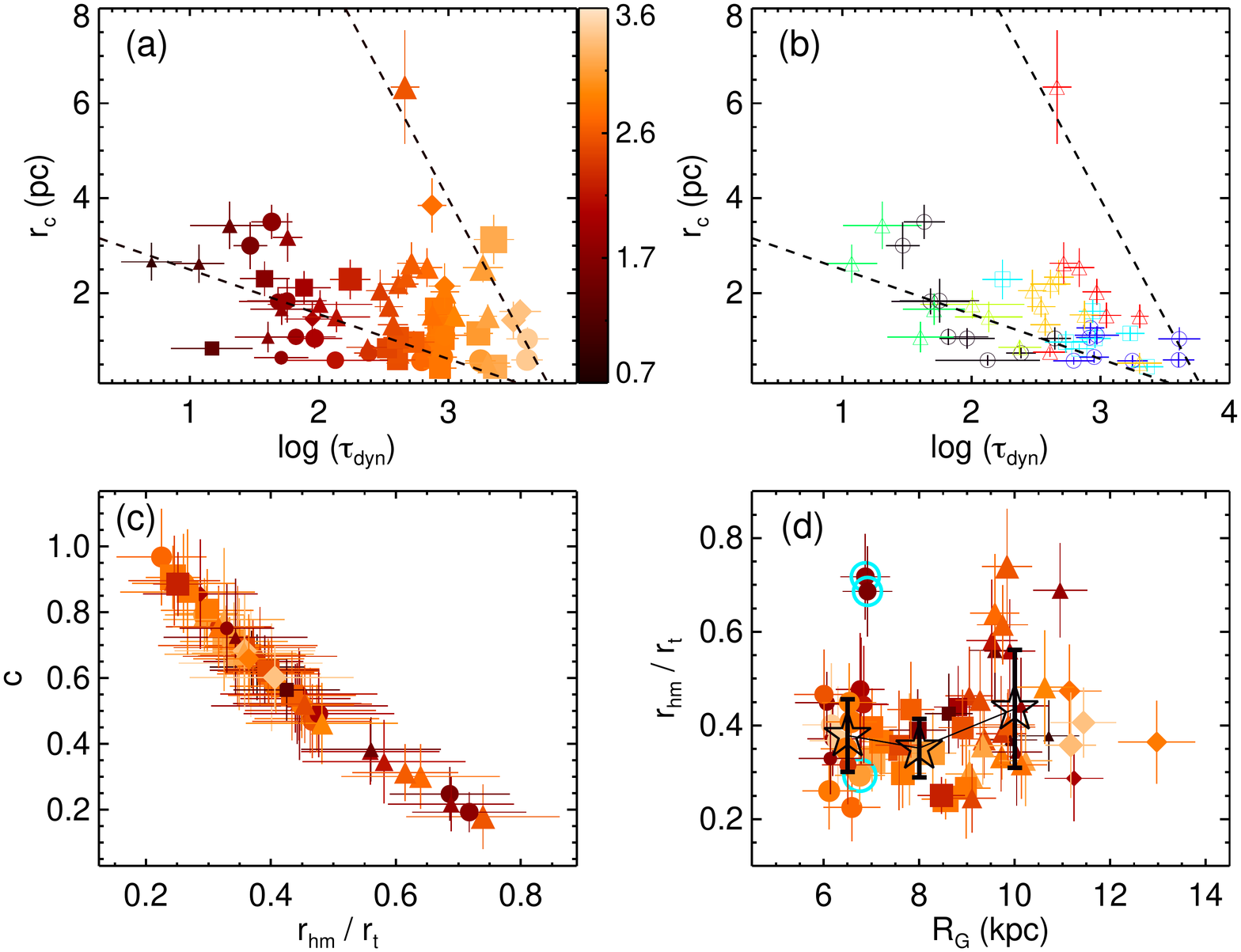}    
    \end{center}    
  }
\caption{ Panel (a): $r_c$ versus logarithmic $\tau_{\textrm{dyn}}$. The dashed lines were plotted just for reference. The colour scale for the symbols was constructed according to the cluster's dynamical ratio. The numerical values in the colourbar indicate log\,($\tau_{\textrm{dyn}}$). Panel (b): same as panel (a), but showing the groups of coeval OCs delimited in Fig.~\ref{Fig:R_gal_versus_age_groups}. Panel (c): Concentration parameter versus $r_{hm}/r_t$ plot. Panel (d): $r_{hm}/r_t$ as function of Galactocentric distance. The vertical position of the open black stars and the associated uncertainties represent, respectively, the mean and dispersion (1\,$\sigma$) of the $r_{hm}/r_t$ ratios for 3 regions: $R_G\leq7\,$kpc, $7<R_G\,(\textrm{kpc})\leq9\,$ and $R_G>9\,$kpc. The encircled symbols (see text for details) represent the OCs NGC\,5617, Trumpler\,22 (both with $r_{hm}/r_t\sim0.7$) and Pismis\,19 ($r_{hm}/r_t\sim0.3$). }

\label{Fig:discussions_bloco2}
\end{center}
\end{figure*}

%As two-body encounters conduct the cluster internal relaxation process, higher mass stars tend to sink towards the system inner regions while lower mass ones move to the cluster outskirts and are preferentially lost leading to the cluster evaporation (e.g., \citeauthor{Spitzer:1969}\,\,\citeyear{Spitzer:1969}; \citeauthor{Portegies-Zwart:2010}\,\,\citeyear{Portegies-Zwart:2010}; \citeauthor{de-la-Fuente-Marcos:2013}\,\,\citeyear{de-la-Fuente-Marcos:2013}). Consequently, there are expected trends relating the core radius ($r_c$) and the dynamical ratio $\tau_{\textrm{dyn}}$=age/t$_{\textrm{cr}}$, where the crossing time $t_{\textrm{cr}}=r_{hm}/\sigma_{\textrm{v}}$ is the dynamical timescale for cluster stars to perform an orbit across the system. $\sigma_{\textrm{v}}$ is the 3D velocity dispersion of member stars, which was determined from the dispersion in proper motions (Table \ref{tab:investig_sample}) assuming that velocity components relative to each cluster centre are isotropically distributed. Based on this assumption, $\sigma_{\textrm{v}}=\sqrt{3/2}\,\sigma_{\mu}$, where $\sigma_{\mu}$ is the dispersion of the projected angular velocities $\mu=\sqrt{\mu_{\alpha}^2\textrm{cos}^2\,\delta+\mu_{\delta}^2}$. We have employed the procedure  described in \cite{Sagar:1989} and in section 4 of \cite{van-Altena:2013} to properly take into account the uncertainties in the proper motion components when deriving $\sigma_{\mu}$. 

As two-body encounters conduct the cluster internal relaxation process, higher mass stars tend to sink towards the system inner regions while lower mass ones move to the cluster outskirts and are preferentially lost leading to the cluster evaporation (e.g., \citeauthor{Spitzer:1969}\,\,\citeyear{Spitzer:1969}; \citeauthor{Portegies-Zwart:2010}\,\,\citeyear{Portegies-Zwart:2010}; \citeauthor{de-la-Fuente-Marcos:2013}\,\,\citeyear{de-la-Fuente-Marcos:2013}). Consequently, there are expected trends relating the core radius ($r_c$) and the dynamical ratio $\tau_{\textrm{dyn}}$.

In this sense, the larger the $\tau_{\textrm{dyn}}$, the more the cluster age surpasses $t_{\textrm{cr}}$ and thus the more dynamically evolved a system is. In Fig.~\ref{Fig:discussions_bloco2}, panel (a), we can see a general anticorrelation between $r_c$ and $\tau_{\textrm{dyn}}$. Most of our data are confined between the dashed lines, which were plotted just to guide the eye. There is an apparent shrinking of $r_c$ for the more dynamically evolved clusters regardless the Galactocentric distance, which is expected if the evolution is dominated by the internal relaxation process (e.g., \citeauthor{Heggie:2003}\,\,\citeyear{Heggie:2003}).

%Besides, from the disposal of our data along the $X$-axis in panels (b) and (d) of Fig.~\ref{Fig:discussions_bloco1}, we can see that there is a clear correlation between clusters ages and the dynamical ratio. In fact, older clusters (larger symbols) are plotted with lighter colours, which indicates more advanced dynamical stages.     

%The impact of the external gravitational forces on cluster structure was investigated by \cite{Miholics:2014}, who performed $N-$body simulations of star clusters formed in a dwarf galaxy (for which the model parameters emulate the potential of the Large Magellanic Cloud; LMC) and then accreted by the Milky Way. To examine how changes in the external potential can affect the inner region of a cluster, they also performed a set of simulations where a modeled cluster evolves in orbits at different Galactocentric distances. The same was performed by submitting the modeled cluster to the LMC potential. They saw that there exists little difference between the $r_c$ of the cluster in the LMC and $r_c$ of the cluster that evolves entirely in the Milky Way (their figure 6). Even when the cluster is switched from the LMC to the Milky Way, the $r_c$ does not change appreciably. This indicates that the cluster's inner region is reasonably insensitive to the external tidal conditions in which it lives. Consequently, the system's inner structure is more importantly determined by the internal relaxation process. 

\begin{figure*}
\begin{center}

\parbox[c]{1.0\textwidth}
  {
   \begin{center}
    \includegraphics[width=1.00\textwidth]{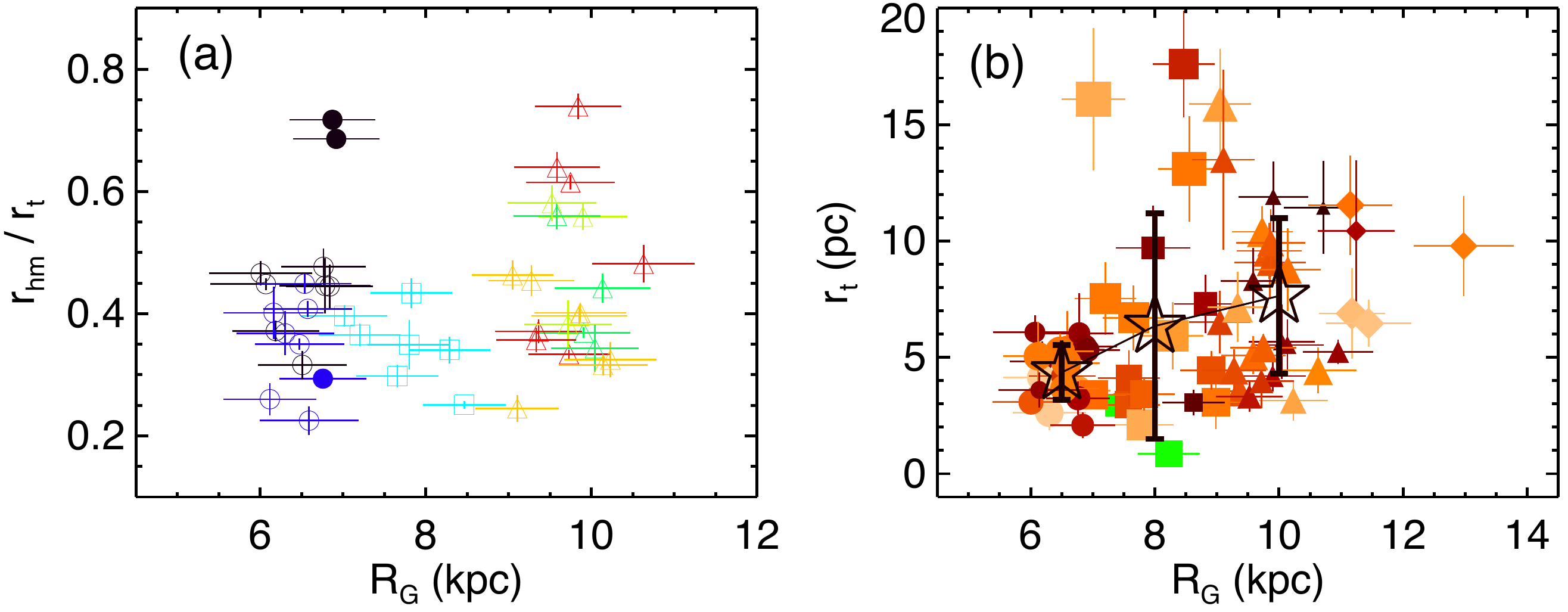}    
    \end{center}    
  }
\caption{  Panel (a): same as panel (d) of Fig.~\ref{Fig:discussions_bloco2}, but showing the groups of coeval OCs delimited in Fig.~\ref{Fig:R_gal_versus_age_groups}. The filled black circles represent the OCs NGC\,5617 and Trumpler\,22. Pismis\,19 is represented by a filled blue circle. Panel (b): $r_t$ versus $R_G$ plot. The mean and dispersion of the $r_t$ values for the same $R_G$ bins of panel (a) are represented, respectively, by the vertical position of the open black stars and the corresponding error bars. }

\label{Fig:discussions_bloco3}
\end{center}
\end{figure*}

%The general decrease in $r_c$ as a stellar group dynamically evolves was also verified by \cite{Ferraro:2019}, who investigated a sample of 5 coeval and old ($t\sim13\,$Gyr) LMC clusters and determined their dynamical states from the level of central segregation of blue stragglers. As shown in their figure 5, clusters with larger $r_c$ are dynamically younger than more compact systems. Analogous trends were also found for a sample of 48 old and coeval Galactic globular clusters (see \citeauthor{Ferraro:2019}\,\,\citeyear{Ferraro:2019} and references therein).   

In this context, panel (b) of Fig.~\ref{Fig:discussions_bloco2} exhibits part of our complete sample divided in 7 groups containing coeval systems sharing nearly the same $R_G$. Symbol colours in this panel were given according to the groups highlighted in Fig.~\ref{Fig:R_gal_versus_age_groups}. For groups of clusters located at compatible $R_G$, and thus subject to almost the same external tidal field, we can note that the data dispersion in Fig.~\ref{Fig:discussions_bloco2}, panel (b), is related to the clusters ages. For example, the black group is composed by dynamically younger clusters compared to the dark blue group. Similar statements can be drawn for the 4 groups located at $\langle R_G\rangle=9.7\pm0.4\,(1\,\sigma)\,$kpc (Fig.~\ref{Fig:R_gal_versus_age_groups}): the dark and light green groups are less evolved than the red and yellow ones.

In this same panel, comparisons between groups of almost the same age, but located at different $R_G$, allow some insights regarding the role of the Galactic tidal field on the cluster's internal structure. Despite the considerable differences in $R_G$, there is not a clear distinction in the dispersion of $r_c$ values between the black and light green groups, for which $\langle \textrm{log}\,t\rangle=8.1\pm0.2\,(1\,\sigma)$ (Fig.~\ref{Fig:R_gal_versus_age_groups}). The yellow, red, dark and light blue groups have reasonably comparable ages: $\langle \textrm{log}\,t\rangle=8.9\pm0.2\,(1\,\sigma)$. We can note that their data in panel (b) are not clearly segregated according to their $R_G$. The range in $r_c$ covered by the light blue, red and yellow groups is almost the same (except for Collinder\,110, which is part of the red group and presents $r_c\sim6.3\,$pc), even with a difference in $R_G$ of $\sim2\,$kpc among them. It is possible to state that, in fact, the anticorrelation between $r_c$ and $\tau_{\textrm{dyn}}$ presented in panels (a) and (b) of Fig.~\ref{Fig:discussions_bloco2} is mainly consequence of internal interactions.

%In this same panel, comparisons between groups of almost the same age, but located at different $R_G$, allow some insights regarding the role of the Galactic tidal field on the cluster's internal structure. Despite the considerable differences in $R_G$, there is not a clear distinction in the dispersion of $r_c$ values between the black and light green groups, for which $\langle \textrm{log}\,t\rangle=8.1\pm0.2\,(1\,\sigma)$ (Fig.~\ref{Fig:R_gal_versus_age_groups}). The yellow, red, dark and light blue groups have reasonably comparable ages: $\langle \textrm{log}\,t\rangle=8.9\pm0.2\,(1\,\sigma)$. We can note that their data in panel (b) are not clearly segregated according to their $R_G$. The range in $r_c$ covered by the light blue, red and yellow groups is almost the same (except for Collinder\,110, which is part of the red group and presents $r_c\sim6.3\,$pc), even with a difference in $R_G$ of $\sim2\,$kpc among them. To the light of the results presented by \cite{Miholics:2014}, it is possible to state that, in fact, the anticorrelation between $r_c$ and $\tau_{\textrm{dyn}}$ presented in panels (a) and (b) of Fig.~\ref{Fig:discussions_bloco2} is mainly consequence of internal interactions. 

Complementary to the above statements, we can not completely disregard that the Galactic tidal field could also play a role. OCs in the dark blue group (panel (b) of Fig.~\ref{Fig:discussions_bloco2}) tend to have smaller $r_c$ values ($r_c\lesssim1.3\,$pc). These objects are among the more evolved ones in our sample and, since they are subject to more intense tidal stresses (Fig.~\ref{Fig:R_gal_versus_age_groups}), their more compact internal structures favour their survival against tidal disruption. Compared to this group, other groups with compatible ages (that is, the red, yellow and light blue groups; Fig.~\ref{Fig:R_gal_versus_age_groups}) are located at larger $R_G$ and therefore the internal interactions can lead these OCs to relax their central stellar content across larger $r_c$ without being tidally disrupted. This is the case of Collinder\,110, which is located at a relatively large $R_G$ ($\sim9.8\,$kpc). Additionally we can infer that, within each group of coeval OCs highlighted in panel (b), differences in the  dynamical states among OCs may be due to different initial conditions at clusters formation.

%The half-mass radius are clearly correlated with the core radius (Peason's correlation coefficient $\rho=97\%$), as shown in panel (c) of Fig.~\ref{Fig:discussions_bloco2}. The linear fit (continuous line) in this log-log plane resulted $\textrm{log}\,(r_{hm})=(0.22\pm0.01)+(0.86\pm0.03)\,\times\,\textrm{log}\,(r_c)$. The uncertainties are represented by the dashed lines in the plot. This was the same relation derived in paper I, but now with a larger sample. It is noticeable that our complete sample has been employed in the fit, which means that the derived relation is not restricted to particular values of $R_G$, age or dynamical state. The considerable correlation between both radii seems determined by internal interactions. \citeauthor{Miholics:2014}\,\,(\citeyear{Miholics:2014}, their figure 1) found that differences in the Galactic potential for $R_G$ between 6 and 10\,kpc result in variations of up to $\sim$20\% in $r_{hm}$ for clusters younger than 5\,Gyr and up to $\sim10\%$ for clusters younger than 4\,Gyr. Since the investigated sample present uncertainties that are typically 10\% of $r_{hm}$ (Table\,~\ref{tab:investig_sample}) and has ages typically younger than 4\,Gyr (the oldest clusters are M\,67 ($t\sim4\,$Gyr), Collinder\,261 ($t\sim5\,$Gyr) and NGC\,188 ($t\sim5\,$Gyr); see Table \ref{tab:investig_sample}), we advocate that variations in the external tidal field present minor impact on the derived relation between $r_{hm}$ and $r_c$.     

Panel (c) of Fig.~\ref{Fig:discussions_bloco2} shows that the concentration parameter for the investigated sample is negatively correlated with the $r_{hm}/r_t$ ratio. We can note that the more compact systems (larger $c$) are less influenced by the external tidal field (smaller $r_{hm}/r_t$), that is, less subject to disruption due to tidal effects. In turn, panel (d) of Fig.~\ref{Fig:discussions_bloco2} allows to verify how the $r_{hm}/r_t$ ratio is affected by the strength of the external tidal field. Although there is not a clear trend between the plotted quantities, the $r_{hm}/r_t$ values are less dispersed for cluster located at $R_G\lesssim9\,$kpc compared to those at larger $R_G$. The position of the black open stars in the plot and the associated error bars correspond, respectively, to the mean and standard deviation of $r_{hm}/r_t$ ratios for OCs in three $R_G$ bins: $R_G\leq7\,$kpc, $7<R_G\,(\textrm{kpc})\leq9$ and $R_G>9\,$kpc. Since OCs located in the larger $R_G$ bin are subject to less intense external gravitational forces, their internal stellar content can be distributed across larger fractions of $r_t$ without being tidally disrupted, which explains their larger dispersion in the plot. On the other side, smaller $r_{hm}/r_t$ values favour the survival of clusters located closer to the Galactic centre, since they are subject to a more intense gravitational field. Similar statements were drawn in Paper I, but with a less robust OCs sample. 

%The $r_c/r_{hm}$ values are also positively correlated with $r_c$ (panel (a) of Fig.~\ref{Fig:discussions_bloco3}), although this is not a tight relation; in fact, the smaller the $r_c/r_{hm}$ of an OC, the smaller its core tend to be.   

%The OCs NGC\,5617 and Trumpler\,22 (marked with light blue circles at $r_{hm}/r_t\sim0.7$ in panel (d) of Fig.~\ref{Fig:discussions_bloco2}) seem to contradict these findings, since they present considerably larger $r_{hm}/r_t$ ratios compared to the bulk of clusters located at $R_G\leq7\,$kpc. However, as discussed in Appendix\,A, these two clusters constitute a probable binary system and have been excluded from the calculation of the mean $r_{hm}/r_t$ for clusters with $R_G < 7\,$kpc. The close gravitational interactions between these 2 objects may have perturbed their internal structures, thus leading to larger $r_{hm}/r_t$. Pismis\,19 (highlighted in panel (d) at $r_{hm}/r_t\sim0.3$) is projected in the same area of the above two OCs (Fig.\,A5 in Appendix\,A). In turn, it is in a considerably more evolved dynamical state than the other two (log\,($\tau_{\textrm{dyn,P19}}$)=3.2; log\,($\tau_{\textrm{dyn,N5617}}$)=1.6; log\,($\tau_{\textrm{dyn,Tr22}}$)=1.5), presenting a much more compact internal structure. As shown in Appendix\,A, Pismis\,19 is probably not in close gravitational interaction with NGC\,5617 and Trumpler\,22.     

The OCs NGC\,5617 and Trumpler\,22 (marked with light blue circles at $r_{hm}/r_t\sim0.7$ in panel (d) of Fig.~\ref{Fig:discussions_bloco2}) seem to contradict these findings, since they present considerably larger $r_{hm}/r_t$ ratios compared to the bulk of clusters located at $R_G\leq7\,$kpc. However, these two clusters constitute a probable binary system \citep{de-La-Fuente-Marcos:2009} and therefore have been excluded from the calculation of the mean $r_{hm}/r_t$ for clusters with $R_G < 7\,$kpc. The close gravitational interactions between these 2 objects may have perturbed their internal structures, thus leading to larger $r_{hm}/r_t$. Pismis\,19 (highlighted in panel (d) at $r_{hm}/r_t\sim0.3$) is also projected in the same area of the above two OCs but is probably not in close gravitational interaction with them. It is in a considerably more evolved dynamical state than the other two (log\,($\tau_{\textrm{dyn,P19}}$)=3.2; log\,($\tau_{\textrm{dyn,N5617}}$)=1.6; log\,($\tau_{\textrm{dyn,Tr22}}$)=1.5), presenting a much more compact internal structure. Our results have confirmed that NGC\,5617 and Trumpler\,22 present almost the same heliocentric distance, age, reddening and metallicity, thus suggesting a common origin. In turn, Pismis\,19 was found to be a background object subjected to a considerable larger interstellar reddening. See Appendix A for further details.

The plot in panel (a) of Fig.~\ref{Fig:discussions_bloco3} is analogous to panel (d) of Fig.~\ref{Fig:discussions_bloco2}, but highlighting only the groups identified in Fig.~\ref{Fig:R_gal_versus_age_groups}. For groups of clusters located at similar $R_G$, there is no trend between $r_{hm}/r_t$ and the clusters ages. For example, the dispersions of $r_{hm}/r_t$ values for the black and dark blue groups (smaller $R_G$ bin; Fig.~\ref{Fig:R_gal_versus_age_groups}) are similar. Furthermore, the $r_{hm}/r_t$ dispersion for these two groups is very similar to that presented by clusters of the light blue group ($\langle R_G\rangle=7.8\pm0.5\,(1\,\sigma)\,$kpc). Analogous statements can be drawn for the 4 groups at $R_G\sim9.7\,$kpc. Again, their data are not age segregated along the $r_{hm}/r_t$ axis in this plot. These results imply that, for a given $R_G$, the internal interactions conduct the OC to relax its stellar content across the allowed volume in a way that is initially set by the conditions at cluster formation.

The effect of variations in the Galactic tidal field on the clusters' external structure can be inferred from panel (b) of Fig.~\ref{Fig:discussions_bloco3}. Mean $r_t$ values have been derived for the same $R_G$ bins employed in panel (d) of Fig.~\ref{Fig:discussions_bloco2}. OCs closer to the Galactic centre ($R_G<7\,$kpc) present smaller and considerably less dispersed $r_t$ values compared to other objects. Their more compact external structures result from the fact that they are submitted to a stronger Galactic potential. On average, as found in Paper I, it is seen an increase in the $r_t$ as we move to regions with less intense external gravitational field.

\section{Summary and concluding remarks}
\label{sec:conclusions}

In this work, we investigated the dynamical states of a set of 65 Galactic OCs: 27 of them were explored in a previous paper and 38 in the present one. Similar methods were employed in both studies. Most OCs in our sample are located at low Galactic latitudes ($\vert b\vert\lesssim10\degr$) and typically projected against dense stellar fields. For each object, the high precision astrometric and photometric data from the \textit{Gaia} DR2 catalogue allowed a proper disentanglement between both cluster and field populations. Our analysis procedure consisted in 3 basic steps: (1) a preliminary analysis, in which we search for a cluster signature in the VPD, after applying proper photometric filters to the high quality data; (2) construction of clusters' RDP, based on proper motions filtered data, and fit of King's profile to derive $r_c$ and $r_t$. Plummer's profile is employed to derive $r_{hm}$; (3) statistical analysis of the cluster and control field data in the astrometric space and construction of decontaminated CMDs.

%Our decontamination algorithm scans the 3D parameters space ($\varpi$, $\mu_{\alpha}\textrm{cos}\,\delta$, $\mu_{\delta}$) looking for overdensities in the cluster region that are statistically distinguishable from fluctuation in the control field data. Based on this strategy, membership likelihoods are attributed to stars in the cluster region. Restricting the clusters CMDs to those stars with higher likelihoods, unambigous evolutionary sequences could be recognized.

Theoretical isochrones were fitted with a semiautomated method, in order to determine the OCs' fundamental parameters ($E(B-V)$, $(m-M)_0$, log\,$t$ and $[Fe/H]$). Updated lists of member stars were also built. Although there is a somewhat agreement with previous literature studies, for some OCs we have found severe discrepancies, which suggests the need of critical review of their parameters. In turn, regarding more recent works also employing \textit{Gaia} DR2, we found consistent results. 

The investigated OCs span wide ranges in age (7.0\,$\lesssim$\,log\,$t$\,$\lesssim$\,9.7), Galactocentric distances (6\,$\lesssim$\,$R_G$\,(kpc)\,$\lesssim\,13\,$) and dynamical ratios (0.7$\,\lesssim$\,log\,($\tau_{\textrm{dyn}})\,\lesssim\,$3.6). The joint analysis of structural and time-related parameters revealed that the core radius tend to decrease with $\tau_{\textrm{dyn}}$. We noted a convergence towards smaller $r_c$ values for dynamically older OCs. Comparisons between groups of OCs presenting compatible $R_G$ and similar ages suggest that the anticorrelation between $r_c$ and $\tau_{\textrm{dyn}}$ is determined by internal forces and modulated by the external tidal field.

%The considerable correlation between $r_{hm}$ and $r_c$ seems mainly consequence of internal dynamical relaxation. Variations in the external potential cause minor impact on the derived relation between both radii. Additionally, the positive correlation verified in the $r_c/r_{hm}$ versus $r_{hm}/r_t$ plot denotes that OCs with more dynamically relaxed structures (i.e., smaller $r_c/r_{hm}$ values) tend to suffer less intense episodes of mass loss due to tidal effects, as inferred from their smaller $r_{hm}/r_t$ ratios. 

The anticorrelation between the concentration parameter ($c$) and the $r_{hm}/r_t$ ratio denotes that OCs with more compact structures (i.e., larger $c$ values) tend to suffer less intense episodes of mass loss due to tidal effects, as inferred from their smaller $r_{hm}/r_t$ ratios. In turn, the $r_{hm}/r_t$ ratio does not present noticeable trends with $R_G$, but its dispersion is greater for those OCs located at $R_G>9\,$kpc. Among the studied sample, we only found $r_{hm}/r_t>0.5$ for OCs in this $R_G$ interval (except for NGC\,5617 and Trumpler\,22, which form a probable physical binary). Since these systems are subject to less tidal stresses due to the Galactic gravitational field, the internal processes can drive stars to occupy larger fractions of $r_t$ without the clusters being tidally dirupted. In this context, differences found among objects that share almost the same age and $R_G$ show that the initial conditions at clusters formation (e.g., initial mass function, mass profile, velocities dispersion, etc) also play a role.

Unlike the clusters' core, their external structure is more affected by the external gravitational potential. Clusters at $R_G<7\,$kpc present significantly smaller and less dispersed $r_t$ values. On average, $r_t$ increases with $R_G$. This result is physically consistent, as clusters subject to weaker external tidal fields can extend their gravitational influence over greater distances compared to OCs closer to the Galactic centre.

The high precision and spatial coverage of data from the \textit{Gaia} mission allows the proper characterization of an increasingly large number of Galactic OCs. This way, we can expect sucessively better observational inputs to evolutionary models dedicated to trace a more precise scenario for the OCs evolution.

%%%%%%%%%%%%%%%%%%%%%%%%%%%%%%%%%%%%%%%%%%%%%%%

\section{Acknowledgments}

The authors thank the anonymous referee for useful suggestions, which helped improving the clarity of the paper. This research has made use of the VizieR catalogue access tool, CDS, Strasbourg, France. This research has made use of the SIMBAD database, operated at CDS, Strasbourg, France. This work has made use of data from the European Space Agency (ESA) mission \textit{Gaia} (https://www.cosmos.esa.int/gaia), processed by the \textit{Gaia} Data Processing and Analysis Consortium (DPAC, https://www.cosmos.esa.int/web/gaia/dpac/consortium). Funding for the DPAC has been provided by national institutions, in particular the institutions participating in the \textit{Gaia} Multilateral Agreement. This research has made use of \textit{Aladin sky atlas} developed at CDS, Strasbourg Observatory, France. The authors thank the Brazilian financial agencies CNPq and FAPEMIG. This study was financed in part by the Coordena\c{c}\~ao de Aperfei\c{c}oamento de Pessoal de N\'ivel Superior $-$ Brazil (CAPES) $-$ Finance Code 001.

\subsection*{Data availability}
\textit{The data underlying this article are available in the article and in its online supplementary material.}

%{\footnotesize
\bibliographystyle{mn2e}
\bibliography{referencias}

\begin{thebibliography}{}

\bibitem[\protect\citeauthoryear{{Angelo}, {Santos} \& {Corradi}}{{Angelo}
  et~al.}{2020}]{Angelo:2020}
{Angelo} M.~S.,  {Santos} J.~F.~C.,    {Corradi} W.~J.~B.,  2020, \mnras, 493,
  3473 (Paper I)

\bibitem[\protect\citeauthoryear{{Angelo}, {Santos}, {Corradi} \&
  {Maia}}{{Angelo} et~al.}{19a }]{Angelo:2019a}
{Angelo} M.~S.,  {Santos} J.~F.~C.,  {Corradi} W.~J.~B.,    {Maia} F.~F.~S.,  {
  2019a }, \aap, 624, A8

\bibitem[\protect\citeauthoryear{{Arenou}, {Luri}, {Babusiaux}, {Fabricius} \&
  {et al.}}{{Arenou} et~al.}{2018}]{Arenou:2018}
{Arenou} F.,  {Luri} X.,  {Babusiaux} C.,  {Fabricius} C.,    {et al.} 2018,
  \aap, 616, A17

\bibitem[\protect\citeauthoryear{{Bica}, {Bonatto} \& {Dutra}}{{Bica}
  et~al.}{2004}]{Bica:2004}
{Bica} E.,  {Bonatto} C.,    {Dutra} C.~M.,  2004, \aap, 422, 555

\bibitem[\protect\citeauthoryear{{Bonatto} \& {Bica}}{{Bonatto} \&
  {Bica}}{2007}]{Bonatto:2007}
{Bonatto} C.,  {Bica} E.,  2007, \mnras, 377, 1301

\bibitem[\protect\citeauthoryear{{Bonatto} \& {Bica}}{{Bonatto} \&
  {Bica}}{2008}]{Bonatto:2008}
{Bonatto} C.,  {Bica} E.,  2008, \aap, 491, 767

\bibitem[\protect\citeauthoryear{{Bonatto} \& {Bica}}{{Bonatto} \&
  {Bica}}{2010}]{Bonatto:2010}
{Bonatto} C.,  {Bica} E.,  2010, \mnras, 407, 1728

\bibitem[\protect\citeauthoryear{{Bonatto}, {Bica} \& {Pavani}}{{Bonatto}
  et~al.}{2004}]{Bonatto:2004a}
{Bonatto} C.,  {Bica} E.,    {Pavani} D.~B.,  2004, \aap, 427, 485

\bibitem[\protect\citeauthoryear{{Bonatto}, {Kerber}, {Bica} \&
  {Santiago}}{{Bonatto} et~al.}{2006}]{Bonatto:2006}
{Bonatto} C.,  {Kerber} L.~O.,  {Bica} E.,    {Santiago} B.~X.,  2006, \aap,
  446, 121

\bibitem[\protect\citeauthoryear{{Bonfanti}, {Ortolani} \&
  {Nascimbeni}}{{Bonfanti} et~al.}{2016}]{Bonfanti:2016}
{Bonfanti} A.,  {Ortolani} S.,    {Nascimbeni} V.,  2016, \aap, 585, A5

\bibitem[\protect\citeauthoryear{{Bressan}, {Marigo}, {Girardi}, {Salasnich},
  {Dal Cero}, {Rubele} \& {Nanni}}{{Bressan} et~al.}{2012}]{Bressan:2012}
{Bressan} A.,  {Marigo} P.,  {Girardi} L.,  {Salasnich} B.,  {Dal Cero} C.,
  {Rubele} S.,    {Nanni} A.,  2012, \mnras, 427, 127

\bibitem[\protect\citeauthoryear{{Caetano}, {Dias}, {L{\'e}pine}, {Monteiro} \&
  {et al.}}{{Caetano} et~al.}{2015}]{Caetano:2015}
{Caetano} T.~C.,  {Dias} W.~S.,  {L{\'e}pine} J.~R.~D.,  {Monteiro} H.~S.,
  {et al.} 2015, \na, 38, 31

\bibitem[\protect\citeauthoryear{{Camargo}, {Bonatto} \& {Bica}}{{Camargo}
  et~al.}{2009}]{Camargo:2009}
{Camargo} D.,  {Bonatto} C.,    {Bica} E.,  2009, \aap, 508, 211

\bibitem[\protect\citeauthoryear{{Camargo}, {Bonatto} \& {Bica}}{{Camargo}
  et~al.}{2010}]{Camargo:2010}
{Camargo} D.,  {Bonatto} C.,    {Bica} E.,  2010, \aap, 521, A42

\bibitem[\protect\citeauthoryear{{Cantat-Gaudin}, {Jordi}, {Vallenari},
  {Bragaglia}, {Balaguer-N{\'u}{\~n}ez}, {Soubiran}, {Bossini}, {Moitinho},
  {Castro-Ginard}, {Krone-Martins}, {Casamiquela}, {Sordo} \&
  {Carrera}}{{Cantat-Gaudin} et~al.}{018b}]{Cantat-Gaudin:2018b}
{Cantat-Gaudin} T.,  {Jordi} C.,  {Vallenari} A.,  {Bragaglia} A.,
  {Balaguer-N{\'u}{\~n}ez} L.,  {Soubiran} C.,  {Bossini} D.,  {Moitinho} A.,
  {Castro-Ginard} A.,  {Krone-Martins} A.,  {Casamiquela} L.,  {Sordo} R.,
  {Carrera} R.,  2018b, \aap, 618, A93 (CJV2018)

\bibitem[\protect\citeauthoryear{{Cantat-Gaudin}, {Vallenari}, {Sordo},
  {Pensabene}, {Krone-Martins}, {Moitinho}, {Jordi}, {Casamiquela},
  {Balaguer-N{\'u}nez}, {Soubiran} \& {Brouillet}}{{Cantat-Gaudin}
  et~al.}{018a}]{Cantat-Gaudin:2018a}
{Cantat-Gaudin} T.,  {Vallenari} A.,  {Sordo} R.,  {Pensabene} F.,
  {Krone-Martins} A.,  {Moitinho} A.,  {Jordi} C.,  {Casamiquela} L.,
  {Balaguer-N{\'u}nez} L.,  {Soubiran} C.,    {Brouillet} N.,  2018a, \aap,
  615, A49

\bibitem[\protect\citeauthoryear{{Carraro} \& {Chiosi}}{{Carraro} \&
  {Chiosi}}{1994}]{Carraro:1994}
{Carraro} G.,  {Chiosi} C.,  1994, \aap, 288, 751

\bibitem[\protect\citeauthoryear{{Carraro}, {Janes}, {Costa} \&
  {M{\'e}ndez}}{{Carraro} et~al.}{2006}]{Carraro:2006}
{Carraro} G.,  {Janes} K.~A.,  {Costa} E.,    {M{\'e}ndez} R.~A.,  2006,
  \mnras, 368, 1078

\bibitem[\protect\citeauthoryear{{Carraro}, {Janes} \& {Eastman}}{{Carraro}
  et~al.}{2005}]{Carraro:2005a}
{Carraro} G.,  {Janes} K.~A.,    {Eastman} J.~D.,  2005, \mnras, 364, 179

\bibitem[\protect\citeauthoryear{{Castro-Ginard}, {Jordi}, {Luri}, {{\'A}lvarez
  Cid-Fuentes}, {Casamiquela}, {Anders}, {Cantat-Gaudin}, {Mongui{\'o}},
  {Balaguer-N{\'u}{\~n}ez}, {Sol{\`a}} \& {Badia}}{{Castro-Ginard}
  et~al.}{2020}]{Castro-Ginard:2020}
{Castro-Ginard} A.,  {Jordi} C.,  {Luri} X.,  {{\'A}lvarez Cid-Fuentes} J.,
  {Casamiquela} L.,  {Anders} F.,  {Cantat-Gaudin} T.,  {Mongui{\'o}} M.,
  {Balaguer-N{\'u}{\~n}ez} L.,  {Sol{\`a}} S.,    {Badia} R.~M.,  2020, \aap,
  635, A45

\bibitem[\protect\citeauthoryear{{Castro-Ginard}, {Jordi}, {Luri}, {Julbe},
  {Morvan}, {Balaguer-N{\'u}{\~n}ez} \& {Cantat-Gaudin}}{{Castro-Ginard}
  et~al.}{2018}]{Castro-Ginard:2018}
{Castro-Ginard} A.,  {Jordi} C.,  {Luri} X.,  {Julbe} F.,  {Morvan} M.,
  {Balaguer-N{\'u}{\~n}ez} L.,    {Cantat-Gaudin} T.,  2018, \aap, 618, A59

\bibitem[\protect\citeauthoryear{{de La Fuente Marcos}}{{de La Fuente
  Marcos}}{1997}]{de-La-Fuente-Marcos:1997}
{de La Fuente Marcos} R.,  1997, \aap, 322, 764

\bibitem[\protect\citeauthoryear{{de La Fuente Marcos} \& {de La Fuente
  Marcos}}{{de La Fuente Marcos} \& {de La Fuente
  Marcos}}{2009}]{de-La-Fuente-Marcos:2009}
{de La Fuente Marcos} R.,  {de La Fuente Marcos} C.,  2009, \aap, 500, L13

\bibitem[\protect\citeauthoryear{{de la Fuente Marcos}, {de la Fuente Marcos},
  {Moni Bidin}, {Carraro} \& {Costa}}{{de la Fuente Marcos}
  et~al.}{2013}]{de-la-Fuente-Marcos:2013}
{de la Fuente Marcos} R.,  {de la Fuente Marcos} C.,  {Moni Bidin} C.,
  {Carraro} G.,    {Costa} E.,  2013, \mnras, 434, 194

\bibitem[\protect\citeauthoryear{{Dias}, {Alessi}, {Moitinho} \&
  {L{\'e}pine}}{{Dias} et~al.}{2002}]{Dias:2002}
{Dias} W.~S.,  {Alessi} B.~S.,  {Moitinho} A.,    {L{\'e}pine} J.~R.~D.,  2002,
  \aap, 389, 871 (DAML02)

\bibitem[\protect\citeauthoryear{{Dias}, {Monteiro}, {L{\'e}pine} \&
  {Barros}}{{Dias} et~al.}{2019}]{Dias:2019}
{Dias} W.~S.,  {Monteiro} H.,  {L{\'e}pine} J.~R.~D.,    {Barros} D.~A.,  2019,
  \mnras, 486, 5726

\bibitem[\protect\citeauthoryear{{Dias}, {Monteiro}, {L{\'e}pine}, {Prates},
  {Gneiding} \& {Sacchi}}{{Dias} et~al.}{2018}]{Dias:2018}
{Dias} W.~S.,  {Monteiro} H.,  {L{\'e}pine} J.~R.~D.,  {Prates} R.,  {Gneiding}
  C.~D.,    {Sacchi} M.,  2018, \mnras, 481, 3887

\bibitem[\protect\citeauthoryear{{Evans}, {Riello}, {De Angeli}, {Carrasco} \&
  {et al.}}{{Evans} et~al.}{2018}]{Evans:2018}
{Evans} D.~W.,  {Riello} M.,  {De Angeli} F.,  {Carrasco} J.~M.,    {et al.}
  2018, \aap, 616, A4

\bibitem[\protect\citeauthoryear{{Ferreira}, {Santos}, {Corradi}, {Maia} \&
  {Angelo}}{{Ferreira} et~al.}{2019}]{Ferreira:2019}
{Ferreira} F.~A.,  {Santos} J.~F.~C.,  {Corradi} W.~J.~B.,  {Maia} F.~F.~S.,
  {Angelo} M.~S.,  2019, \mnras, 483, 5508

\bibitem[\protect\citeauthoryear{{Gaia Collaboration}, {Brown}, {Vallenari},
  {Prusti}, {de Bruijne}, {Babusiaux}, {Bailer-Jones}, {Biermann}, {Evans} \&
  {Eyer}}{{Gaia Collaboration} et~al.}{2018}]{Gaia-Collaboration:2018}
{Gaia Collaboration} {Brown} A.~G.~A.,  {Vallenari} A.,  {Prusti} T.,  {de
  Bruijne} J.~H.~J.,  {Babusiaux} C.,  {Bailer-Jones} C.~A.~L.,  {Biermann} M.,
   {Evans} D.~W.,    {Eyer} L.,  2018, \aap, 616, A1

\bibitem[\protect\citeauthoryear{{Gilmore}, {Randich}, {Asplund}, {Binney},
  {Bonifacio} \& {et al.}}{{Gilmore} et~al.}{2012}]{Gilmore:2012}
{Gilmore} G.,  {Randich} S.,  {Asplund} M.,  {Binney} J.,  {Bonifacio} P.,
  {et al.} 2012, The Messenger, 147, 25

\bibitem[\protect\citeauthoryear{{Girardi}, {Bertelli}, {Bressan}, {Chiosi},
  {Groenewegen}, {Marigo}, {Salasnich} \& {Weiss}}{{Girardi}
  et~al.}{2002}]{Girardi:2002}
{Girardi} L.,  {Bertelli} G.,  {Bressan} A.,  {Chiosi} C.,  {Groenewegen}
  M.~A.~T.,  {Marigo} P.,  {Salasnich} B.,    {Weiss} A.,  2002, \aap, 391, 195

\bibitem[\protect\citeauthoryear{{Heggie} \& {Hut}}{{Heggie} \&
  {Hut}}{2003}]{Heggie:2003}
{Heggie} D.,  {Hut} P.,  2003, {The Gravitational Million-Body Problem: A
  Multidisciplinary Approach to Star Cluster Dynamics}.
Cambridge University Press

\bibitem[\protect\citeauthoryear{{Kharchenko}, {Piskunov}, {Schilbach},
  {R{\"o}ser} \& {Scholz}}{{Kharchenko} et~al.}{2013}]{Kharchenko:2013}
{Kharchenko} N.~V.,  {Piskunov} A.~E.,  {Schilbach} E.,  {R{\"o}ser} S.,
  {Scholz} R.-D.,  2013, \aap, 558, A53

\bibitem[\protect\citeauthoryear{{King}}{{King}}{1962}]{King:1962}
{King} I.,  1962, Astronomical Journal, 67, 471

\bibitem[\protect\citeauthoryear{{Koposov}, {Glushkova} \&
  {Zolotukhin}}{{Koposov} et~al.}{2008}]{Koposov:2008}
{Koposov} S.~E.,  {Glushkova} E.~V.,    {Zolotukhin} I.~Y.,  2008, \aap, 486,
  771

\bibitem[\protect\citeauthoryear{{Krone-Martins} \& {Moitinho}}{{Krone-Martins}
  \& {Moitinho}}{2014}]{Krone-Martins:2014}
{Krone-Martins} A.,  {Moitinho} A.,  2014, \aap, 561, A57

\bibitem[\protect\citeauthoryear{{Krumholz}, {McKee} \& {Bland
  -Hawthorn}}{{Krumholz} et~al.}{2019}]{Krumholz:2019}
{Krumholz} M.~R.,  {McKee} C.~F.,    {Bland -Hawthorn} J.,  2019, \araa, 57,
  227

\bibitem[\protect\citeauthoryear{{Liu} \& {Pang}}{{Liu} \&
  {Pang}}{2019}]{Liu:2019}
{Liu} L.,  {Pang} X.,  2019, \apjs, 245, 32

\bibitem[\protect\citeauthoryear{{Lloyd}}{{Lloyd}}{1982}]{Lloyd:1982}
{Lloyd} S.,  1982, IEEE Transactions on Information Theory, 28, 129

\bibitem[\protect\citeauthoryear{{Luri}, {Brown}, {Sarro}, {Arenou},
  {Bailer-Jones}, {Castro-Ginard}, {de Bruijne}, {Prusti}, {Babusiaux} \&
  {Delgado}}{{Luri} et~al.}{2018}]{Luri:2018}
{Luri} X.,  {Brown} A.~G.~A.,  {Sarro} L.~M.,  {Arenou} F.,  {Bailer-Jones}
  C.~A.~L.,  {Castro-Ginard} A.,  {de Bruijne} J.,  {Prusti} T.,  {Babusiaux}
  C.,    {Delgado} H.~E.,  2018, \aap, 616, A9

\bibitem[\protect\citeauthoryear{{Maia}, {Corradi} \& {Santos} Jr.}{{Maia}
  et~al.}{2010}]{Maia:2010}
{Maia} F.~F.~S.,  {Corradi} W.~J.~B.,    {Santos} Jr. J.~F.~C.,  2010, \mnras,
  407, 1875

\bibitem[\protect\citeauthoryear{{Netopil}, {Paunzen}, {Heiter} \&
  {Soubiran}}{{Netopil} et~al.}{2016}]{Netopil:2016}
{Netopil} M.,  {Paunzen} E.,  {Heiter} U.,    {Soubiran} C.,  2016, \aap, 585,
  A150

\bibitem[\protect\citeauthoryear{{Ostriker}, {Spitzer} Lyman \&
  {Chevalier}}{{Ostriker} et~al.}{1972}]{Ostriker:1972}
{Ostriker} J.~P.,  {Spitzer} Lyman J.,    {Chevalier} R.~A.,  1972, \apjl, 176,
  L51

\bibitem[\protect\citeauthoryear{{Overbeek}, {Friel}, {Donati}, {Smiljanic},
  {Jacobson} \& {et al.}}{{Overbeek} et~al.}{2017}]{Overbeek:2017}
{Overbeek} J.~C.,  {Friel} E.~D.,  {Donati} P.,  {Smiljanic} R.,  {Jacobson}
  H.~R.,    {et al.} 2017, \aap, 598, A68

\bibitem[\protect\citeauthoryear{{Perren}, {V{\'a}zquez} \& {Piatti}}{{Perren}
  et~al.}{2015}]{Perren:2015}
{Perren} G.~I.,  {V{\'a}zquez} R.~A.,    {Piatti} A.~E.,  2015, \aap, 576, A6

\bibitem[\protect\citeauthoryear{{Piatti}, {Dias} \& {Sampedro}}{{Piatti}
  et~al.}{2017}]{Piatti:2017a}
{Piatti} A.~E.,  {Dias} W.~S.,    {Sampedro} L.~M.,  2017, \mnras, 466, 392

\bibitem[\protect\citeauthoryear{{Plummer}}{{Plummer}}{1911}]{Plummer:1911}
{Plummer} H.~C.,  1911, \mnras, 71, 460

\bibitem[\protect\citeauthoryear{{Portegies Zwart}, {McMillan} \&
  {Gieles}}{{Portegies Zwart} et~al.}{2010}]{Portegies-Zwart:2010}
{Portegies Zwart} S.~F.,  {McMillan} S.~L.~W.,    {Gieles} M.,  2010, \araa,
  48, 431

\bibitem[\protect\citeauthoryear{{Randich}, {Sestito}, {Primas}, {Pallavicini}
  \& {Pasquini}}{{Randich} et~al.}{2006}]{Randich:2006}
{Randich} S.,  {Sestito} P.,  {Primas} F.,  {Pallavicini} R.,    {Pasquini} L.,
   2006, \aap, 450, 557

\bibitem[\protect\citeauthoryear{{Reid}}{{Reid}}{1993}]{Reid:1993a}
{Reid} M.~J.,  1993, \araa, 31, 345

\bibitem[\protect\citeauthoryear{{Rieke} \& {Lebofsky}}{{Rieke} \&
  {Lebofsky}}{1985}]{Rieke:1985}
{Rieke} G.~H.,  {Lebofsky} M.~J.,  1985, \apj, 288, 618

\bibitem[\protect\citeauthoryear{{Ryu} \& {Lee}}{{Ryu} \&
  {Lee}}{2018}]{Ryu:2018}
{Ryu} J.,  {Lee} M.~G.,  2018, \apj, 856, 152

\bibitem[\protect\citeauthoryear{{Sagar} \& {Bhatt}}{{Sagar} \&
  {Bhatt}}{1989}]{Sagar:1989}
{Sagar} R.,  {Bhatt} H.~C.,  1989, \mnras, 236, 865

\bibitem[\protect\citeauthoryear{{Sim}, {Lee}, {Ann} \& {Kim}}{{Sim}
  et~al.}{2019}]{Sim:2019}
{Sim} G.,  {Lee} S.~H.,  {Ann} H.~B.,    {Kim} S.,  2019, Journal of Korean
  Astronomical Society, 52, 145

\bibitem[\protect\citeauthoryear{{Skrutskie}, {Cutri}, {Stiening}, {Weinberg}
  \& {et al.}}{{Skrutskie} et~al.}{2006}]{Skrutskie:2006}
{Skrutskie} M.~F.,  {Cutri} R.~M.,  {Stiening} R.,  {Weinberg} M.~D.,    {et
  al.} 2006, \aj, 131, 1163

\bibitem[\protect\citeauthoryear{{Smith}}{{Smith}}{2014}]{Smith:2014}
{Smith} N.,  2014, \araa, 52, 487

\bibitem[\protect\citeauthoryear{{Spitzer} Jr.}{{Spitzer}}{1958}]{Spitzer:1958}
{Spitzer} Jr. L.,  1958, \apj, 127, 17

\bibitem[\protect\citeauthoryear{{Spitzer} Jr.}{{Spitzer}}{1969}]{Spitzer:1969}
{Spitzer} Jr. L.,  1969, \apjl, 158, L139

\bibitem[\protect\citeauthoryear{{Theuns}}{{Theuns}}{1991}]{Theuns:1991}
{Theuns} T.,  1991, \memsai, 62, 909

\bibitem[\protect\citeauthoryear{{Torrealba}, {Belokurov} \&
  {Koposov}}{{Torrealba} et~al.}{2019}]{Torrealba:2019}
{Torrealba} G.,  {Belokurov} V.,    {Koposov} S.~E.,  2019, \mnras, 484, 2181

\bibitem[\protect\citeauthoryear{{Vall{\'e}e}}{{Vall{\'e}e}}{2008}]{Vallee:2008}
{Vall{\'e}e} J.~P.,  2008, \aj, 135, 1301

\bibitem[\protect\citeauthoryear{{van Altena}}{{van
  Altena}}{2013}]{van-Altena:2013}
{van Altena} W.~F.,  2013, {Astrometry for Astrophysics}.
Cambridge University Press

\bibitem[\protect\citeauthoryear{{Vink}, {de Koter} \& {Lamers}}{{Vink}
  et~al.}{2001}]{Vink:2001}
{Vink} J.~S.,  {de Koter} A.,    {Lamers} H.~J.~G.~L.~M.,  2001, \aap, 369, 574

\end{thebibliography}
%}

%\newpage

\appendix

%%%%%%%%%%%%%%%%%%%%%%%%%%%%%%%%%%%
\section{Individual comments on some studied OCs}
\label{comments_some_clusters}
%%%%%%%%%%%%%%%%%%%%%%%%%%%%%%%%%%%
%We dedicate this Appendix to highlight particular comments on some OCs in our main sample (Table\,1): (i) Ruprecht\,26, for which we found results that are discrepant (see Fig.\,7) in relation to recent studies; (ii) NGC\,5617, Pismis\,19 and Trumpler\,22, which may constitute a multiple physical system; (iii) Ruprecht\,152 and the massive cluster NGC\,2477, which are projected close to each other; (iv) Lynga\,9, which was previously characterized as an asterism; (v) Trumpler\,23, which was observed spectroscopically in a previous study. Appendix A is available only in the supplementary material.

We dedicate this Appendix to highlight particular comments on some OCs: (i) Ruprecht 26, for which we found results that are discrepant (see Fig.~\ref{fig:compara_com_CGaudin}) in relation to recent studies; (ii) NGC\,5617, Pismis\,19 and Trumpler\,22, which may constitute a multiple physical system. Appendix A is available in the online supplementary material.

%%%%%%%%%%%%%%%%%%%%%%%%%%%%%%%%%%%%%%%%%%%
%\section{Previous literature information on the investigated sample}
%\label{appendix:previous_lit_information}
%%%%%%%%%%%%%%%%%%%%%%%%%%%%%%%%%%%%%%%%%%%

%In this Appendix, we list the fundamental parameters ($E(B-V)$, $(m-M)_0$ and log\,$t$) found in the literature (Bica, Bonatto \& Camargo's series of papers) for OCs in our main sample (see also Table\,1 and Fig.\,8). In Table\,B1, we inform the specific reference for each cluster and if it has been analysed in Paper I or in the present one. Appendix B is available only in the supplementary material.

%%%%%%%%%%%%%%%%%%%%%%%%%%%%%%%%%%%%%%%%%%%
\section{Supplementary figures}
\label{appendix:suppl_figures}
%%%%%%%%%%%%%%%%%%%%%%%%%%%%%%%%%%%%%%%%%%%

This appendix (available in the online supplementary material) contains the whole sets of plots (RDPs, CMDs, VPDs and $\varpi$ versus $G$ mag plots) not shown in the main text and neither in Appendix A.

%This Appendix contains the whole sets of plots (RDPs, CMDs, VPDs and $\varpi$\,versus $G\,$mag plots; Figs.\,B1 to B24) not shown in the main text and neither in the previous appendix. Appendix B is available only in the supplementary material.

\bsp

\label{lastpage}

\end{document}